\documentclass[a4paper,11pt]{article}
\pdfoutput=1
\usepackage[T1]{fontenc}
\usepackage{graphicx}
\usepackage{amsmath}
\usepackage{xspace}
\usepackage[numbers,sort&compress]{natbib}
\usepackage{subfigure}
\usepackage[shortlabels]{enumitem}
\usepackage{slashed}
\usepackage{array}

\usepackage{pstricks}

\newcommand\wpl{\ell^+\nu_{\sss\ell}}
\newcommand\wml{\ell^-\bar\nu_{\sss\ell}}
\newcommand\zll{\ell^+\ell^-}

\newcommand\sw{s_{\mathrm{w}}}
\newcommand\cw{c_{\mathrm{w}}}
\newcommand\f{\varphi}

\newcommand\Cew{C^{\rm ew}}
\newcommand\Iz{I^{\sss Z}}
\newcommand\aofpi{\frac{\aem}{4\pi}}
\newcommand\de{\delta}
\newcommand\ka{\kappa}

\usepackage{jheppub}
\hypersetup{
  pdfauthor={F. Granata, J. M. Lindert, S. Pozzorini, C. Oleari},
}

\newcommand\NLOEW{NLO EW}

\newcommand\sNLOQCD{\s_{\sss \rm QCD}}
\newcommand\sNLOQCDEW{\s_{\sss \rm QCD+EW}}
\newcommand\sNLOQCDEWNLL{\s_{\sss \rm QCD+NLL\,EW}}

\newcommand\HW{\ensuremath{HW}}
\newcommand\HWJ{\ensuremath{HW\!j}}
\newcommand\HZ{\ensuremath{H\hspace{-0.7pt}Z}}
\newcommand\HZJ{\ensuremath{H\hspace{-0.7pt}Zj}}
\newcommand\HV{\ensuremath{HV}}
\newcommand\HVJ{\ensuremath{HV\!j}}
\newcommand\HWm{\ensuremath{HW^-}}
\newcommand\HWp{\ensuremath{HW^+}}
\newcommand\HWpm{\ensuremath{HW^\pm\!}}
\newcommand\HWmJ{\ensuremath{HW^-\!j}}
\newcommand\HWpJ{\ensuremath{HW^+\!j}}
\newcommand\HWpmJ{\ensuremath{HW^\pm\!j}}
\newcommand\leplv{$\ensuremath{\ell\nu_{\sss\ell}}$}
\newcommand\lepll{$\ensuremath{\ell^+\ell^-}$}

\newcommand\Hleplvj{$\ensuremath{H\ell\nu_{\sss \ell}+{\rm jet}}$}
\newcommand\Hlepllj{$\ensuremath{H\ell^+\ell^-+{\rm jet}}$}
\newcommand\wsmall{0.495\textwidth}
\newcommand\wmedium{0.6\textwidth}

\newcommand\CKM{V^{\sss\rm CKM}}

\newcommand\HAWK{{\tt HAWK}}
\newcommand\POWHEGBOX{{\tt POWHEG~\!BOX}}
\newcommand\POWHEG{{\tt POWHEG}}
\newcommand\MCatNLO{{\tt MC@NLO}}

\newcommand\PythiaEightPone{{\tt Pythia~8.1}}

\newcommand\Py{{\tt Pythia}}
\newcommand\MadGraphFour{{\tt MadGraph4}}
\newcommand\RES{{\tt POWHEG~\!BOX~\!RES}}

\newcommand\VTWO{{\tt POWHEG~\!BOX~\!V2}}

\newcommand{\OpenLoops}{{\tt OpenLoops}\xspace}

\newcommand\Collier{{\tt COLLIER}\xspace}

\newcommand{\GeV}{\text{GeV}\xspace}

\newcommand{\GF}{{G_{\sss\mu}}}

\newcommand{\kt}{\ensuremath{k_{\rm\sss T}}\xspace}
\newcommand\sss{\mathchoice%
{\displaystyle}%
{\scriptstyle}%
{\scriptscriptstyle}%
{\scriptscriptstyle}%
}
\newcommand{\pt}{\ensuremath{p_{\sss \rm T}}\xspace}

\newcommand{\ord}{\mathcal{O}}

\newcommand{\Pt}{\ensuremath{t}\xspace}

\def\sw{s_{\mathrm{w}}}
\def\cw{c_{\mathrm{w}}}

\def\beq{\begin{equation}}
\def\beqn{\begin{eqnarray}}
\def\eeq{\end{equation}}
\def\eeqn{\end{eqnarray}}

\def\({\left(} 
\def\){\right)} 
 
\newcommand     \MSB            {\ifmmode {\overline{\rm MS}} \else
                                 $\overline{\rm MS}$\fi}

\newcommand\as{\alpha_{\sss\rm S}}

\newcommand\aem{\alpha_{\sss\rm EM}}

\newcount\minutes 
\newcount\scratch 
\def\timestamp{%
\scratch=\time 
\divide\scratch by 60 
\edef\hours{\the\scratch} 
\multiply\scratch by 60 
\minutes=\time 
\advance\minutes by -\scratch 
---$\,$\hours:\null 
\ifnum\minutes< 10 0\fi 
\the\minutes}

\def\refeq#1{\mbox{Eq.~(\ref{#1})}}

\def\reffi#1{\mbox{Fig.~\ref{#1}}}

\def\refse#1{\mbox{Sec.~\ref{#1}}}

\def\refapp#1{\mbox{App.~\ref{#1}}}

\def\citeres#1{\mbox{Refs.~\cite{#1}}}

\def\rd{\mathrm{d}}

\definecolor{mygray}{gray}{0.5}

\def\MZ{M_{\sss Z}}
\def\MW{M_{\sss W}}

\def\MH{M_{\sss H}}
\def\MV{M_{\sss V}}
\newcommand\MINLO{{\tt MiNLO}}
\newcommand\MiNLO{{\tt MiNLO}}

\newcommand\Sherpa{{\tt Sherpa}}
\newcommand\MEPSatNLO{{\tt MEPS@NLO}}

\def\g{\gamma}
\def\V{\sss V}
\def\M{\mathcal M}

\def\l{\left}
\def\r{\right}
\def\s{\sigma}

\preprint{
\begin{flushright}
IPPP/17/46 \\
ZU-TH 14/17 
\end{flushright}
}

\author[a]{F.~Granata,}
\author[b]{J.~M.~Lindert,}
\author[a]{C.~Oleari,}
\author[c]{and S.~Pozzorini}

\affiliation[a]{Universit\`a di Milano-Bicocca and INFN, Sez.~di
  Milano-Bicocca, Piazza della Scienza 3, 20126 Milano, Italy }

\affiliation[b]{Institute for Particle Physics Phenomenology, 
		Durham University, 
		South Rd, Durham DH1 3LE, 
		UK}

\affiliation[c]{Physik-Institut, Universit\"at Z\"urich, Winterthurerstrasse
  190, CH-8057 Z\"urich, Switzerland }

\emailAdd{federico.granata@unimib.it}
\emailAdd{jonas.m.lindert@durham.ac.uk}
\emailAdd{carlo.oleari@mib.infn.it}
\emailAdd{pozzorin@physik.uzh.ch}

\title{ NLO QCD+EW predictions for $\boldsymbol{\HV}$ and
  $\boldsymbol{\HV}$+jet production including parton-shower effects}

\abstract{We present the first NLO QCD+EW predictions for Higgs boson
production in association with a \leplv{} or \lepll{} pair plus zero or one
jets at the LHC.  Fixed-order NLO QCD+EW calculations are 
combined with a QCD+QED parton shower using the
recently developed resonance-aware method in the \POWHEG{} framework.
Moreover, applying the improved \MINLO{} technique to
\Hleplvj{} and \Hlepllj{} production at NLO QCD+EW, we 
obtain predictions that are 
NLO accurate for observables with both zero or one resolved jet.
This  approach permits also to capture higher-order effects associated
with the interplay of EW corrections and QCD radiation.
The behavior of EW corrections is studied for various kinematic
distributions, relevant for experimental analyses of Higgsstrahlung processes
at the 13\,TeV LHC.  Exact NLO EW corrections are complemented with
approximate analytic formulae that account for the leading and
next-to-leading Sudakov logarithms in the high-energy regime.  In the tails
of transverse-momentum distributions, relevant for analyses in the boosted
Higgs regime, the Sudakov approximation works well, and NLO EW effects can
largely exceed the ten percent level.
Our predictions are based on the \RES{}+\OpenLoops framework
in combination with the \PythiaEightPone{} parton shower.


}

\begin{document}
\maketitle
\flushbottom

\section{Introduction}
\label{sec:intro}

The discovery of the Higgs boson~\cite{Aad:2012tfa, Chatrchyan:2012xdj} has
opened the door to the direct experimental investigation of the Higgs and
Yukawa sectors of the Standard Model.  While present measurements of Higgs
boson properties and interactions are consistent with the Standard
Model~\cite{Khachatryan:2016vau}, the full set of data collected during
Run~II and in subsequent runs of the LHC will provide more and more stringent
tests of the mechanism of electroweak symmetry breaking.

In this context, the associated production of a Higgs and a vector boson,
$pp\rightarrow \HV$ with $V=W$ and $Z$, plays a prominent role.  In spite of
the fact that the total cross sections for these so-called Higgsstrahlung
processes are subleading as compared to Higgs boson production via gluon
fusion and vector-boson fusion, the possibility to reconstruct the full \HV{}
final state and the clean signatures that result from leptonically decaying
vector bosons offer unique opportunities of testing Higgs boson interactions
with vector bosons and heavy quarks  (see~\citeres{Butterworth:2008iy,Ellis:2014dva,Greljo:2015sla} and references therein).
The associated \HV{} production makes it possible to disentangle Higgs boson
couplings to $W$ and $Z$ bosons from one another and to measure them in a
broad kinematic range.  In addition, the presence of the associated vector
boson allows for an efficient suppression of QCD backgrounds.
In particular, $pp\to \HV$ is the most favorable channel for measurements of
the $H\to b\bar b$ branching ratio, and thus for determinations of the bottom
Yukawa coupling.
In \HV{} production with $H\to b\bar b$ decay, the boosted
region, with Higgs boson transverse momentum above 200~GeV, plays a
particularly important role, both in order to achieve an improved control of
the QCD backgrounds~\cite{Butterworth:2008iy} and for the sensitivity to
possible anomalies in the $HVV$ couplings.
Higgsstrahlung processes permit also to probe invisible Higgs boson decays,
both through direct measurements of $pp\to \HZ$ with invisible Higgs decays
and through indirect bounds based on measurements of the $H\to b\bar b$
branching ratio.

The accuracy of present and future measurements of \HV{} production, at the
level of both fiducial cross sections and differential distributions, calls
for increasingly accurate theoretical predictions.  The inclusion of
higher-order QCD corrections is crucial, both for total rates and for a
precise description of the QCD radiation that accompanies the production of
the \HV{} system.  The role of QCD corrections can be particularly important
in the boosted regime or in the presence of cuts and for observables that are
sensitive to QCD radiation.

In general, in order to account for experimental cuts and observables,
higher-order QCD and EW predictions should be available for arbitrary
differential distributions, and experimental analyses require particle-level
Monte Carlo generators where state-of-the-art theoretical calculations are
matched to parton showers.  Finally, when QCD and EW higher-order effects are
both sizable, also their combination needs to be addressed.

Theoretical calculations for the associated-production processes are widely
available in the literature.  
Among the numerous studies on \HV{} production at next-to-leading order~(NLO) 
QCD we quote here Refs.~\cite{Arnold:2008rz, MCFM, Campanario:2014lza}.
Predictions for inclusive \HZ{} and \HW{} production at
next-to-next-to-leading order~(NNLO) QCD have first been obtained in
Refs.~\cite{Brein:2003wg, Brein:2011vx} and are implemented in the {\tt
  VH@NNLO} program~\cite{Brein:2012ne}.
Besides contributions of Drell--Yan~(DY) type, where the Higgs boson results
from an $s$-channel $V^*\to \HV$ subtopology, Higgsstrahlung at NNLO QCD
involves also extra $\mathcal{O}(\as^2)$ contributions where the Higgs boson
couples to heavy-quark loops.  Such non-DY contributions arise via squared
one-loop amplitudes in the $gg\to \HZ$ channel~\cite{Kniehl:1990iva} and
through the interference of one-loop and tree amplitudes in the $qg\to \HV q$
and crossing-related channels.  Studies of possible anomalous coupling in the
$gg\to \HZ$ channel can be found in Ref.~\cite{Nishiwaki:2013cma,
  Hespel:2015zea}. 
Heavy-quark loop contributions to the $gg\to \HZ$ channel are known up to
$\mathcal{O}(\as^3)$ in the limit of the mass of the bottom quark going to
zero, and the mass of the top quark going to
infinity~\cite{Altenkamp:2012sx}. Their impact, especially in the boosted
regime, can be quite significant~\cite{Englert:2013vua}.

Fully differential NNLO calculations for \HV{} production with off-shell
vector-boson decays were first presented in Refs.~\cite{Ferrera:2011bk,
  Ferrera:2013yga, Ferrera:2014lca}, including all DY contributions plus
heavy-quark-loop contributions to $gg\to \HZ$.
More recently, a NNLO QCD calculation that includes also the small
heavy-quark loop contributions in the $q g\to \HV q$ channel and in the
crossing-related $\bar q g$ and $q\bar q$ channels became
available~\cite{Campbell:2016jau} and also $\HV\to b\bar{b} V $ production
with NNLO QCD corrections both in the production and in the decay part of the
process~\cite{Ferrera:2017zex}.

Analytic resummations have been discussed in Refs.~\cite{Dawson:2012gs,
  Shao:2013uba, Li:2014ria, Harlander:2014wda}, while leading-logarithmic
resummation can be routinely achieved through the matching of NLO QCD
calculations to parton showers~(PS).
The first NLO+PS generators in the \MCatNLO{}~\cite{Frixione:2002ik} and
\POWHEG{} frameworks~\cite{Nason:2004rx, Frixione:2007vw, Alioli:2010xd} have
been presented in Refs.~\cite{Frixione:2005gz} and~\cite{Hamilton:2009za},
respectively.  More recently, new generators that provide an NLO
accurate description of $\HV$ and $\HV$ + jet production became available.
The first generator of this kind was presented in Ref.~\cite{Luisoni:2013cuh}
based on the \MINLO{} method~\cite{Hamilton:2012np, Hamilton:2012rf}, while a
simulation of $pp\to \HZ$~+~0~and~1~jet, based on the \MEPSatNLO{} multijet
merging technique~\cite{Hoeche:2012yf, Gehrmann:2012yg}, was presented in
Ref.~\cite{Goncalves:2015mfa}.
Concerning fermion loops, the \POWHEGBOX{} generator of
Ref.~\cite{Luisoni:2013cuh} can account for all $\mathcal{O}(\as^2)$ NLO
contributions of DY and non-DY type to $pp \to \HV$~+~jet and also for the finite
$gg\to \HZ$ loop-induced contributions, with the possibility of studying
anomalous couplings in the ``kappa framework''.
A more general study, which uses an effective field theory approach and
introduces generic six-dimensional operators, can be found in
Ref.~\cite{Mimasu:2015nqa}.

The heavy-quark loop-mediated production $gg\to \HZ g$ was first studied in
Ref.~\cite{Agrawal:2014tqa}. More recently, the \Sherpa{} generator of
Ref.~\cite{Goncalves:2015mfa} has included also NNLO-type squared quark-loop
contributions in the $gg\to \HZ$, $gg\to \HZ g$, and $gq\to \HZ q$ plus
crossing-related channels.
Lately, a NNLO+PS generator for $pp\to \HW$~\cite{Astill:2016hpa} that
combines the NNLO QCD calculation of Ref.~\cite{Ferrera:2011bk} with the
parton shower using the method of Refs.~\cite{Hamilton:2012rf,
  Hamilton:2013fea} was presented.

Electroweak corrections to $pp\to \HV$, including off-shell $W$- and
$Z$-boson decays, are known at NLO~\cite{Ciccolini:2003jy, Denner:2011id} and
are implemented in the parton-level Monte Carlo program
\HAWK~\cite{Denner:2014cla}.  These corrections are at the level of 5\% for
inclusive quantities, but in the high-energy regime they can reach various
tens of percent due to the presence of Sudakov
logarithms~\cite{Ciafaloni:1998xg, Kuhn:1999de,Fadin:1999bq, Denner:2000jv,Beenakker:2000kb,Melles:2000gw,Denner:2003wi,Chiu:2008vv}.
For this reason, especially in boosted searches, the inclusion of EW corrections is
mandatory.  An interesting aspect of these corrections in \HV{} production is
that they induce also a dependence on the Higgs sector, and in particular on
the trilinear coupling $\lambda_{\sss HHH}$.  Thus, precise measurements of
Higgsstrahlung processes can be exploited for setting limits on
$\lambda_{\sss HHH}$~\cite{Degrassi:2016wml, Bizon:2016wgr, Degrassi:2017ucl,
  DiVita:2017eyz}.
To date, none of the existing NLO+PS generators implement EW corrections.

In this paper, for the first time, we present NLO QCD and NLO EW calculations
for the production of a Higgs boson in conjunction with a \leplv{} or
\lepll{} leptonic pair, plus zero or one jet, at the LHC.  While, for
convenience, the above-mentioned processes will often be denoted as
\HV/\HVJ{} production (with $V=W^\pm$ and $Z$) in the rest of the paper, all
the results we are going to present always correspond to the complete decayed
final-state processes, with spin effects, off-shell and non-resonant
contributions taken into account.

At NLO QCD we include the full set of $\mathcal{O}(\as)$ contributions to
$pp\to \HV$ and $\mathcal{O}(\as^2)$ contributions to $pp\to \HVJ$. Although
terms of non-DY type are implemented in our codes, we have not included them
in our simulations. In addition, we do not include NNLO-like loop-induced
contributions to \HZ{} plus 0 and 1~jet production.

Besides showing fixed-order NLO QCD+EW predictions at parton level for
typical observables, we also present full NLO+PS simulations for \HV{} and
\HVJ{} production. To this end, we have implemented our NLO calculations for
\HV{} and \HVJ{} production into four separate codes (\HWpm, \HWpmJ, \HZ{}
and \HZJ) in the \POWHEGBOX{} framework. In this way, we have consistently
combined the radiation emitted at NLO QCD+EW level with a QCD+QED parton
shower.
In this context, photon radiation from the charged leptons can lead to severe
unphysical distortions of the $Z$- and $W$-boson line shapes, if not properly
treated.  This problem was first pointed out in the context of NLO QCD+PS
simulations of off-shell top-quark production and decay, and was solved in
the context of the \POWHEGBOX{} framework by means of the so-called
resonance-aware method~\cite{Jezo:2015aia}.
The first application of this method and its variants, in the context
  of electroweak corrections, has appeared in
  Refs.~\cite{CarloniCalame:2016ouw, Muck:2016pko}.
In this paper, we exploit the flexibility of the resonance-aware method to
perform a fully consistent NLO QCD+EW matching in the presence of non-trivial
EW resonances.  To this end, our NLO calculations and generators are
implemented in the new version of the \POWHEGBOX{} framework, known as \RES.
In this recent version, the hardest radiation generated by \POWHEG{}
preserves the resonance virtualities present at the underlying-Born level. At
the same time, the resonance information can be passed on to the parton
shower, which in turn preserves the virtualities of intermediate resonances of
the hard process in subsequent emissions.

Similarly to what was done in Ref.~\cite{Luisoni:2013cuh} for \HVJ{}
production at NLO QCD, we have applied the improved
\MiNLO~\cite{Hamilton:2012np, Hamilton:2012rf} approach to \HVJ{} production
in order to get a sample of events that has simultaneously NLO QCD accuracy
for \HV{} plus 0 and 1 jet.
In the \MINLO{} framework, also the NLO EW corrections to \HV{} and \HVJ{}
production have been consistently combined in the same inclusive sample.
This can be regarded as an approximate treatment of $\mathcal{O}(\as\,\aem)$
corrections in observables that are very sensitive to QCD radiation and
receive, at the same time, large EW corrections.
Moreover, although we do not present a rigorous proof, based on
considerations related to unitarity and factorization of soft and collinear
QCD radiation, we will argue that our \MINLO{} predictions should preserve
full NLO QCD+EW accuracy in the phase space with zero or one resolved jets.
As we will see, this conclusion is supported by our numerical results.

While our NLO EW results are exact (apart from the treatment of
photon-initiated contributions), we also present approximate NLO EW
predictions in the so-called Sudakov limit, where all kinematic
invariants are well above the electroweak scale. Specifically, based on the
general results of Refs.~\cite{Denner:2000jv, Denner:2001gw}, we provide
explicit analytic expressions for all logarithmic EW corrections to $pp\to
\HV$ + 0~and~1~jet in next-to-leading-logarithmic~(NLL) approximation.  Based
on the observed accuracy of the NLL Sudakov formulas, this approximation can
be exploited both in order to speed up the evaluation of EW corrections at
NLO and in order to predict the dominant EW effects beyond NLO.

All needed matrix elements for $pp\to \HV$ + 0 and 1~jet at NLO EW have been
generated using the \OpenLoops{} program~\cite{Cascioli:2011va, OLhepforge},
which supports the automated generation of NLO QCD+EW scattering amplitudes
for Standard Model processes~\cite{Kallweit:2014xda, Kallweit:2015dum,
  Kallweit:2017khh}.  The implementation in the \RES{} framework was achieved
exploiting the generic interface developed in Ref.~\cite{Jezo:2016ujg}.  For
what concerns NLO QCD corrections, on the one hand we implemented in-house
analytic expressions for the virtual corrections. On the other hand,
following the approach of Ref.~\cite{Luisoni:2013cuh}, for real-emission
contributions we used \MadGraphFour{}~\cite{Alwall:2007st} matrix elements,
via the interface described in Ref.~\cite{Campbell:2012am}.

The paper is organized as follows.
In \refse{sec:NLO} we introduce the various ingredients of \HV{} and \HVJ{}
production at NLO QCD+EW.  In particular, in Sec.~\ref{se:MINLO} we present a
schematic proof of the NLO QCD+EW accuracy of \MINLO{} predictions for
inclusive observables.  Further technical aspects of the calculation as well
as input parameters and cuts are specified in \refse{sec:ingredients}.
Fixed-order NLO QCD+EW predictions are discussed in \refse{sec:foresults},
while in Secs.~\ref{sec:results_hv} and~\ref{sec:results_hvj} we present
NLO+PS QCD+EW results for \HV{} production and \MINLO{} QCD+EW results for
\HVJ{} production, respectively.
The predictions of the NLO+PS \HV{} and \MINLO{}+PS \HVJ{} generators are
compared in \refse{sec:generators}.
Our main findings are summarized in \refse{sec:conclusions}. 
In the appendices we document the validation of EW corrections in \HV{}
production against \HAWK{}~(\refapp{sec:validation}), detailed NLO EW
formulas in the Sudakov approximation~(\refapp{sec:sudakov}), a reweighting
approach that we employ in order to speed up the evaluation of EW
corrections~(\refapp{sec:fast_ew}), and technical aspects of the interface
between the \RES{} and \PythiaEightPone~(\refapp{sec:py8_interface}).

\section{NLO QCD and EW corrections to $\boldsymbol{\HV}$ and
  $\boldsymbol{\HVJ}$ production} 
\label{sec:NLO}

In this section we describe the QCD and EW NLO corrections to the production
of a Higgs boson in association with a \leplv{} or \lepll{} leptonic pair
plus zero or one additional jets. For convenience, these Higgsstrahlung processes 
will be denoted as associated \HV{} and \HVJ{} production, with $V=W^\pm$ or 
$Z$. However, all results presented in this paper 
correspond to the complete processes
\begin{eqnarray}
\label{eq:fullprocs}
  &&pp \to \HWp \, (j) \to H \,\wpl \,(j)\,, \nonumber\\
  &&pp \to  \HWm\, (j) \to H \,\wml\, (j)\,, \\
  && pp \to \HZ \, (j) \to H \,\zll\, (j)\,,\nonumber
\end{eqnarray}
including all spin-correlation and off-shell effects.  The combination of
\HWp/\HWpJ{} and \HWm/\HWmJ{} Higgsstrahlung will be denoted as \HW/\HWJ{}
production.
In our calculations, we have considered only one leptonic generation, and all
leptons are treated as massless.

\subsection{NLO QCD+EW matrix elements}
In this section we describe the various tree and one-loop amplitudes that
have been assembled to form a NLO QCD+EW Monte Carlo program based on the
\POWHEGBOX{} framework~\cite{Alioli:2010xd}.

\begin{figure}[htb]
  \begin{center}
  \subfigure[]
   {\includegraphics[width=0.32\textwidth]{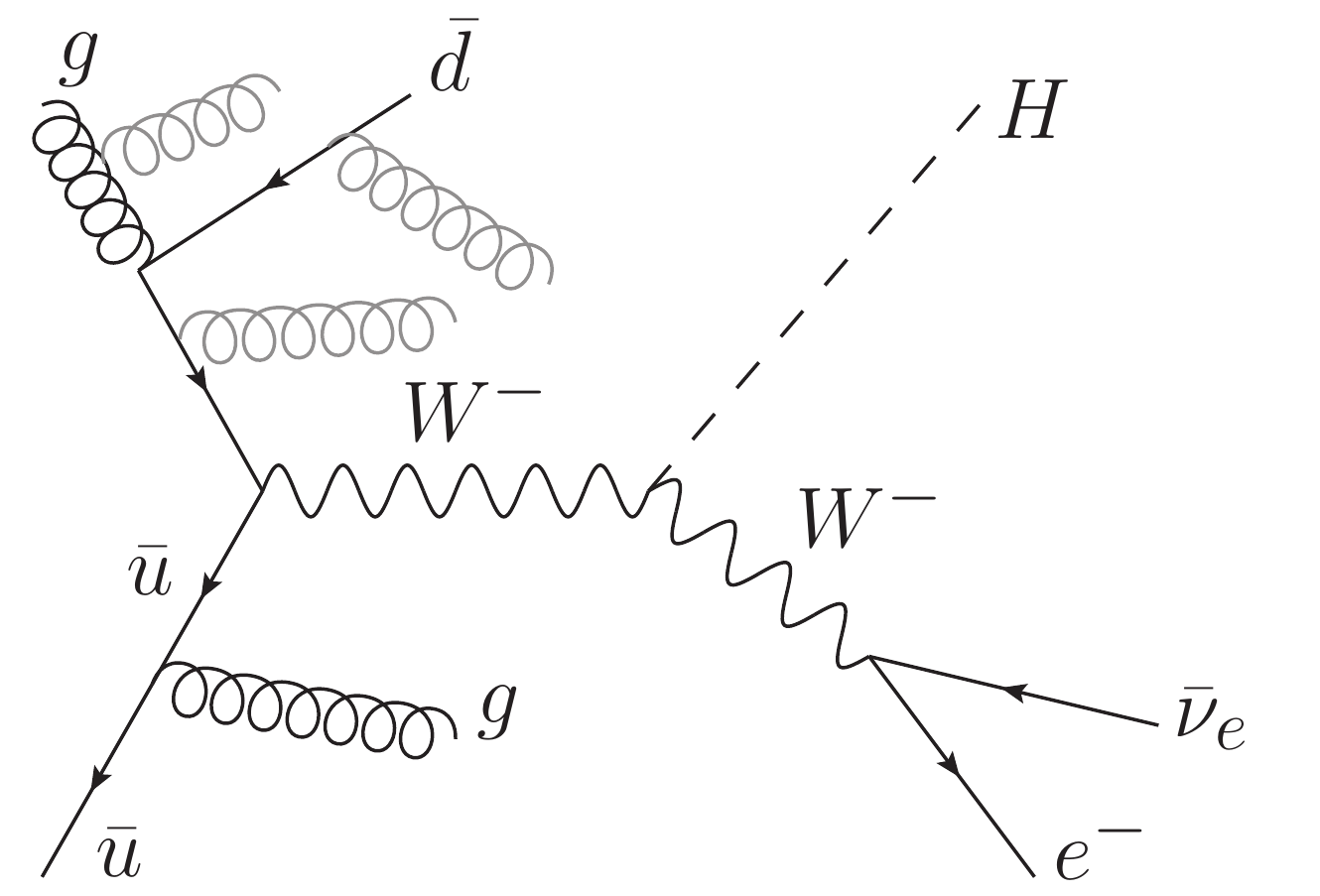}}
  \subfigure[]
   {\includegraphics[width=0.32\textwidth]{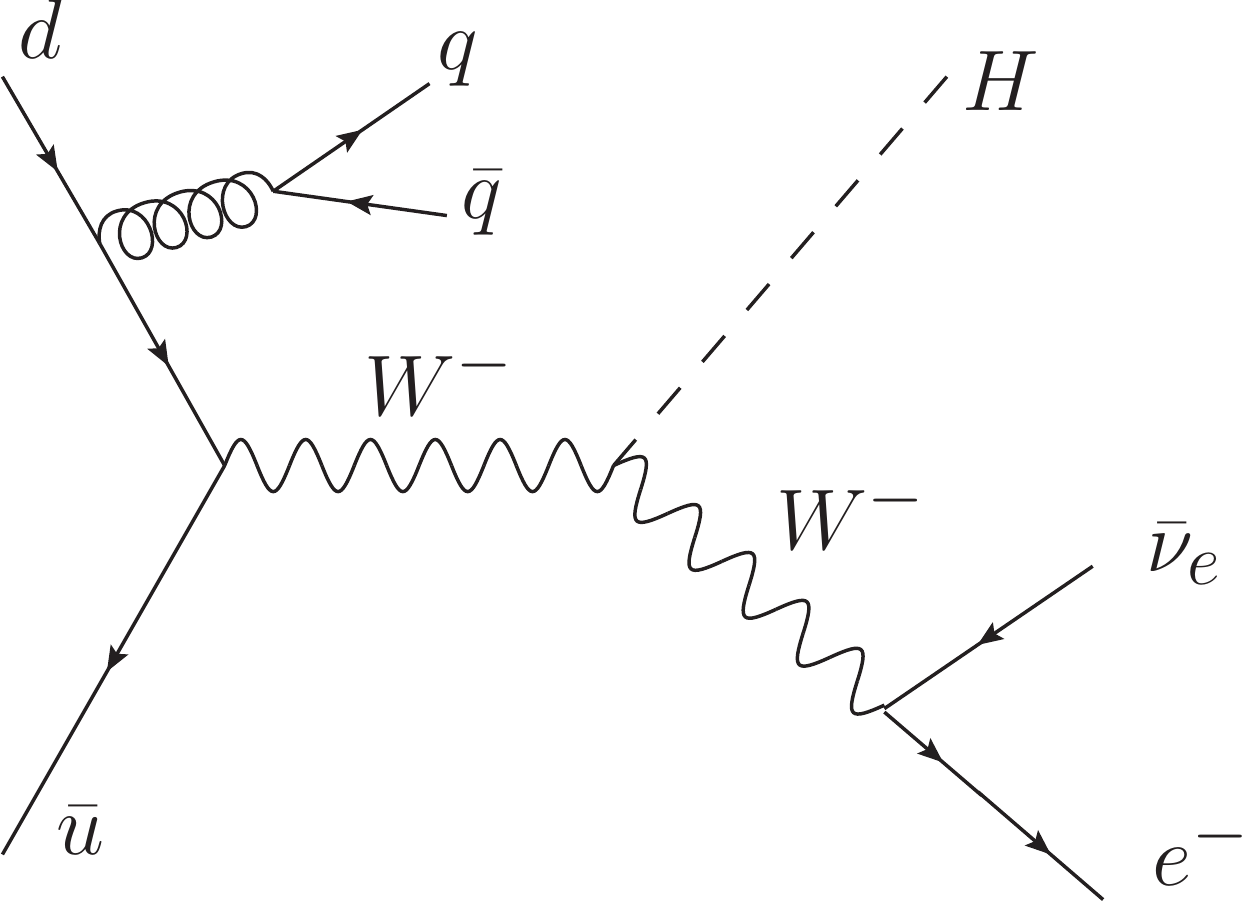}}
    \subfigure[]
   {\includegraphics[width=0.32\textwidth]{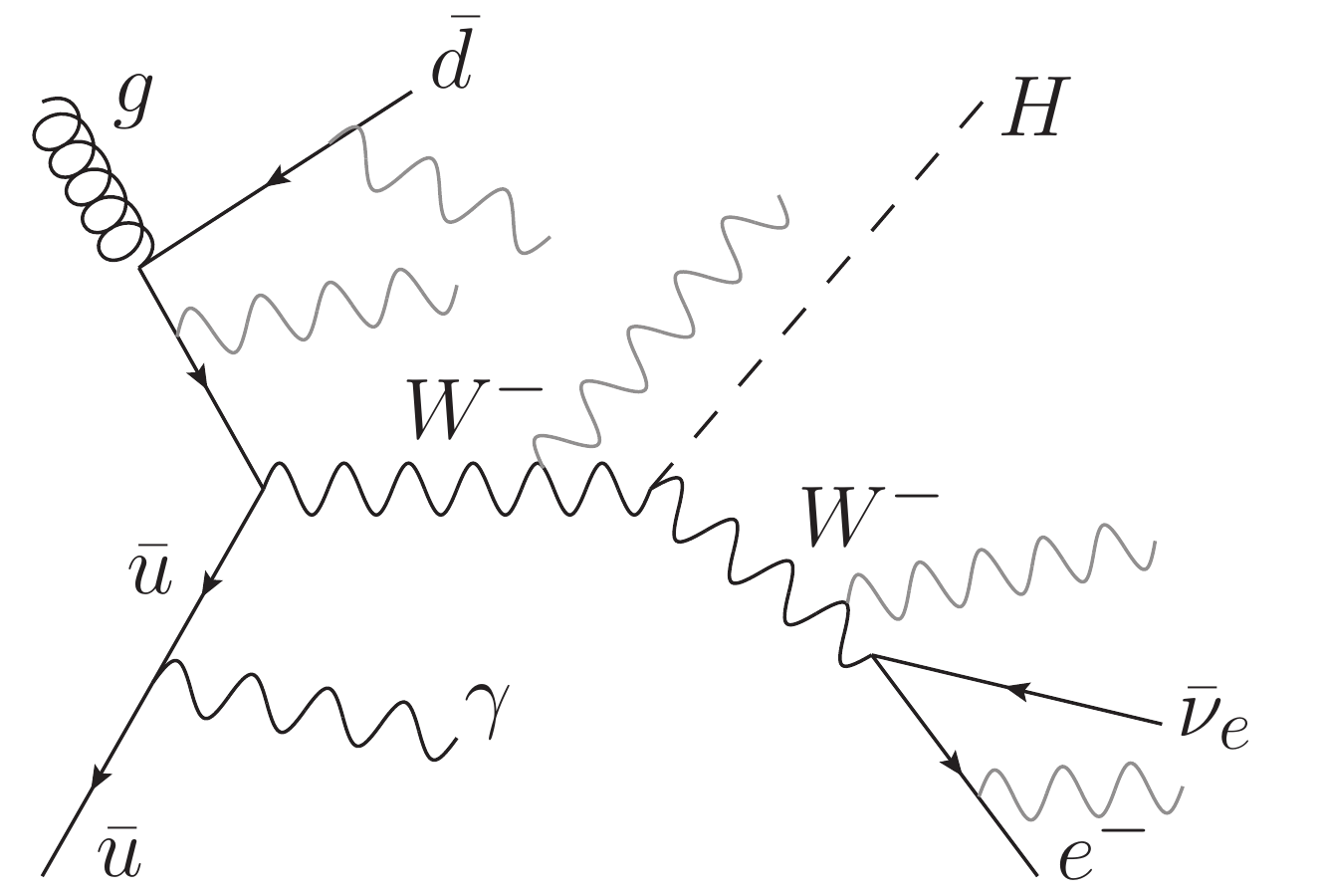}}
  \caption{A sample of QCD~(a, b), and EW~(c), real radiation diagrams
    contributing to \HWmJ{} production.  While only one photon or gluon at a
    time is present at fixed order, for illustration purpose, in~(a) and~(c),
    we have shown various possible gluon and photon emissions.
\label{fig:HWJ_real}}
  \end{center}
 \end{figure}

Associated \HV{} production proceeds through quark--antiquark annihilation
at leading order, which corresponds to $\ord(\aem^3)$.  In \HVJ{} production,
where the leading order corresponds to $\ord(\as\,\aem^3)$, additional
(anti)quark--gluon initiated processes contribute.
All $\ord(\as\,\aem^3)$ NLO QCD corrections to \HV{} production have been
computed analytically, since they simply affect the $Vq \bar q'$ vertex, and
the calculation of the real and virtual corrections is trivial.
In \HVJ{} production, the virtual $\ord(\as^2\,\aem^3)$ NLO QCD corrections
have been computed analytically~\cite{Granata_PhD}.
The color- and spin-correlated Born amplitudes and the real contributions at
$\ord(\as^2\,\aem^3)$ have been computed using the automated
interface~\cite{Campbell:2012am} between the \POWHEGBOX{} and
\MadGraphFour{}~\cite{Alwall:2007st}.  The real contributions involve tree
diagrams with either an additional gluon or an external gluon replaced with a
$q\bar q$-pair. Example diagrams are shown in Fig.~\ref{fig:HWJ_real}~(a, b).

\begin{figure}[htb]
  \begin{center}
  \subfigure[]
   {\includegraphics[width=0.32\textwidth]{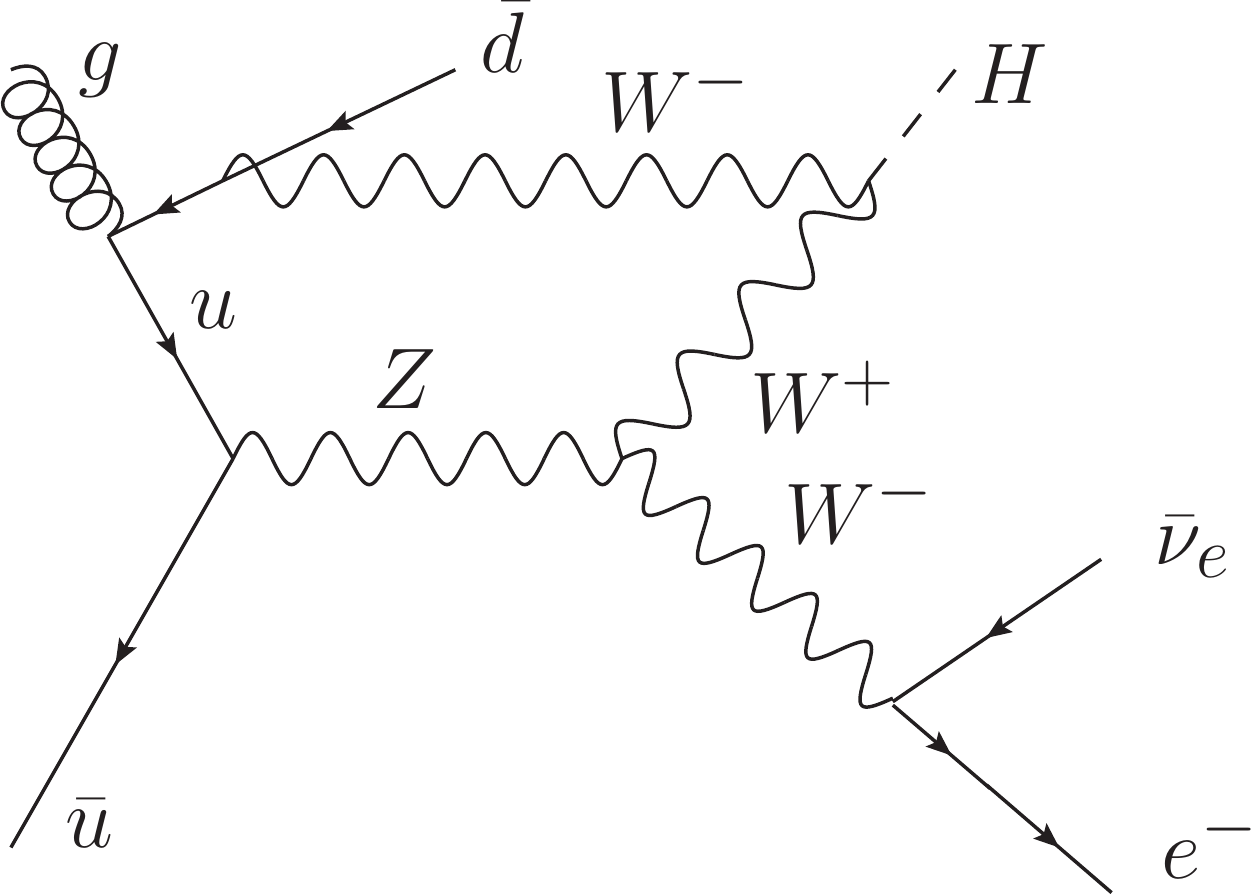}}
  \subfigure[]
   {\includegraphics[width=0.32\textwidth]{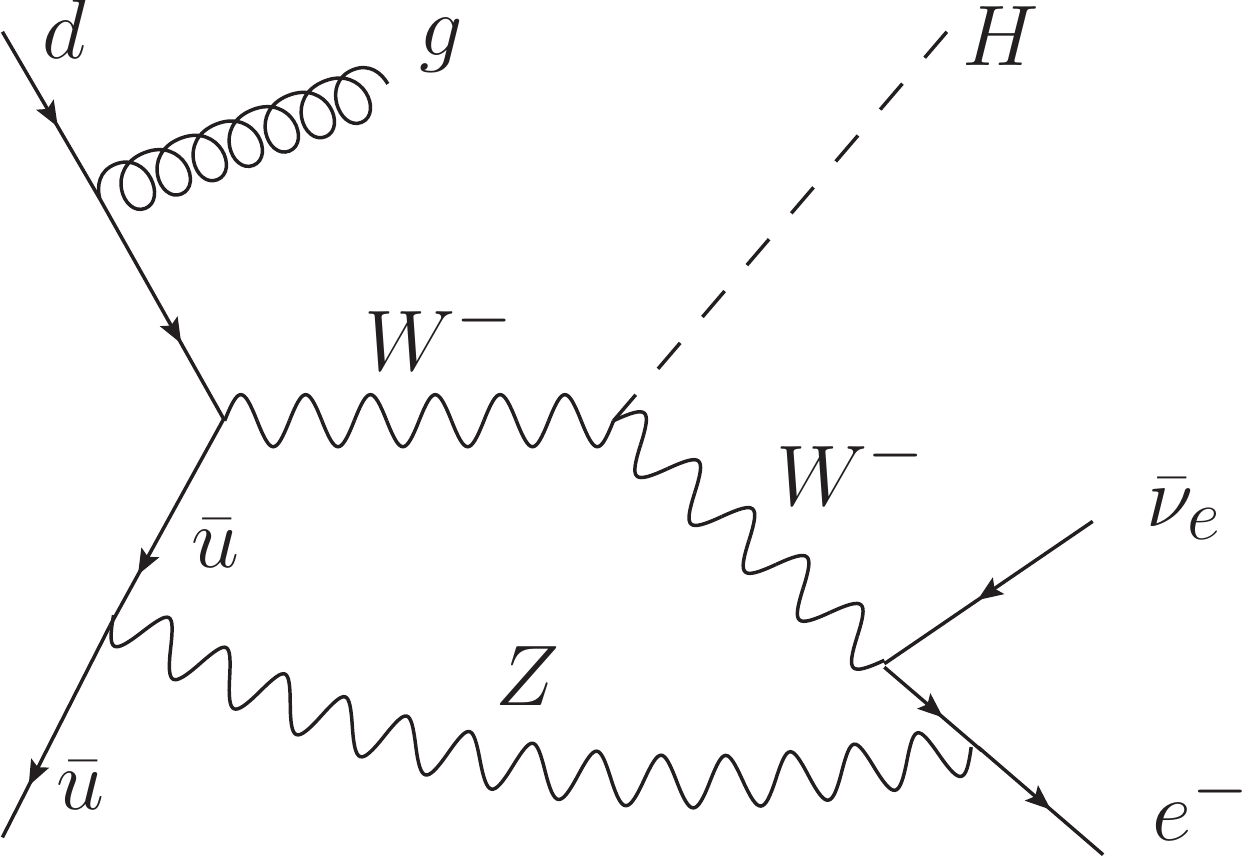}}
  \subfigure[]
   {\includegraphics[width=0.32\textwidth]{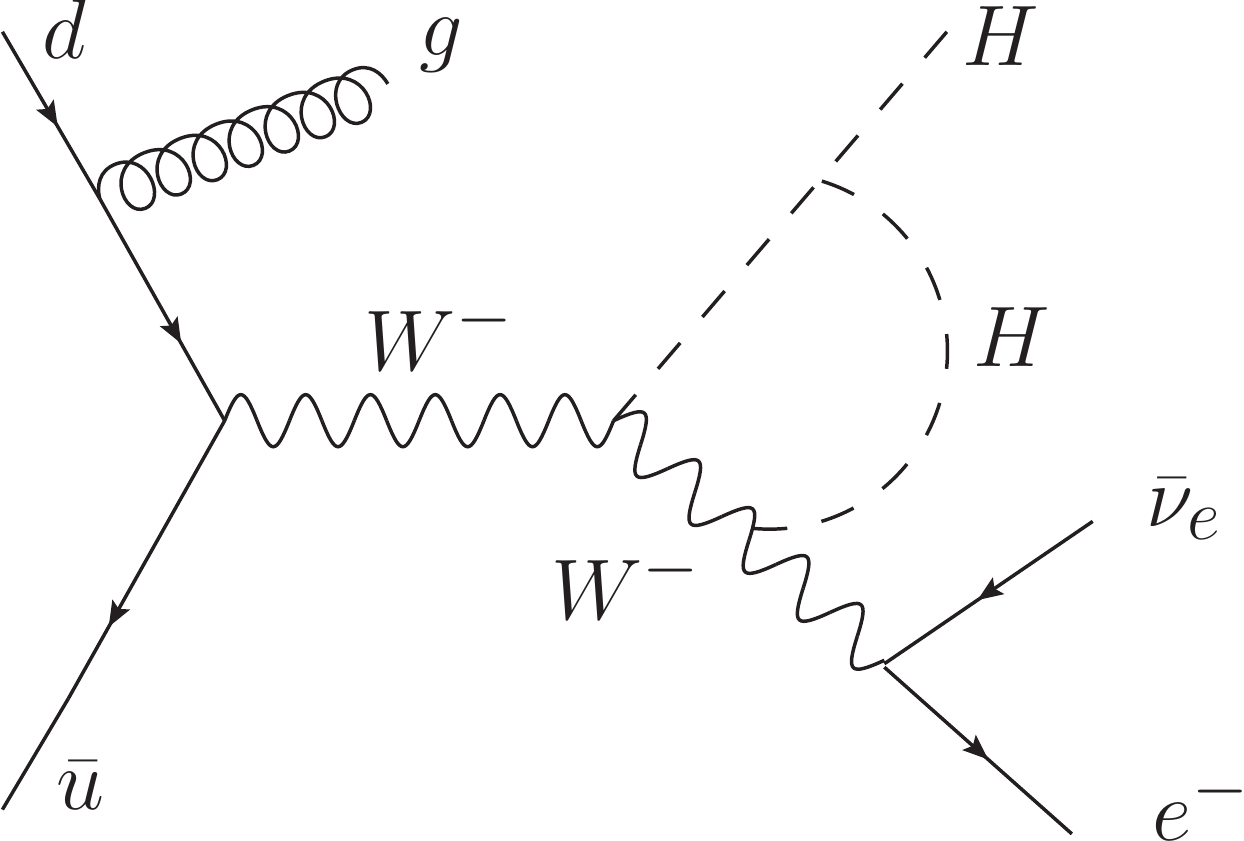}}
  \caption{A sample of virtual EW diagrams contributing to \HWmJ{}
    production. \label{fig:HWJ_ew_virt}}
  \end{center}
\end{figure}

The virtual EW corrections to \HV{} and \HVJ{} production comprise loop
amplitudes up to pentagon and hexagon configurations, respectively. Example
diagrams for \HVJ{} production are shown in \reffi{fig:HWJ_ew_virt}.  All the
internal resonances have been treated in the complex-mass
scheme~\cite{Denner:2005fg,Denner:2006ic} throughout.

As pointed out in the Introduction and illustrated in
\reffi{fig:HWJ_ew_virt}~(c), the virtual NLO EW amplitudes induce a
dependence on the Higgs trilinear coupling $\lambda_{\sss HHH}$.  This
dependence arises both from the bare virtual amplitudes and from the Higgs
boson self-energy entering the Higgs boson wave-function renormalization.  In
view of the possibility of exploiting precision measurements of
Higgsstrahlung processes for an indirect determination of $\lambda_{\sss
  HHH}$, we allow $\lambda_{\sss HHH}$ to be set independently of the Higgs
boson mass.\footnote{The corresponding parameter can directly be set in the
  \RES{} input file.}

The real NLO EW corrections to \HV{} and \HVJ{} production comprise QED
radiation off all charged particles, i.e.~they have an additional photon in
the final state, as illustrated in~\reffi{fig:HWJ_real}~(c).  Photon-induced
real radiation contributions, where the photon is crossed to the initial
state, are, on the other hand, not considered here, as they are suppressed by
the small photon density in the proton.
These corrections for \HV{} production have been computed for the
  first time in Ref.~\cite{Denner:2011id} and are included in the
  \HAWK~\cite{Denner:2014cla} Monte Carlo generator.
Interestingly, they reach several
percent for inclusive \HW{} production, but remain at the $2\%$ level when
leptonic selection cuts are applied, and are negligible for \HZ{}
production~\cite{deFlorian:2016spz}. For \HVJ{} production, photon-induced
contributions enter already at Born level: however, they are of
$\ord(\aem^4)$ and thus formally subleading with respect to the
$\ord(\as\,\aem^3)$ leading order. Still, the NLO QCD corrections to these
photon-induced processes are of $\ord(\as\,\aem^4)$ and thus formally of the
same order as the NLO EW corrections to the quark--antiquark and
(anti)quark--gluon initiated channels in \HVJ{} production.  Also not
considered here are mixed QCD-EW bremsstrahlung contributions to \HVJ{}
production at $\ord(\as\,\aem^4)$.  These tree-level contributions are finite
and can easily be investigated separately. Similar contributions in the NLO
EW corrections to $V$+jet production are known to yield relevant
contributions only in jet observables at very large transverse
momentum~\cite{Denner:2009gj, Kallweit:2015dum}.  Finally, also virtual QCD corrections to
$\HV\gamma$ production contribute formally at $\ord(\as\,\aem^4)$ and are
thus of the same perturbative order as the NLO EW corrections to \HVJ{}
production. However, if a photon isolation is applied,  as is done in 
this paper (see \refse{sec:cuts}),
$\HV\gamma$ production can be considered as a separate process and thus excluded 
from the definition of \HVJ{} production.

All the electroweak real and virtual corrections have been computed using a
recent interface of the \RES{} to \OpenLoops~\cite{Jezo:2016ujg}.

In this study we combine NLO QCD and EW corrections in an additive way,
i.e.~corresponding perturbative contributions are simply added. At fixed
order, an improved description can easily be obtained via a factorized
ansatz, where differential NLO QCD cross sections are multiplied with
relative EW correction factors.  Such a multiplicative combination can be
motivated from the factorization of soft QCD radiation and EW Sudakov
logarithms, which can be tested comparing relative NLO EW corrections for
\HV{} and \HVJ{} production.

\subsection[\MINLO{} approach for \HVJ{} production  at NLO QCD+EW]{$\boldsymbol{\MINLO}$
  approach for $\boldsymbol{\HVJ}$ production at NLO QCD+EW} 
\label{se:MINLO}

\def\LO{\mathrm{\sss LO}}
\def\EW{\mathrm{\sss EW}}
\def\MILO{\mathrm{\sss MiLO}}
\def\MINLOEW{\sss \mathrm{MiNLO\,EW}}
\def\NLOEW{\mathrm{\sss NLO\,EW}}
\def\PhiHV{\Phi_{\HV}}
\def\PhiHVJ{\Phi_{\HVJ}}
\def\PhiJ{\Phi_{j}}
\def\Bbar{\bar{B}}
\def\fin{\sss \mathrm{fin}}

In order to obtain an optimal description of QCD radiation, both in the hard
and soft regime, all NLO QCD+EW calculations for $pp\to\HVJ{}$ have been
performed using the ``Multiscale improved
NLO''~(\MINLO)~\cite{Hamilton:2012np} method.  This approach effectively
resums logarithmic singularities of soft and collinear type to
NLL accuracy, thereby ensuring a finite \HVJ{}
cross section in all regions of phase space, even when the extra jet becomes
unresolved.  In the \MINLO{} approach, NLL resummation is achieved by means
of a CKKW scale setting~\cite{Catani:2001cc,Krauss:2002up} for the strong
coupling factors associated with each QCD vertex, together with an
 appropriate factorization-scale choice and NLL QCD Sudakov form factors. These
are applied to all internal and external lines corresponding to the
underlying-Born skeleton of each event.  In addition, improving the \MINLO{}
resummation as described in Ref.~\cite{Hamilton:2012rf}, we have obtained a
fully inclusive description of \HV{} production with NLO QCD accuracy in all
phase space regions.
In other words, besides providing \HVJ{} kinematic distributions that are NLO
accurate and also finite when the hardest jet goes unresolved, the improved
\MINLO{} predictions for $pp\to \HVJ$ are NLO accurate also for distributions
in inclusive variables such as the rapidity or the transverse momentum of the
\HV{} pair.

All NLO QCD+EW predictions for $pp\to \HVJ$ presented in this paper, both at
fixed order and including matching to the parton shower, are based on the
\MINLO{} approach, which is applied to all contributions of NLO QCD and NLO
EW type.  Technically, the \MINLO{} Sudakov form factors and scale choices
are implemented at the level of $pp\to \HVJ$ underlying-Born events that
correspond to so-called $\bar B$ terms in the \POWHEG{}
jargon.\footnote{Real-emission events of NLO QCD and NLO EW type are related
  to underlying-Born events of type $pp\to \HVJ$ via FKS
  mappings~\cite{Frixione:1995ms}.}  Note that the \MINLO{} procedure resums
only logarithms associated with soft and collinear QCD singularities that
result form the presence of QCD radiation at Born level, while QED radiation
is not present at Born level.  Thus there is no need to introduce NLO EW
effects in the \MINLO{} Sudakov form factors.
This implies that, in contrast to the case of NLO QCD, the NLO EW
corrections to $pp\to \HVJ$ do not need to be matched to the \MINLO{} form
factors. In practice, for what concerns the EW corrections,  the \MINLO{} procedure 
is applied in a way that is equivalent to Born level. 

For observables where QCD radiation is integrated out, the \MINLO{} improved
NLO EW contributions assume the form
\begin{equation}
\label{eq:proof2}
\frac{\rd\sigma_{\HVJ}^{\MINLOEW}}{\rd\PhiHV} =
\int\rd\PhiJ\,\Bbar_{\HVJ}^\EW(\PhiHV,\PhiJ)\,
\Delta\big(\kt(\PhiJ)\big),
\end{equation}
where $\Phi_{\HV}$ and $\PhiJ$ denote the factorized phase spaces of the
$\HV$ system and the jet, respectively. The term
$\Bbar_{\HVJ}^\EW(\PhiHV,\PhiJ)$ includes $\mathcal{O}(\aem)$
corrections\footnote{Since Born contributions are part of the usual QCD
  $\Bbar$ term, in the $\Bbar^\EW$ term we only include $\mathcal{O}(\aem)$
  corrections.}  of virtual and real type, and the latter are integrated over
the corresponding emission phase space.  The \MINLO{} approach is implemented
through an implicitly understood CKKW scale choice for the $\as$ term in
$\Bbar_{\HVJ}^\EW$, and through the NLL Sudakov from factor
$\Delta\big(\kt(\PhiJ)\big)$ in \refeq{eq:proof2}.  For later convenience,
together with the Sudakov form factor, we introduce a corresponding emission
kernel $K(\PhiJ)$ that is formally related to $\Delta$ via
\begin{equation}
\label{eq:proof3}
\Delta\big(\pt\big)
= \exp\left[-\int\rd\PhiJ\,K(\PhiJ)\Theta\big(\kt(\PhiJ)-\pt\big)\right].
\end{equation}
In the following, based on the factorization properties of soft and collinear
QCD radiation, encoded in the kernel $K(\PhiJ)$, and using the unitarity
relation
\begin{equation}
\label{eq:proof4}
\int\rd\PhiJ\,K(\PhiJ)\, \Delta\big(\kt(\PhiJ)\big) = 1,
\end{equation}
we will argue that the inclusive \MINLO{} predictions of \refeq{eq:proof2}
are not only NLO QCD accurate, but also NLO EW accurate.  More precisely, we
will prove (in a schematic way) that
\begin{equation}
\label{eq:proof5}
\frac{\rd\sigma_{\HVJ}^{\MINLOEW}}{\rd\PhiHV}=
\frac{\rd\sigma_{\HV}^{\NLOEW}}{\rd\PhiHV}
+\mathcal{O}(\aem\,\as)\,,
\end{equation}
where
\begin{equation}
\label{eq:proof1}
\frac{\rd\sigma_{\HV}^{\NLOEW}}{\rd\PhiHV}=
\Bbar_{\HV}^\EW(\PhiHV)\, .
\end{equation}
We first demonstrate the Born-level version of \refeq{eq:proof5}, 
which corresponds to
\begin{equation}
\label{eq:proof5b}
\frac{\rd\sigma_{\HVJ}^{\MILO}}{\rd\PhiHV}= 
\frac{\rd\sigma_{\HV}^{\LO}}{\rd\PhiHV}
+\mathcal{O}(\as)\, ,
\end{equation}
where {\tt MiLO} denotes the Born (or LO) version of the \MINLO{} approach.
The above identity can be written as
\begin{equation}
\label{eq:proof7b}
\int\rd\PhiJ\,B_{\HVJ}(\PhiHV,\PhiJ) \, \Delta\big(\kt(\PhiJ)\big)
=
B_{\HV}(\PhiHV)+\mathcal{O}(\as)\, ,
\end{equation}
where $B_{\HVJ}$ and $B_{\HV}$ are the Born counterparts of the
$\Bbar_{\HVJ}^\EW$ and $\Bbar_{\HV}^\EW$ terms in Eqs.~(\ref{eq:proof2})
and~(\ref{eq:proof3}).  The meaning of Eqs.~(\ref{eq:proof5b})
and~(\ref{eq:proof7b}) is that the \MINLO{} approach at Born level guarantees
LO accuracy for observables that are inclusive with respect to the extra jet.
In order to demonstrate this property, we split the $pp\to \HVJ$ Born term
$B_{\HVJ}$ into an IR divergent and a finite part,
\begin{equation}
\label{eq:proof6}
B_{\HVJ}(\PhiHV,\PhiJ)
=
B_{\HV}(\PhiHV)K(\PhiJ) +B^{\fin}_{\HVJ}(\PhiHV,\PhiJ)\,.
\end{equation}
Here the singularities associated with QCD radiation in the soft and
collinear limits are factorized\footnote{In this schematic derivation we
  assume a simple factorization of multiplicative type, while the
  factorization of initial-state collinear singularities takes the form of a
  convolution.}  into the $pp\to \HV{}$ Born term times the NLL kernel
$K(\PhiJ)$, while the $B^{\fin}_{\HVJ}$ remainder is free from singularities.
Thus, upon integration over the jet phase space, the $B^{\fin}_{\HVJ}$
remainder yields only $\mathcal{O}(\as)$ suppressed contributions with
respect to $B_{\HV}$, while using the unitarity relation~(\ref{eq:proof4}) it
is easy to show that the singular term in \refeq{eq:proof6} leads to
\refeq{eq:proof7b}.

Thanks to the fact that applying the \MINLO{} approach to NLO EW
contributions is largely equivalent to applying \MINLO{} at Born level, the
NLO EW accuracy property~(\ref{eq:proof5}) can be proven along the same lines
as for the LO accuracy property~(\ref{eq:proof5b}).  As sole additional
ingredient, the NLO EW proof requires certain factorization properties of
soft and collinear QCD radiation.  More precisely, the factorization
properties of~\refeq{eq:proof6} must hold also in the presence of EW
corrections, i.e.
\begin{equation}
\label{eq:proof8}
\Bbar^\EW_{\HVJ}(\PhiHV,\PhiJ)
=
\Bbar^\EW_{\HV}(\PhiHV)K(\PhiJ) +\Bbar^{\EW,\,\fin}_{\HVJ}(\PhiHV,\PhiJ)\,.
\end{equation}
Here the remainder $\Bbar^{\EW,\,\fin}_{\HVJ}$ should be free from QCD
singularities, so that it yields only $\mathcal{O}(\as)$-suppressed
contributions relative to $\Bbar^{\EW}_{\HV}$, when the extra jet is
integrated out.  Based on this natural assumption, in full
analogy with the LO case, we easily arrive at
\begin{eqnarray}
\label{eq:proof9}
\int\rd\PhiJ\,\Bbar_{\HVJ}^\EW(\PhiHV,\PhiJ)\, 
\Delta\big(\kt(\PhiJ)\big)
&=&
\Bbar_{\HV}^\EW(\PhiHV)\int\rd\PhiJ\,K(\PhiJ) \,\Delta\big(\kt(\PhiJ)\big)  +\mathcal{O}(\aem\as)\nonumber\\
&=&
\Bbar_{\HV}^\EW(\PhiHV) +\mathcal{O}(\aem\,\as)\,,
\end{eqnarray}
which is equivalent to the hypothesis~(\ref{eq:proof5}).

In summary, based on unitarity and factorization properties of QCD radiation,
we expect that the improved \MINLO{} procedure applied to NLO QCD+EW matrix
elements for $pp\to \HVJ$ should preserve its full QCD+EW accuracy when the
jet is integrated out. As we will see, this conclusion is well supported by
our numerical findings in \mbox{Secs.~\ref{sec:foresults}--\ref{sec:generators}.}
Nevertheless, due to the schematic nature of the presented derivations and
related assumptions, the above conclusions should be regarded as an educated
guess that deserves further investigation.

\subsection{Sudakov approximation at NLO EW}
In the Sudakov high-energy regime, where all kinematic invariants are of the
same order and much larger than the electroweak scale, the NLO EW corrections
are dominated by soft and collinear logarithms of Sudakov type.  Based on the
general results of Refs.~\cite{Denner:2000jv, Denner:2001gw} we have
derived analytic expressions for the NLO EW corrections to \HV{} and \HVJ{}
production in NLL approximation.  Details and scope of this approximation are
discussed in App.~\ref{app:sudakov}.

The Sudakov approximation at NLO provides us with qualitative and
quantitative insights into the origin of the dominant NLO EW effects.
Moreover, it can be easily extended to the two-loop
level~\cite{Melles:2000gw,Denner:2003wi}, thereby opening the door to
approximate NNLO EW predictions based on the combination of exact NLO EW
corrections with Sudakov logarithms at two loops.  From the practical point
of view, the Sudakov approximation at NLO permits to obtain the bulk of the
EW virtual corrections at much higher computational speed as compared to an
exact NLO EW calculation.

In \refse{sec:results_hv}, we will assess the quality of the Sudakov
approximation\footnote{ As explained in more detail in
  App.~\ref{app:sudakov}, the Sudakov approximation is applied only to the
  virtual part of EW corrections, while real QED radiation is always treated
  exactly.}  through a detailed comparison against exact NLO EW corrections.
Finally, in App.~\ref{sec:fast_ew}, we show how the NLL EW approximation can
be used in order to speed up the Monte Carlo integration, while keeping full
NLO EW accuracy in the final predictions.

\section{Technical aspects and setup of the simulations}
\label{sec:ingredients}

\subsection{The \RES{} framework at NLO QCD+EW }
\label{sec:RES}

The QCD+EW NLO calculations for \HV{} and \HVJ{} production have been matched
to parton showers using the \POWHEG{} method. To this end, we used the
recently-released version of the \POWHEGBOX{} framework, called \RES.  The
major novelty of this new version is the resonance-aware
approach~\cite{Jezo:2015aia}, which guarantees a consistent treatment of
intermediate resonances at NLO+PS level.  This is achieved by generating the
hardest radiation in a way that preserves the virtuality of resonances
present at the underlying-Born level.  At the same time, the resonance
information can be passed on to the parton shower, which in turn preserves the
virtuality of intermediate resonances of the hard process in subsequent
emissions.  This method was introduced in order to address the combination of
NLO QCD corrections with parton showers in the presence of top-quark
resonances.  However, since it is based only on general properties of
resonances and infrared singularities, the resonance-aware approach is
applicable also to the combination of EW corrections with QED parton showers.
In fact, this method has already been applied in the
  context of electroweak corrections in Refs.~\cite{CarloniCalame:2016ouw,
    Muck:2016pko}.

In the \RES{} jargon~\cite{Jezo:2015aia}, a radiated parton (or photon) can
be associated to one or more ``resonances'' present in the process, or to the
``production'' part, if it cannot be associated to a particular
resonance. The \RES{} framework automatically finds all the possible so-called
``resonance histories'' for a given partonic process.
For the processes at hand, considering QED radiation, only two resonance
histories are detected: a production history, where the photon can be emitted
by any quark (both in the initial and in the final state), and a vector-boson
decay history, where the photon is radiated off a final-state charged lepton,
and the virtuality of the intermediate vector boson needs to be preserved.
Soft photons that are radiated from a $W$ resonance are 
attributed either to the production subprocess or to the $W$ decay, consistently with the
virtualities of the quasi-resonant $W$ propagators ``before'' and ``after'' the
photon emission.

The treatment of QED radiation was first introduced in the \POWHEGBOX{} for
the calculation of the EW corrections to Drell--Yan
processes~\cite{Barze:2012tt,Barze':2013yca, CarloniCalame:2016ouw, Muck:2016pko}.
In this context, leptons were
considered as massive particles, and QED subtraction in the \POWHEGBOX{} was
implemented accordingly.  In the study at hand, leptons are treated as
massless, and we have implemented the treatment of photon radiation off
massless charged particles (both leptons and quarks). To this end, we have
adapted the QCD soft and virtual counterterms already present in the
\POWHEGBOX{} to the QED case. Moreover, we have computed a new upper-bounding
function for the generation of photon radiation with the highest-bid method,
as described in Ref.~\cite{Frixione:2007vw}.

By default, in the \RES{} framework, only the hardest radiation out of all
singular regions is kept, before passing the event to shower Monte Carlo
programs like \Py{} or {\tt Herwig}. In this way, for each event, at most one
of the decaying resonances (or the production part of the process) includes
an NLO-accurate radiation.  Moreover, in case of combined QCD and EW
corrections, QED emission occurs in competition with the QCD one.  The \RES{}
uses the highest-bid method to decide what kind of radiation~(QED or QCD,
initial- or final-state) is generated. Due to the larger center-of-mass
energy available in the production stage, initial-state radiation is enhanced
with respect to final-state radiation, and since the QCD coupling is larger
than the QED one, initial-state quarks tend to radiate gluons rather than
photons.
Thus, QED emission from the decay of a resonance would hardly be kept at the
Les~Houches event~(LHE) level, and the QED radiation would mainly be
generated by the shower Monte Carlo program.

The resonance-aware formalism implemented in the \RES{} framework offers the
opportunity to further improve the \POWHEG{} radiation formula.  With this
improvement, first introduced in Ref.~\cite{Campbell:2014kua}, radiation from
each singular region is generated and, instead of keeping only the hardest
overall one, the hardest from each resonance is stored.  As a result, the LHE
file contains a radiated particle for each decaying resonance, plus possibly
one emission from the production stage. In this way NLO+LL accuracy is
ensured for radiation off each resonance.  The subsequent shower from each
resonance generated by the Monte Carlo shower program has to be softer than
each corresponding \POWHEG{} radiation.\footnote{This multiple-radiation mode
  can be activated by setting the flag {\tt allrad} to~1 in the input file.}
All NLO+PS results presented in this paper are based on this
multiple-radiation scheme.

As a final remark, we note that in the \RES{} framework both the \HV{} and
\HVJ{} processes can be computed at NLO or NLO+PS level with only QCD
corrections, with only EW corrections, or with combined NLO QCD+EW
corrections.\footnote{The flag {\tt qed\_qcd} controls this behavior in the
  input file. The values it can assume are: 0, to compute only
  QCD corrections, 1, to compute only EW corrections or 2, for both.}

\subsection{\OpenLoops tree and one-loop amplitudes}
\label{sec:OL}

All needed amplitudes at NLO EW have been generated with
\OpenLoops{}~\cite{Cascioli:2011va,OLhepforge} and implemented in the \RES{}
framework through the general interface introduced in
Ref.~\cite{Jezo:2016ujg}. Thanks to the recursive numerical approach of
Ref.~\cite{Cascioli:2011va} combined with the \Collier tensor reduction
library~\cite{Denner:2016kdg}, or with {\sc CutTools}~\cite{Ossola:2007ax},
the \OpenLoops{} program permits to achieve high CPU performance and a high
degree of numerical stability.
The amplitudes employed for the EW corrections in this paper are based on the
recently achieved automation of EW corrections in
\OpenLoops~\cite{Kallweit:2014xda,Kallweit:2015dum,Kallweit:2017khh}.

Within \OpenLoops, ultraviolet and infrared divergences are dimensionally
regularized in $D$ dimensions. However, all ingredients of the numerical
recursion are handled in four space-time dimensions.  The missing
$(4-D)$-dimensional contributions, called $R_2$ rational terms, are universal
and can be restored from process-independent effective
counterterms~\cite{Ossola:2008xq, Binoth:2006hk, Bredenstein:2008zb}.  The
implementation of the corresponding Feynman rules for the complete EW
Standard Model in \OpenLoops is largely based on Refs.~\cite{Garzelli:2009is,
  Garzelli:2010qm, Garzelli:2010fq, Shao:2011tg}.  Relevant contributions for
\HV{} and \HVJ{} production have been validated against independent algebraic
results in $D=4-2\epsilon$ dimensions.  UV divergences at NLO EW are
renormalized in the on-shell scheme~\cite{Denner:1991kt} extended to complex
masses~\cite{Denner:2005fg}.

\subsection{Input parameters, scales choices and other aspects of the setup}
\label{sec:setup}

In our $pp\to \HV$(+jet)  simulations at NLO QCD+EW, we have set the
gauge-boson masses and widths to the following values~\cite{Agashe:2014kda}
\begin{eqnarray}
&\MZ =91.1876~\GeV,\qquad\qquad  & \!\!\MW =80.385~\GeV,\\
&\Gamma_{\sss Z}=2.4955~\GeV,\qquad \qquad  &\Gamma_{\sss W} =2.0897~\GeV.
  \nonumber
\end{eqnarray}
The latter are obtained from state-of-the-art theoretical calculations.
Assigning a finite width to the Higgs boson in the final state would
invalidate EW Ward identities: we then consider the Higgs boson as on shell
with $\Gamma_{\sss H}=0$ and set its mass to $\MH=125$~GeV.  The top-quark
mass and width are set respectively to $m_t=172.5$~GeV and
$\Gamma_t=1.5083$~GeV.  All other quarks and leptons are treated as massless.
In the EW corrections, the top-quark contribution enters only at loop level,
the dependence of our results on $\Gamma_\Pt$ is thus completely negligible.

For the treatment of unstable particles we employ the complex-mass
scheme~\cite{Denner:2005fg, Denner:2006ic}, where finite-width effects
are absorbed into complex-valued renormalized masses
\begin{equation}
\mu^2_k=M_k^2-i \,\Gamma_k\,M_k \qquad\qquad \mbox{for} \ k=W,\, Z,\, t\,.
\end{equation}
The electroweak couplings are derived
from the gauge-boson masses and the Fermi constant,
$\GF=1.16637\times10^{-5}\,\GeV^{-2}$, and the electromagnetic coupling is
set accordingly to
\begin{equation}
\aem=\left|\frac{\sqrt{2}\, \sw^2\, \mu^2_{\sss W} \,G_\mu}{\pi}\right|,
\end{equation}
where  $\mu^2_{\sss W}$ and the squared sine of the weak mixing angle
\begin{equation}
\sw^2=1-\cw^2=1-\frac{\mu_{\sss W}^2}{\mu_{\sss Z}^2},
\end{equation}
are complex-valued.\footnote{By default we use the $G_\mu$ scheme throughout. However,
in the \RES{} framework, there is the option to evaluate
  the virtual EW corrections using $\aem$ computed in the $\GF$ scheme,
  and use the Thomson value $\aem(0)=1/137.035999$ in the evaluation
  of the contribution due to photon radiation.}

The absolute values of the CKM matrix elements are set to 
\begin{equation}
\begin{array}{c}
\\ |\CKM|=
\end{array}
\begin{array}{c l}
  & \qquad\,\, d\qquad\ \ s\ \ \qquad\quad b \\
\begin{array}{c}
u\\
c\\
t
\end{array} 
&
\!\!\!\l(
\begin{array}{c c c}
0.97428 & 0.2253 & 0.00347\\
0.2252 & 0.97345 & 0.0410\\
0.00862 & 0.0403 & 0.999152\end{array}
\r).
\end{array}
\end{equation}
Our default set of parton-distribution functions (PDF) is the {\tt
  NNPDF2.3\_as\_0119\_qed} set~\cite{Ball:2013hta}, that includes QED
contributions to the parton evolution and a photon density.\footnote{It
  corresponds to the PDF set 244800, in the LHAPDF6~\cite{Buckley:2014ana}
  numbering scheme.}  The value of the strong coupling constant corresponding
to this PDF set is $\as(\MZ)= 0.119$.

Finally, in \HV{} production, the renormalization and factorization scales
are set equal to the invariant mass of the \HV{} pair at the underlying-Born
level,
\begin{equation}\label{eq:mur_muf}
\mu_{\sss R} = \mu_{\sss F} = M_{\sss HV}, \qquad M_{\sss HV}^2 = 
\l(p_{\sss H} + p_{\ell_1} + p_{\bar{\ell}_2} \r)^2,
\end{equation}
where $\ell_1$ and  $\ell_2$ are the final-state leptons,  
while in $pp\to\HVJ$ the improved \MINLO~\cite{Hamilton:2012np, Hamilton:2012rf}
procedure is applied, and the scales are set accordingly.

Predictions at NLO+PS generated with the \POWHEG{} method are combined with
the \PythiaEightPone{} QCD+QED parton shower using the ``Monash 2013''
tune~\cite{Skands:2014pea}.  Effects due to hadronization, multi-particle
interactions and underlying events are not considered in this paper.

\subsection{Physics objects and cuts in NLO+PS simulations}
\label{sec:cuts}

In the following we specify the definition of physics objects and cuts 
that are applied in the phenomenological NLO+PS studies presented 
in Secs.~\ref{sec:results_hv}--\ref{sec:generators}.

All leptonic observables are computed in terms of dressed leptons, which are
constructed by recombining the collinear photon radiation emitted within a
cone (in the $(y,\phi)$ plane) of radius $R_{\g \ell}=0.1$ from charged
leptons, and the recombined photons are treated as unresolved particles.
Observables that depend on the reconstructed vector bosons are defined by
combining the momenta of the dressed charged leptons and the neutrino
associated with their decay. The latter is taken at Monte Carlo truth level.

Jets are constructed with {\tt FastJet} using the anti-$k_{\sss T}$
algorithm~\cite{Cacciari:2005hq, Cacciari:2008gp} with $R=0.5$.  The jet
algorithm is applied in a democratic way to QCD partons and non-recombined
photons, with the exception of photons that fulfill the isolation
criterion of Ref.~\cite{Frixione:1998jh} with a cone of radius $R_0 = 0.4$ and a
maximal hadronic energy fraction $\epsilon_{\sss h} = 0.5$. The hardest of
such isolated photons is excluded from the jet algorithm and is treated as
resolved photon.

The following standard Higgsstrahlung cuts are applied.  For every dressed
charged lepton we require
\begin{equation}\label{eq:cut_lep1}
\pt^\ell \ge 25~\GeV, \qquad\qquad |y^\ell| \le 2.5\,.
\end{equation}
In \HW/\HWJ{} production, we also impose
\begin{equation}
\label{eq:cut_lep2}
\slashed{E}_{\sss\rm T} \ge 25~\GeV\,,
\end{equation}
where $\slashed{E}_{\sss \rm T}$ is the transverse momentum of the neutrino
that results from the $W$-boson decay at Monte Carlo truth level.
In \HZ/\HZJ{} production, the invariant mass of the dressed-lepton
pair is required to satisfy
\begin{equation}\label{eq:cut_lep3} 
60~\GeV \le M^{\ell^+ \ell^-} \le 140~\GeV\,.
\end{equation}
Besides these inclusive selection cuts, we also present more exclusive results 
in the boosted regime. In this case, we impose the following 
additional cuts on the transverse momentum of the Higgs and vector bosons
\begin{equation}\label{eq:boosted_regime}
\pt^{\sss \rm H} \ge 200~\GeV, \qquad\qquad \pt^{\sss \rm V} \ge 190~\GeV\,.
\end{equation}
Such a selection of events with a boosted Higgs boson improves the
signal-over-background ratio in the $H\to b\bar{b}$ decay channel.

\section{Results for $\boldsymbol{\HV}$ and  $\boldsymbol{\HVJ}$  production at fixed NLO QCD+EW}
\label{sec:foresults}

In this section we present fixed-order NLO QCD+EW predictions for $pp\to
\HV{}$ and \mbox{$pp\to \HVJ{}$} at 13~TeV.  For \HVJ{} production the
improved \MINLO{} approach~\cite{Hamilton:2012np, Hamilton:2012rf} is
applied. Higgs boson production in association with $W$ and $Z$ bosons is
discussed in Secs.~\ref{sec:fixed_hw} and~\ref{sec:fixed_hz}, respectively.
Predictions based on exact NLO EW calculations (apart from 
photon-initiated contributions that have been neglected) are compared against the
Sudakov NLL approximation (see App.~\ref{sec:sudakov}), which includes
virtual EW logarithms supplemented by an exact treatment of QED radiation.

The fixed-order results presented in this section are not subject to
the cuts and definitions of Sec.~\ref{sec:cuts}.  No acceptance
cut is applied, and differential observables are defined in terms of
the momenta of the Higgs and vector bosons. 
The latter are
defined in terms of the momenta of their leptonic decay products at the 
level of underlying-Born events, i.e.~before the emission of NLO radiation.
Photons and QCD partons are
clustered in a fully democratic way using the anti-$k_{\sss T}$
algorithm with $R=0.5$. Effectively this procedure corresponds to an
inclusive treatment of QED radiation.

Besides total cross sections, we consider various differential distributions,
focusing on regions of high invariant masses and transverse momenta, where EW
corrections are enhanced by Sudakov logarithms.  Such phase-space regions
play an important role for experimental analyses of \HV{} production in the
boosted regime.
 
An in-depth validation of our fixed-order NLO EW results for \HV{} production
against the ones implemented in the public Monte Carlo program
\HAWK~\cite{Denner:2011id,Denner:2014cla}, is presented in
App.~\ref{sec:validation}.

\subsection[\HW{} and \HWJ{} production]{$\boldsymbol{\HW}$ and
  $\boldsymbol{\HWJ}$ production}
  \label{sec:fixed_hw}

In this section we focus on NLO results for $pp\to \HW{}$ and $pp\to\HWJ{}$.
In Tab.~\ref{tab:sigtot_NLO_W} we report inclusive NLO cross sections. In the
case of \HWJ{} production, the improved \MINLO{} approach yields finite cross
sections without imposing any minimum transverse momentum on the hardest jet.
For comparison, we report also \HWJ{}~\MINLO{} cross sections for the case
where a minimum $\pt$ of 20~GeV is required for the hardest jet.  In this
case, the \MINLO{} Sudakov form factor plays hardly any role, since it damps
the cross section only at $\pt$ of the order of a few GeV, i.e.~far below the
imposed cut. Thus, at fixed order, the \MINLO{} procedure only affects the
choice of scales, as described in Sec.~\ref{sec:setup}.  The EW corrections
lower the inclusive NLO QCD cross section by roughly $-7\%$ for \HWpm{}
production and $-5\%$ for \HWpmJ{} production, while they amount to only
$-2\%$ when a resolved jet with $\pt^{\sss\rm j_1}>20$~GeV is required in the
\HWpmJ{} calculation.
\begin{table}[htb]
\begin{center}
\begin{tabular}{lccc}
& $HW^-$ NLO & \multicolumn{2}{c}{\HWmJ{} \MINLO}  
\\ 
\hline
selection & inclusive & inclusive & $\pt^{\sss\rm j_1}>20$~GeV\\
\hline\hline
$\sNLOQCD\,[{\rm fb}]$ & $59.25 \pm 0.03 $
& $57.46 \pm 0.02$ & $26.720 \pm 0.008$ \\ 
$\sNLOQCDEW\,[{\rm fb}]$ & $55.31 \pm 0.02$
& $55.3 \pm 0.1$ & $26.19 \pm 0.04$ \\ 
$\sNLOQCDEWNLL\,[{\rm fb}]$ & $59.49 \pm 0.01$
& $59.6 \pm 0.1$ & $27.82 \pm 0.04$ \\ \hline
$\sNLOQCDEW / \sNLOQCD\,$ & 0.93  & 0.96 & 0.98 \\
$\sNLOQCDEWNLL / \sNLOQCD\,$ & 1.00 &  1.04 & 1.04 \\
\hline
 &  &  & \\ 
& $HW^+$ NLO & \multicolumn{2}{c}{\HWpJ{} \MINLO}  
\\ 
\hline
selection & inclusive & inclusive & $\pt^{\sss\rm j_1}> 20$~GeV\\
\hline\hline
$\sNLOQCD\,[{\rm fb}]$ & $93.24 \pm 0.05 $
& $ 90.8 \pm 0.2 $ & $ 42.2 \pm 0.1 $ \\ 
$\sNLOQCDEW\,[{\rm fb}]$ & $86.91 \pm 0.02$
& $ 86.2 \pm 0.2$ & $ 41.16 \pm 0.09 $ \\ 
$\sNLOQCDEWNLL\,[{\rm fb}]$ & $93.37 \pm 0.02$
& $ 93.0 \pm 0.2$ & $ 43.74 \pm 0.09$ \\ \hline
$\sNLOQCDEW / \sNLOQCD\,$ & 0.93  & 0.95 & 0.98 \\
$\sNLOQCDEWNLL / \sNLOQCD\,$ & 1.00 &  1.02 & 1.04 \\
\hline
\end{tabular}
\end{center}
\caption{NLO total cross sections for \HV~(second column) and \HVJ~(third and
  fourth column) production with $V=W^-$ (top) and $V=W^+$ (bottom), at a
  center-of-mass energy of $\sqrt{s} = 13$~TeV, at NLO QCD, NLO QCD+EW, and
  in the NLO QCD+NLL EW approximation. The \HVJ{} cross sections are based on
  the improved \MINLO{} procedure (third and fourth column). The effect of a
  cut of $\pt^{\sss\rm j_1}>20$~GeV on the transverse momentum of the hardest
  jet in \HVJ{} production is shown in the last column. Listed uncertainties
  are due to Monte Carlo integration.}
\label{tab:sigtot_NLO_W}
\end{table}  
Inclusive cross sections in the NLO QCD+NLL EW approximation 
differ by several percent from the exact NLO QCD+EW results. This
is expected, since the NLL approximation is only valid in the 
high-energy regime.

In the following we investigate the impact of EW corrections and the 
validity of the NLL approximation in differential distributions
for \HWm{} and \HWmJ{} production. Results for $\HWp(j)$ production 
(not shown) are very similar.

\begin{figure}[htb]
\begin{center}
  \includegraphics[width=\wsmall]{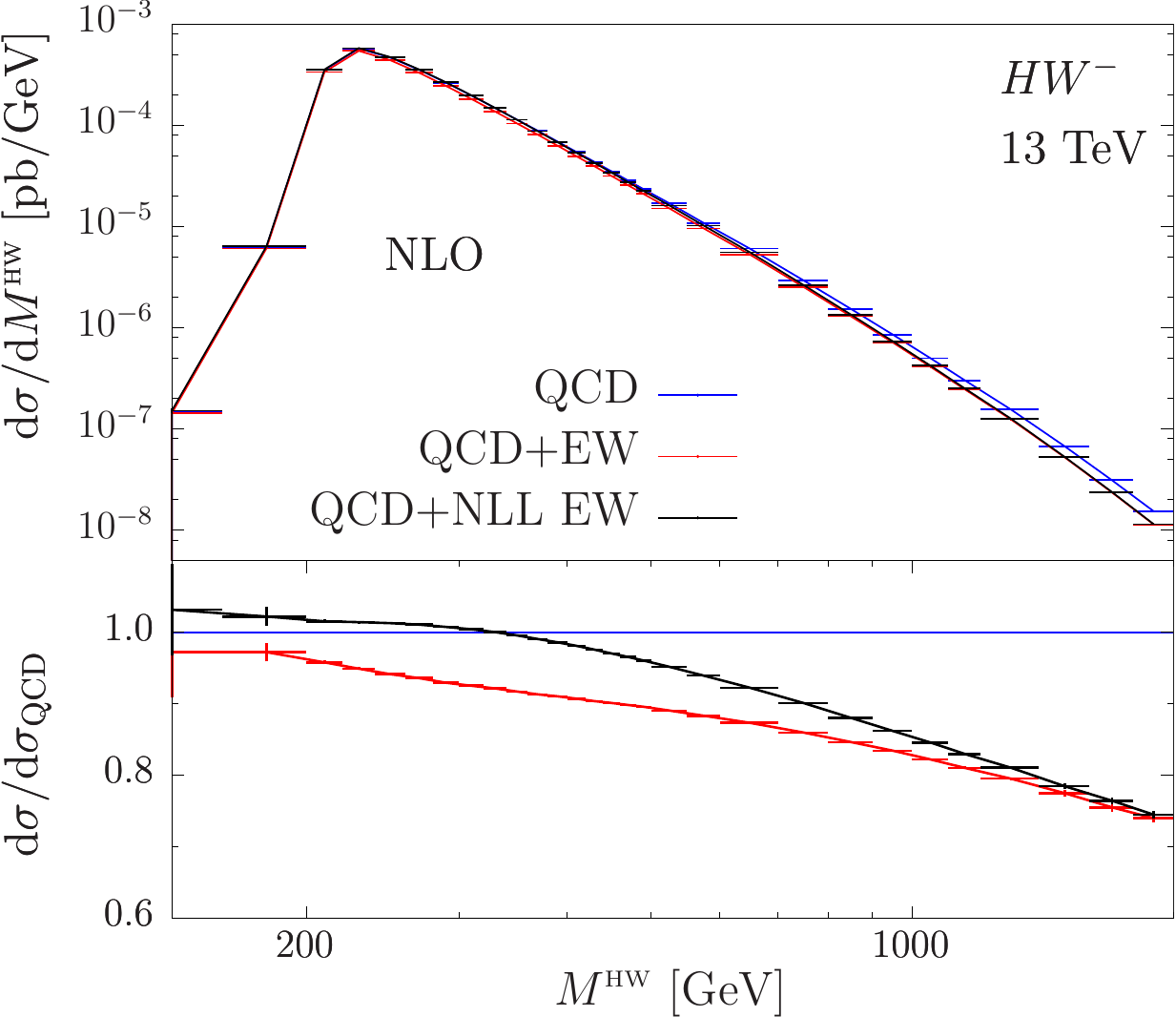}
  \includegraphics[width=\wsmall]{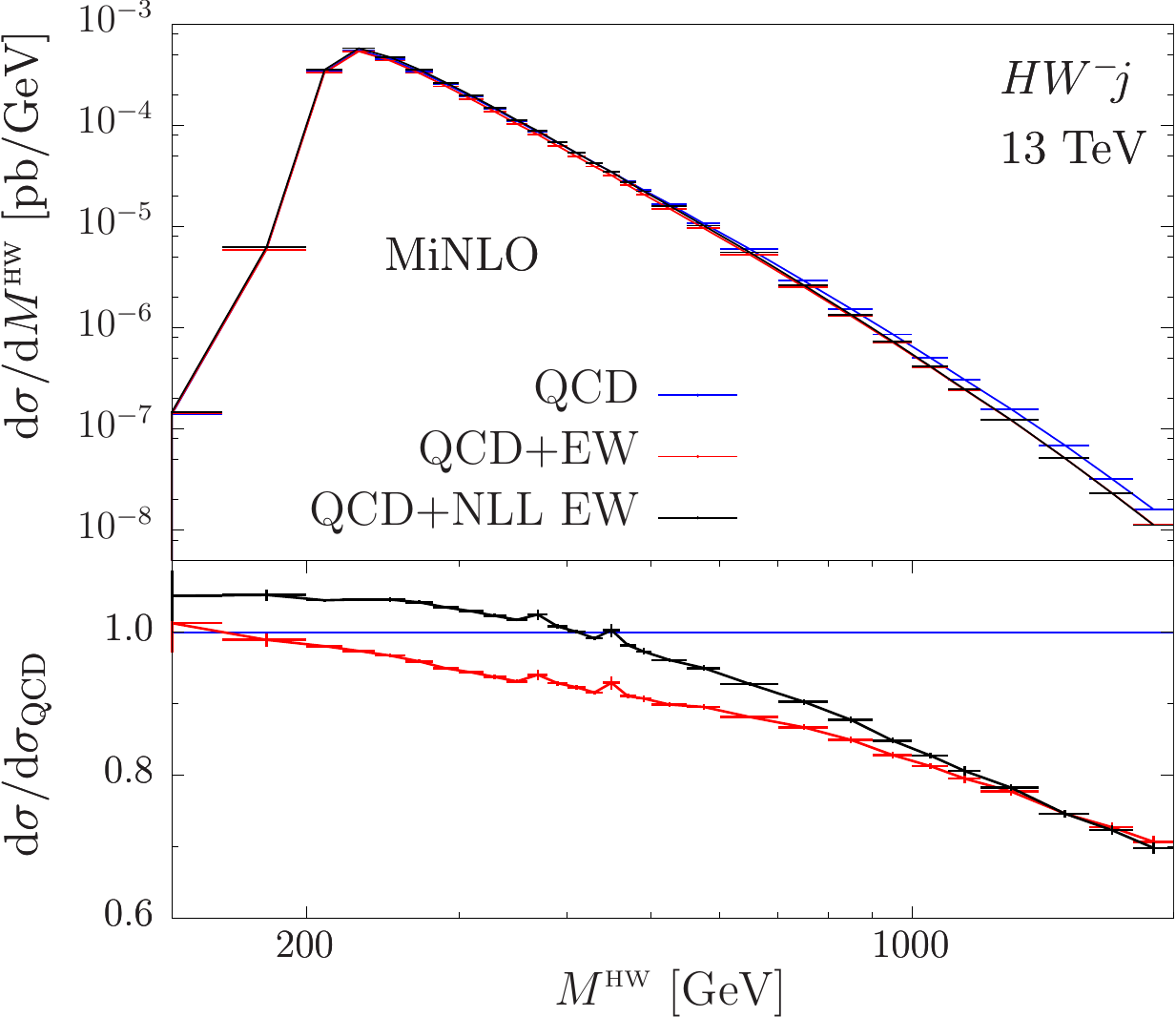}
\end{center}
\caption{NLO predictions for the invariant mass of the \HWm{} pair in
  \HWm{}~(left) and \HWmJ{}~(right) production. Shown are predictions at NLO
  QCD~(blue), NLO QCD+EW~(red) and at NLO QCD+NLL EW~(black). In the lower
  panel we plot the ratio with respect to NLO QCD.}
\label{fig:HW-13TeV_HW-m_NLO_QCD-EW-sud}
\end{figure}
In Fig.~\ref{fig:HW-13TeV_HW-m_NLO_QCD-EW-sud} we plot the invariant mass of
the reconstructed \HWm{} pair, both for \HWm{} and \HWmJ{} production. The
three curves represent predictions at NLO QCD, NLO QCD+EW and in NLO QCD+NLL
EW approximation.  While EW corrections have a moderate impact on the total
cross sections, they affect the tail of the $M^{\sss \rm\!HW}$ distribution
in a substantial way. At large $M^{\sss \rm\!HW}$ we observe the typical
Sudakov behavior, with increasingly large negative EW corrections that reach
the level of $-25\%$ ($-30\%$) for \HV{} (\HVJ{}) production at 2~TeV.
The Sudakov NLL approximation captures the bulk of these large EW corrections
as expected. In the tail it agrees at the percent level with the exact result
for both processes, while for moderate invariant masses it overestimates EW
correction effects by up to 5\%.

\begin{figure}[htb]
\begin{center}
  \includegraphics[width=\wsmall]{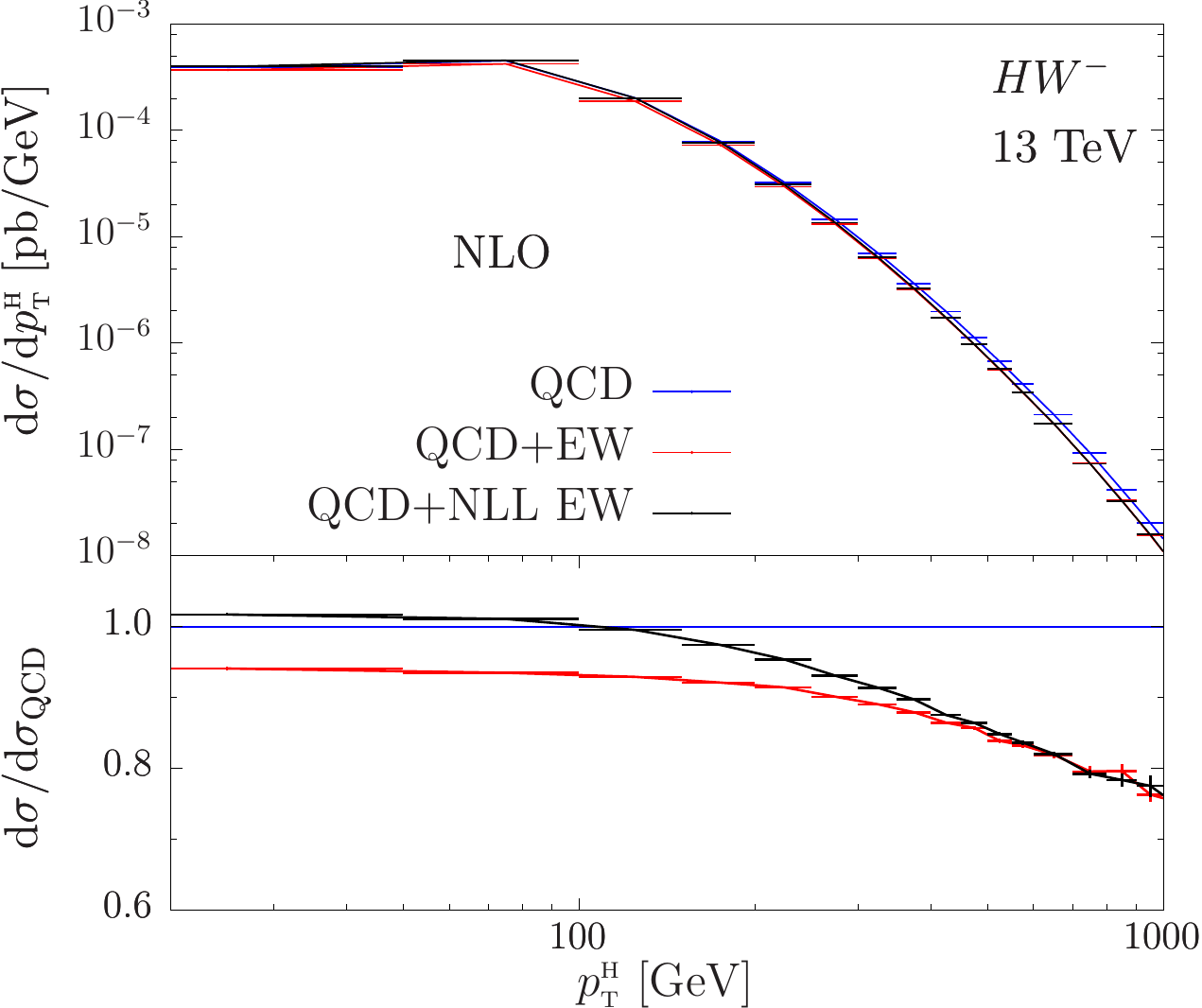}
  \includegraphics[width=\wsmall]{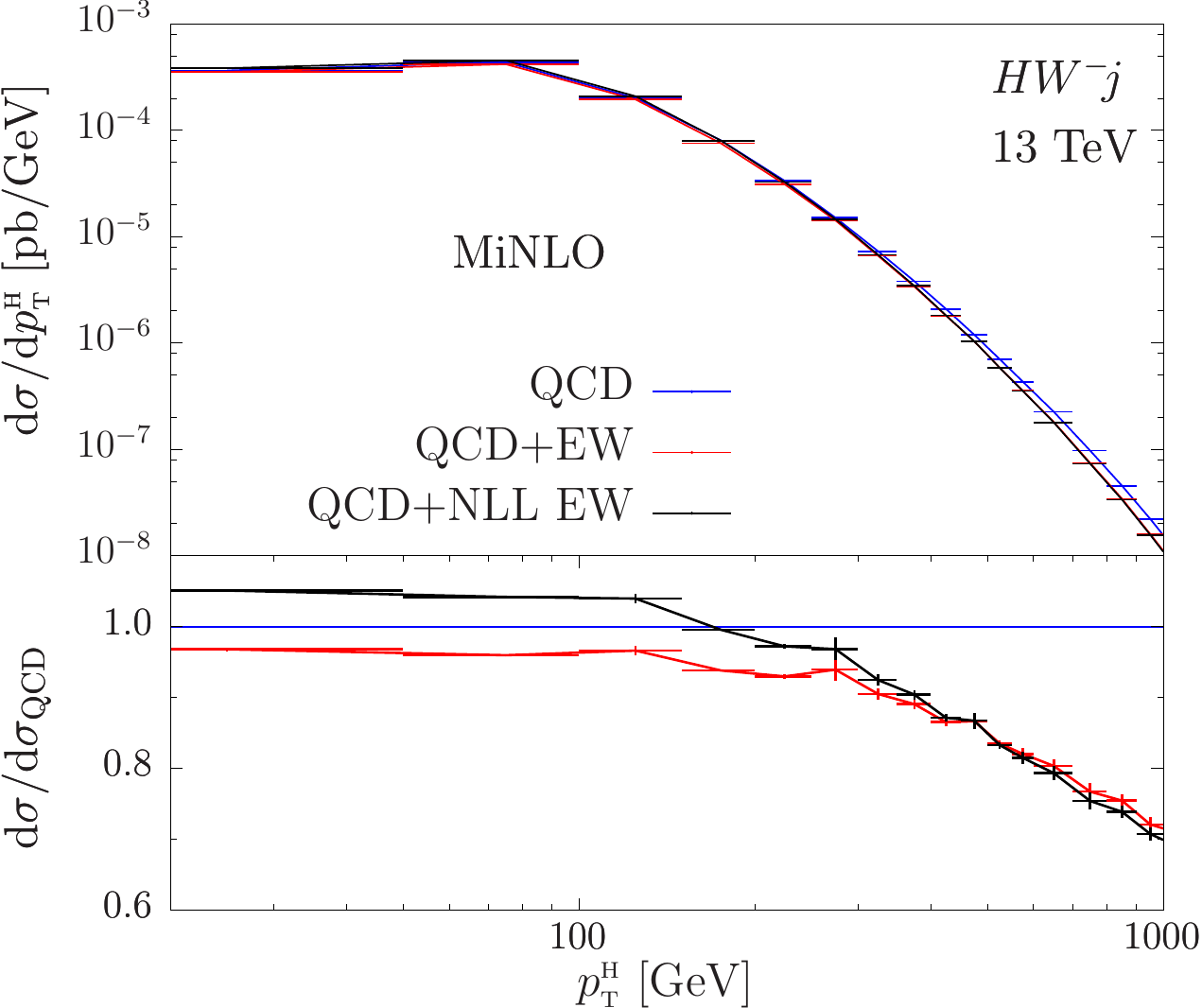}
\end{center}
\caption{NLO predictions for the transverse momentum of the Higgs
  boson in \HWm{}~(left) and \HWmJ{}~(right) production. Predictions and labels
  as in Fig.~\ref{fig:HW-13TeV_HW-m_NLO_QCD-EW-sud}.}
\label{fig:HW-13TeV_H-ptlarge_NLO_QCD-EW-sud}
\end{figure}
In Fig.~\ref{fig:HW-13TeV_H-ptlarge_NLO_QCD-EW-sud} we investigate the
transverse momentum of the Higgs boson. Also in this case EW corrections
become negative and large in the tail, exceeding $-20\%$ in the TeV region.
For both processes the Sudakov approximation agrees at the percent level with
exact NLO EW results for $\pt^{\sss\rm H} > 300$~GeV.

\begin{figure}[htb]
\begin{center}
  \includegraphics[width=\wsmall]{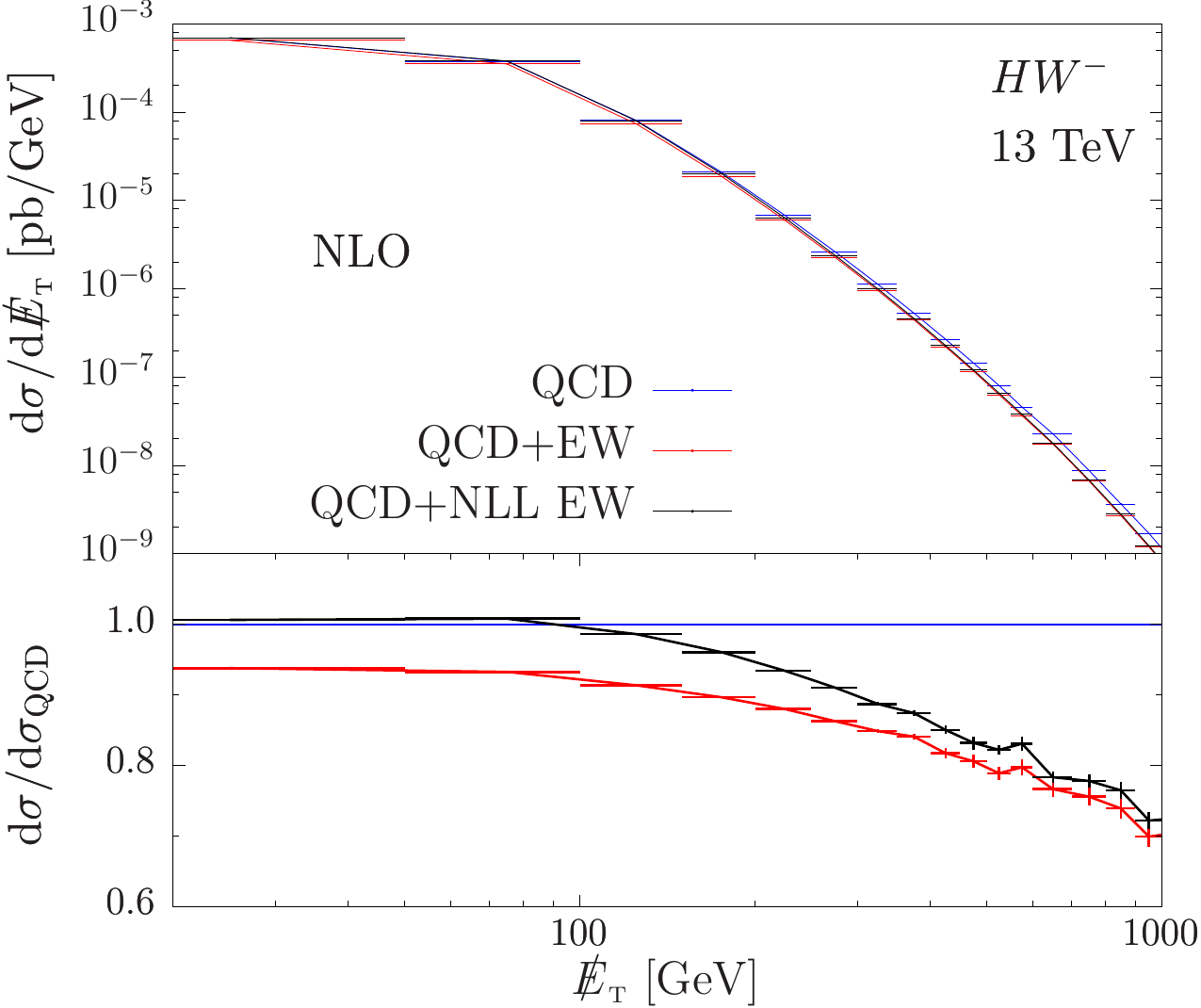}
  \includegraphics[width=\wsmall]{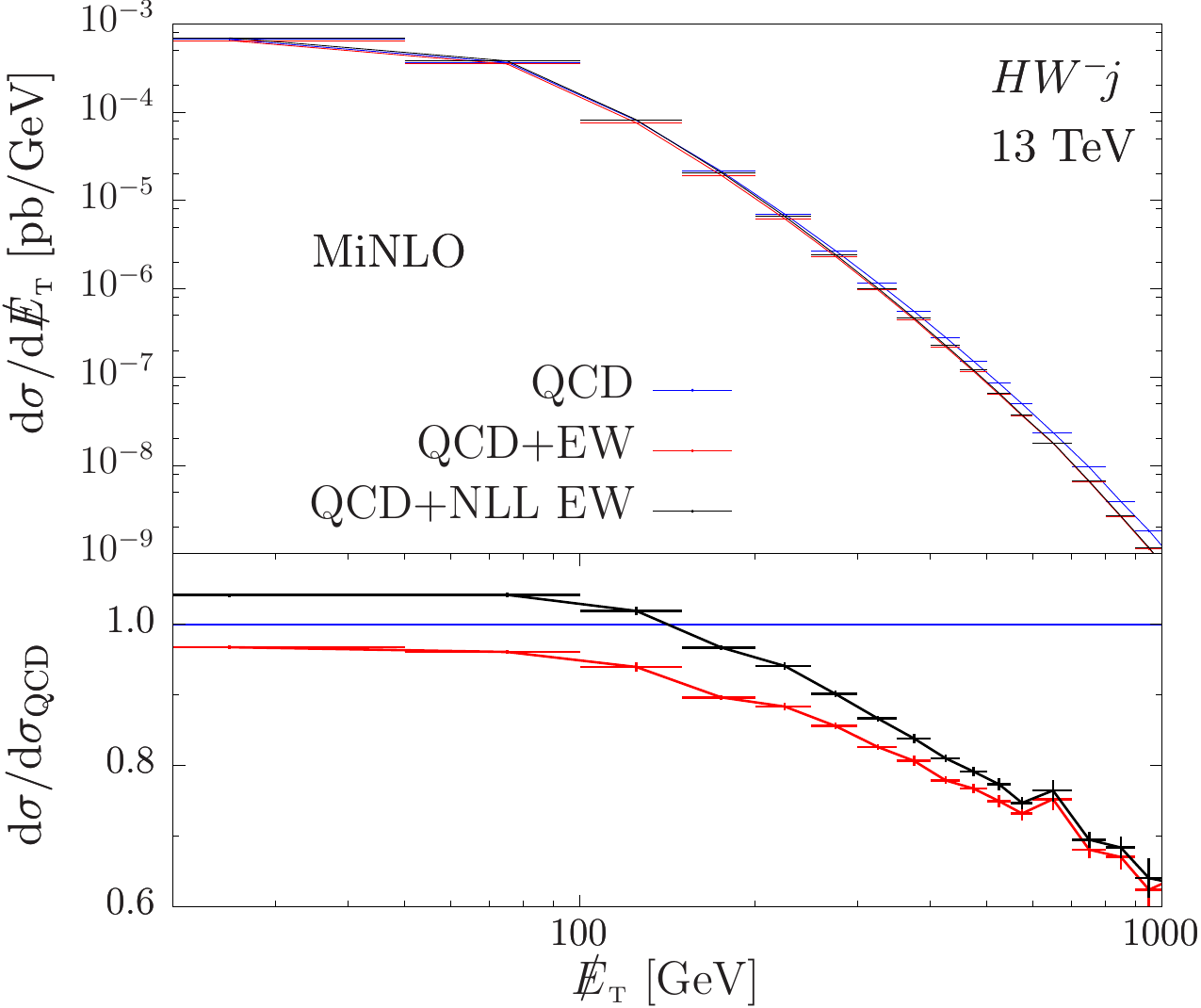}
\end{center}
\caption{NLO predictions for the missing transverse momentum in
  \HWm{}~(left) and \HWmJ{}~(right) production. Predictions and labels as in
  Fig.~\ref{fig:HW-13TeV_HW-m_NLO_QCD-EW-sud}.}
\label{fig:HW-13TeV_miss-pt_NLO_QCD-EW-sud}
\end{figure}
The EW corrections have a sizable impact also on the missing transverse
momentum distribution, shown in
Fig.~\ref{fig:HW-13TeV_miss-pt_NLO_QCD-EW-sud}. Size and shape of these
corrections are very similar to the ones observed for the Higgs boson $\pt$
distribution.

\begin{figure}[htb]
  \begin{center}
    \includegraphics[width=\wsmall]{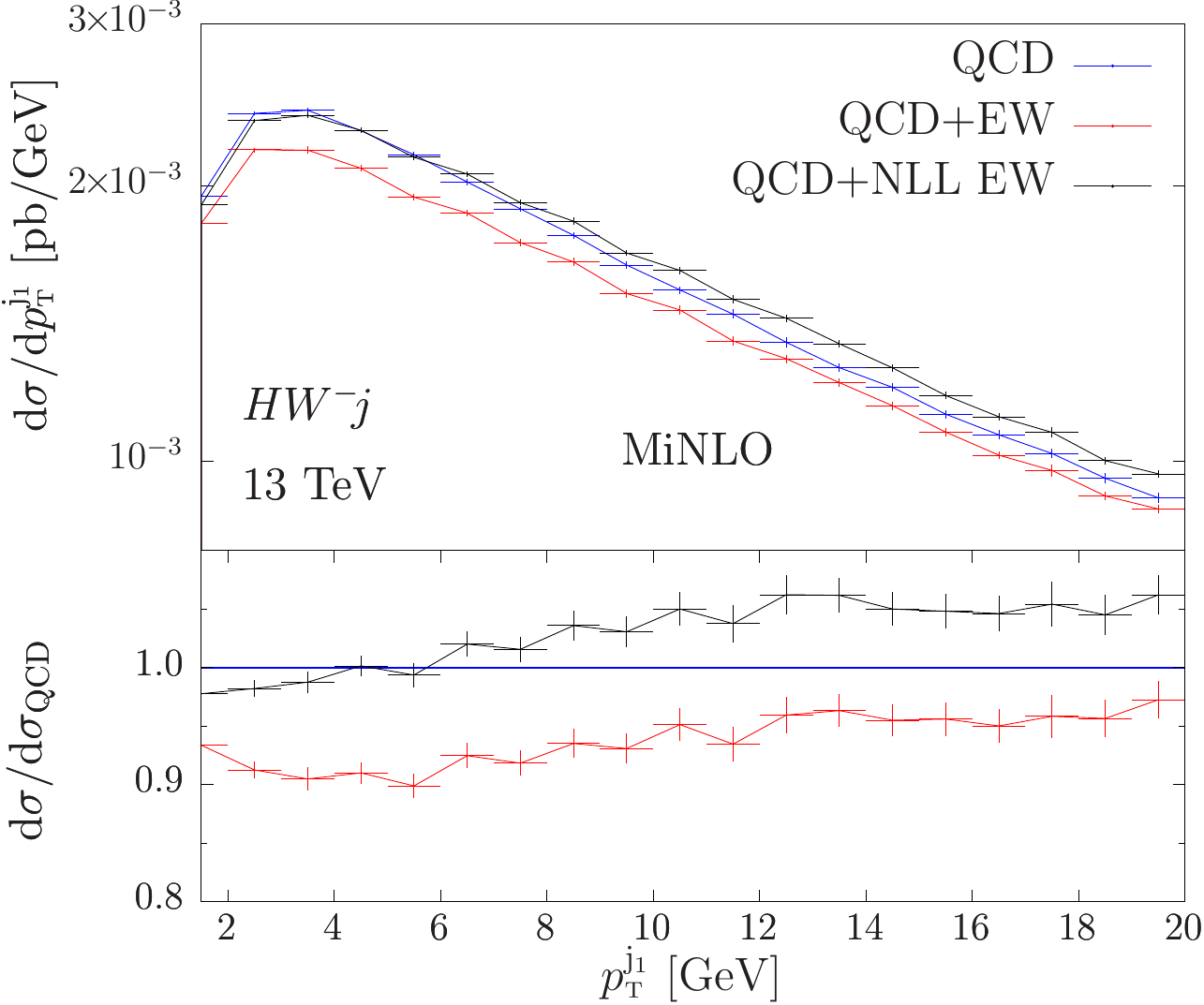}
    \includegraphics[width=\wsmall]{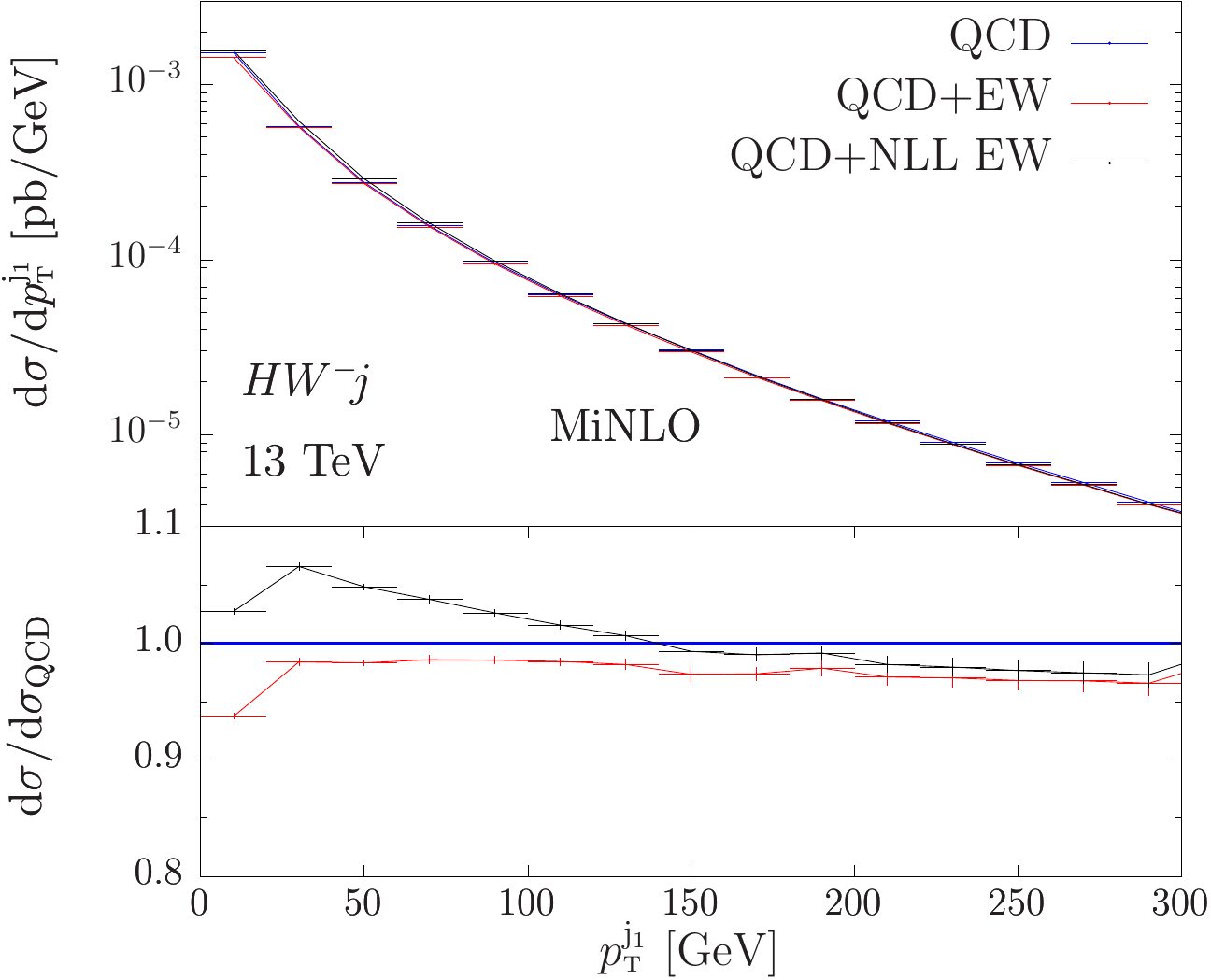}
    \caption{NLO predictions for the transverse momentum of the leading jet
      in \HWmJ{} production for different $\pt^{\rm \sss j_1}$ ranges.  The left plot
      corresponds to the first bin of the right plot.  Predictions and labels
      as in Fig.~\ref{fig:HW-13TeV_HW-m_NLO_QCD-EW-sud}.}
    \label{fig:HWJ-13TeV_j1-pt_NLO_QCD-EW-sud}
  \end{center}
\end{figure}
In Fig.~\ref{fig:HWJ-13TeV_j1-pt_NLO_QCD-EW-sud} we present $\HWmJ{}$
predictions for the distribution in the $\pt$ of the leading jet.  At low
$\pt^{\rm \sss j_1}$ (left plot) the \MINLO{} Sudakov form factor damps soft
and collinear singularities at zero transverse momentum yielding finite cross
sections below the Sudakov peak, which is located around 3~GeV.
Concerning EW effects, the NLL approximation converges to the exact NLO
results already for values of $\pt^{\sss \rm j_1}$ around 200~GeV.  In the
region of moderate transverse momentum, NLO EW corrections are nearly
constant, and in the limit of vanishing jet-$\pt$ they converge towards an EW
$K$-factor that is very close to the one of the NLO QCD+EW calculation for
the inclusive $pp\to \HWm$ cross section (see Tab.~\ref{tab:sigtot_NLO_W}).
This observation is consistent with the theoretical
considerations presented in Sec.~\ref{se:MINLO}, namely with the fact that
EW corrections are insensitive to soft and collinear QCD radiation, and
that \MINLO{} predictions for \HVJ{} production preserve NLO QCD+EW accuracy
when the extra jet is integrated out. In fact, in the inclusive distributions
of
Figs.~\ref{fig:HW-13TeV_HW-m_NLO_QCD-EW-sud}--\ref{fig:HW-13TeV_miss-pt_NLO_QCD-EW-sud},
we observe that the EW corrections obtained from \HWm{} and \HWmJ{}
calculations are very similar, with small differences that can be attributed
to NNLO effects.

\begin{figure}[htb]
  \begin{center}
    \includegraphics[width=\wsmall]{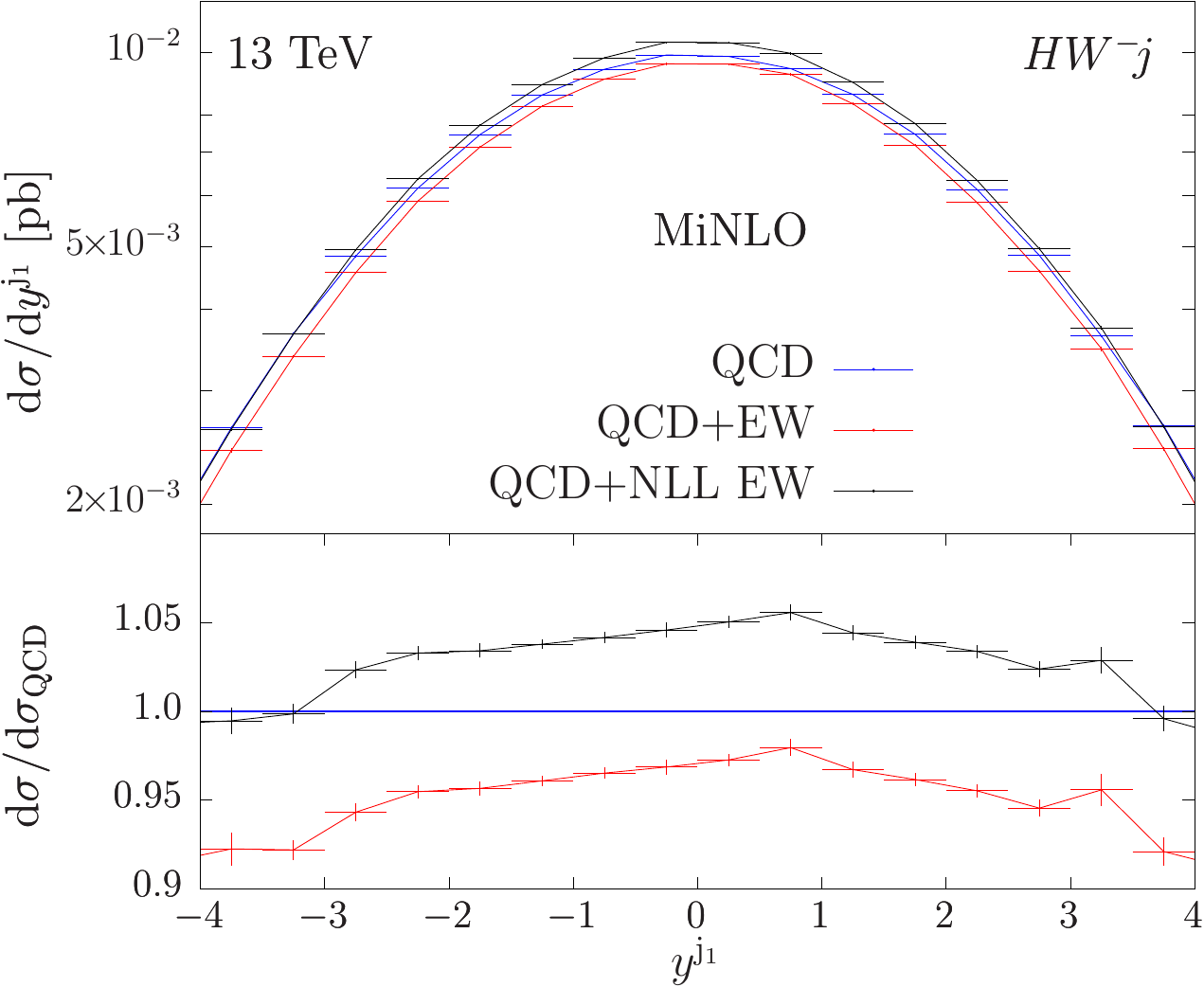}
    \caption{NLO predictions for the rapidity of the leading jet in 
      \HWmJ{} production. Same curves and labels as in
      Fig.~\ref{fig:HW-13TeV_HW-m_NLO_QCD-EW-sud}.}
    \label{fig:HWJ-13TeV_j1-y_NLO_QCD-EW-sud}
  \end{center}
\end{figure}
Finally, in Fig.~\ref{fig:HWJ-13TeV_j1-y_NLO_QCD-EW-sud}
we see that the EW corrections affect the rapidity of the leading jet 
in a rather uniform way over the whole phase space.
We have observed a similar behavior of EW  
corrections in several angular distributions.

\subsection[\HZ{} and \HZJ{} production]{$\boldsymbol{\HZ}$ and $\boldsymbol{\HZJ}$ production}
  \label{sec:fixed_hz}
In line with the discussion of \HW{} and \HWJ{} production, we present in
this section fixed-order results for \HZ{} and \HZJ{} production.  In
Tab.~\ref{tab:sigtot_NLO_Z} we collect the inclusive cross sections at NLO
QCD, NLO QCD+EW and NLO QCD+NLL EW. The EW corrections decrease the total NLO
QCD cross section for \HZ{} production by about 4\%, and by about 1\% for
inclusive \HZJ{} production.  In the presence of a jet threshold of $20$~GeV,
the EW corrections are positive and amount to about $4\%$.

\begin{table}[htb]
\begin{center}
\begin{tabular}{lccc}
& $HZ$ NLO & \multicolumn{2}{c}{\HZJ{} \MINLO}  
\\ 
\hline
selection & inclusive & inclusive & $\pt^{\sss\rm j_1}> 20$~GeV\\
\hline\hline
$\sNLOQCD\,[{\rm fb}]$ &  $25.551 \pm 0.005$
& $24.801 \pm 0.009$ & $11.720 \pm 0.004$ \\ 
$\sNLOQCDEW\,[{\rm fb}]$ & $24.382 \pm 0.008$
& $24.59 \pm 0.07$ & $12.22 \pm 0.02$ \\ 
$\sNLOQCDEWNLL\,[{\rm fb}]$ & $25.457 \pm 0.008$
& $25.84 \pm 0.07$ & $12.69 \pm 0.01$ \\ \hline
$\sNLOQCDEW / \sNLOQCD\,$ & 0.95  & 0.99 & 1.04 \\
$\sNLOQCDEWNLL / \sNLOQCD\,$ & 1.00 &  1.04 & 1.08 \\
\hline
\end{tabular}
\end{center}
\caption{NLO total cross sections for \HZ~(second column) and \HZJ~(third and
  fourth column) production at a center-of-mass energy of $\sqrt{s} =
  13$~TeV. Predictions and labels as in Tab.~\ref{tab:sigtot_NLO_W}.}
\label{tab:sigtot_NLO_Z}
\end{table}

\begin{figure}[htb]
\begin{center}
  \includegraphics[width=\wsmall]{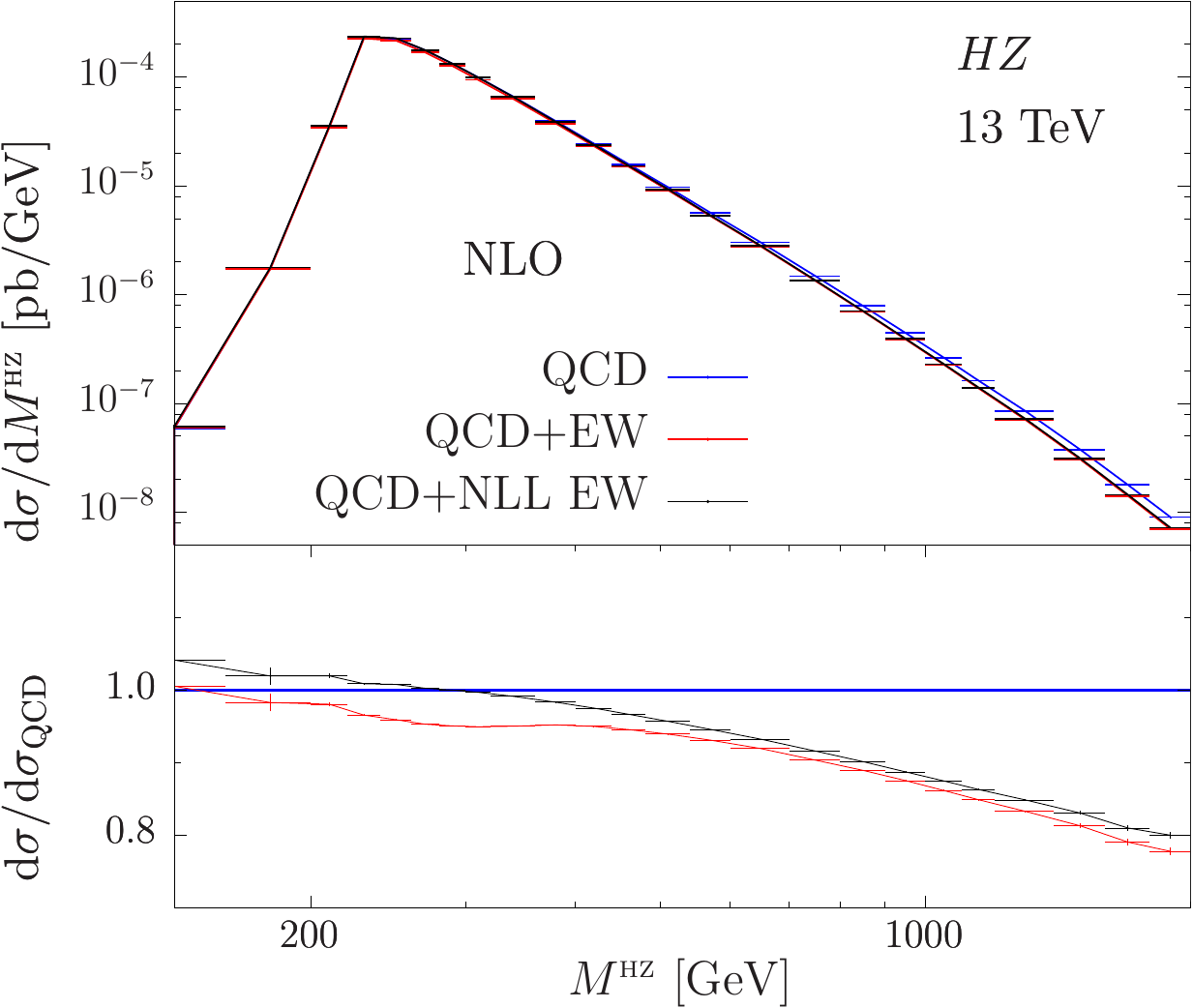}
  \includegraphics[width=\wsmall]{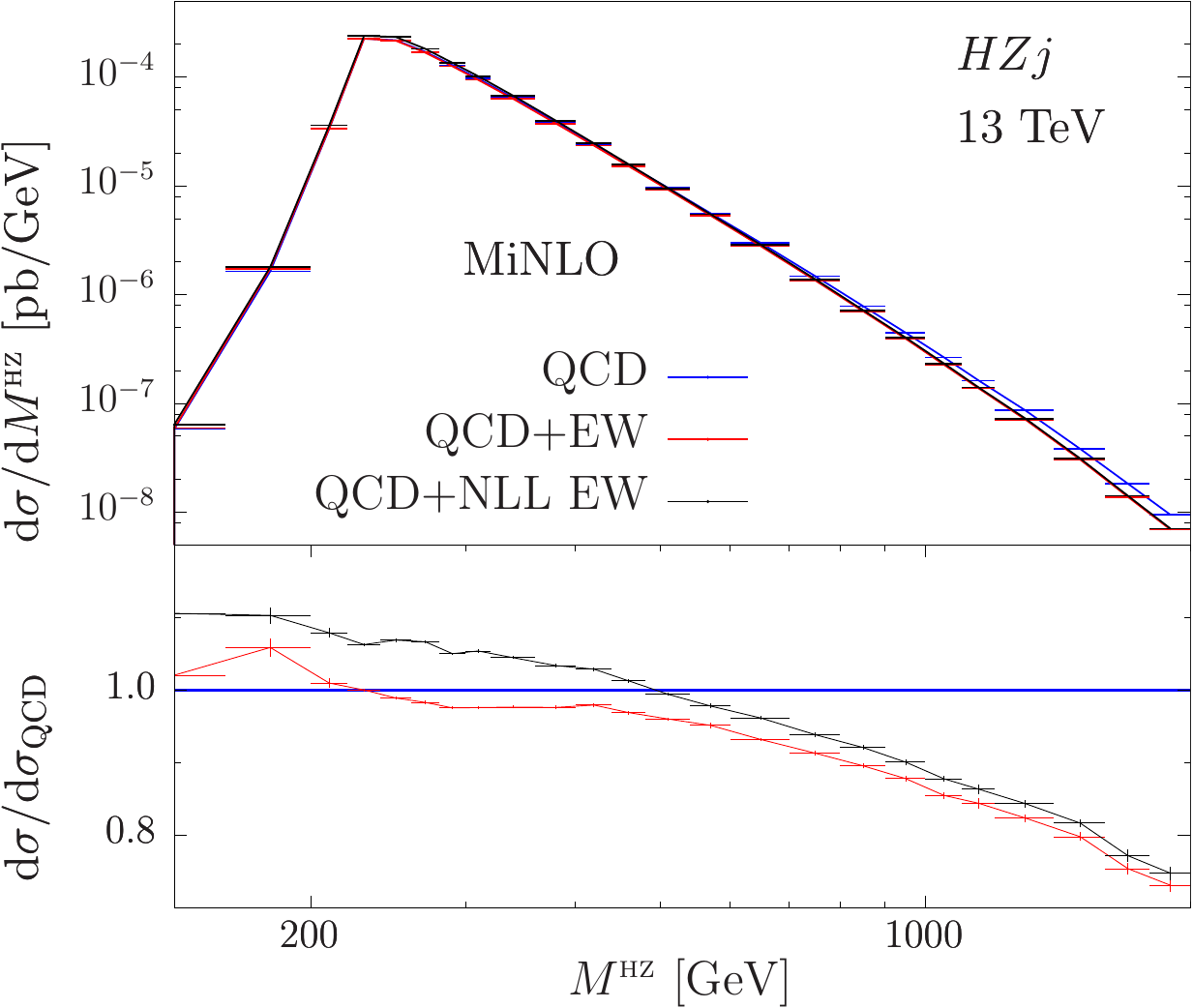}
\end{center}
\caption{NLO predictions for the invariant mass of the \HZ{} pair in
  \HZ{}~(left) and \HZJ{}~(right) production. The three curves represent the
  QCD, QCD+EW and the QCD+NLL EW predictions. The lower panel displays ratios
  with respect to NLO QCD.}
\label{fig:HZ-13TeV_HZ-m_NLO_QCD-EW-sud}
\end{figure}

\begin{figure}[htb]
\begin{center}
  \includegraphics[width=\wsmall]{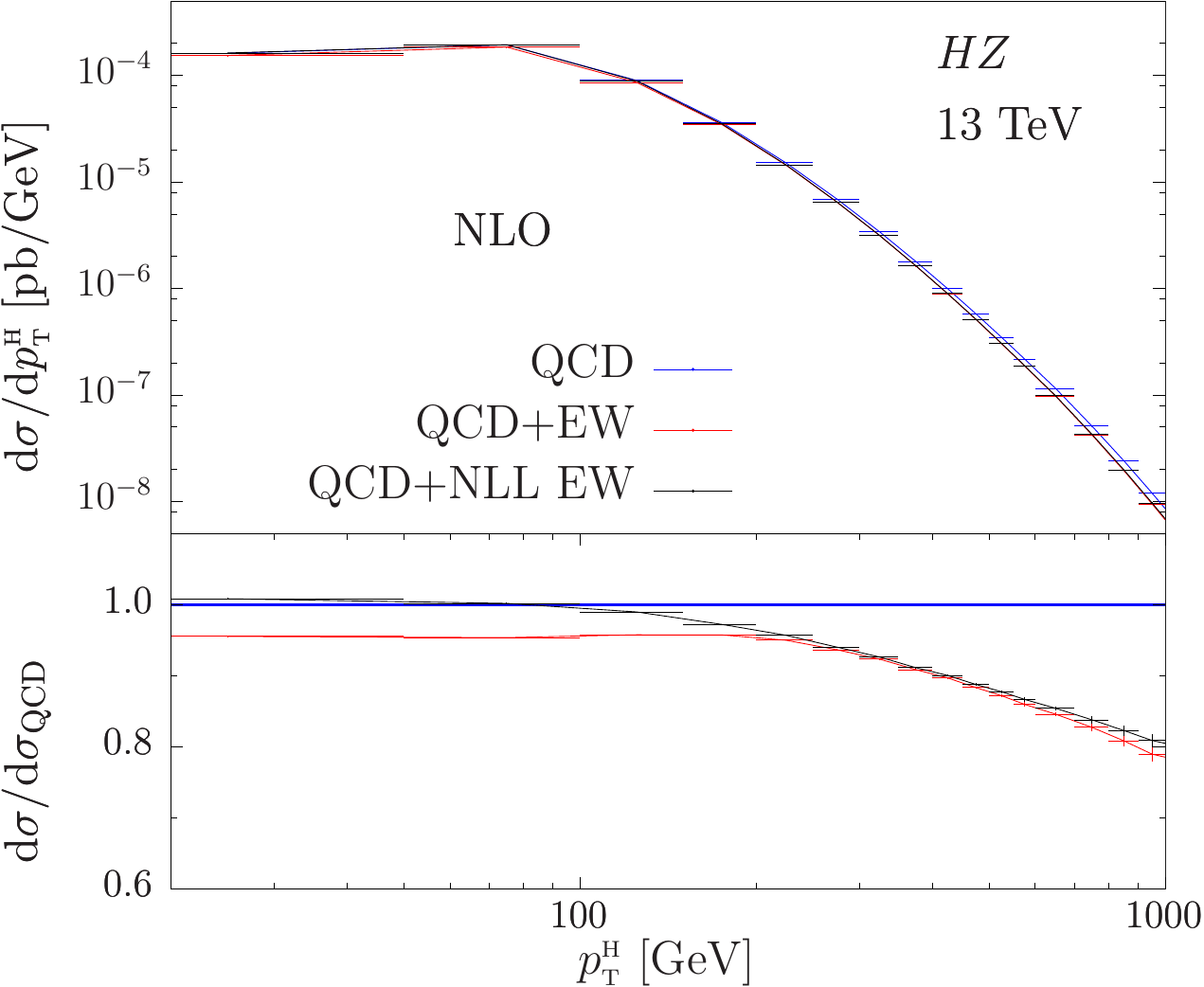}
  \includegraphics[width=\wsmall]{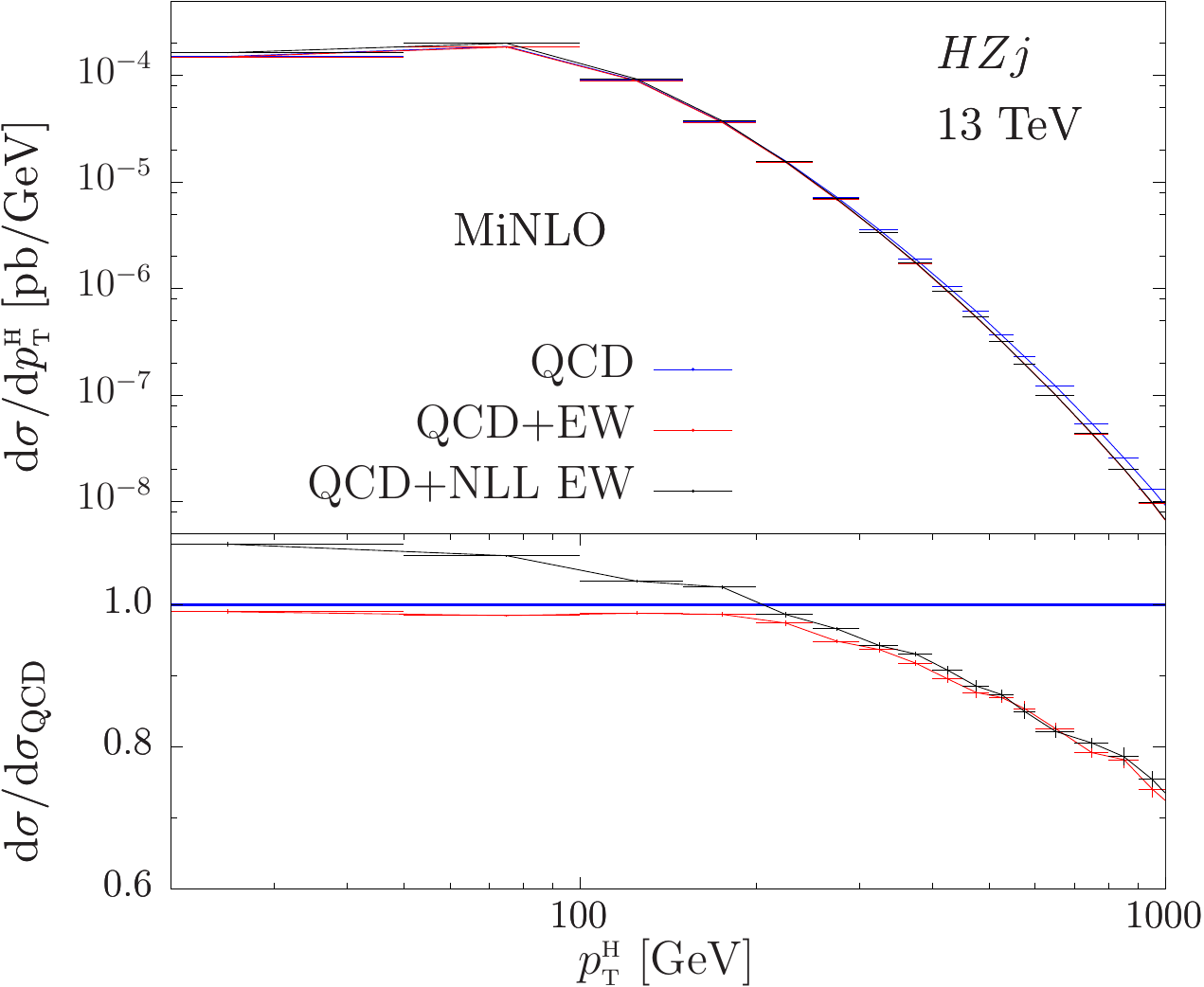}
\end{center}
\caption{NLO predictions for the Higgs boson transverse momentum in 
  \HZ{}~(left) and \HZJ{}~(right) production. Same curves and labels as in
  Fig.~\ref{fig:HZ-13TeV_HZ-m_NLO_QCD-EW-sud}.}
\label{fig:HZ-13TeV_H-ptlarge_NLO_QCD-EW-sud}
\end{figure}

In Figs.~\ref{fig:HZ-13TeV_HZ-m_NLO_QCD-EW-sud}
and~\ref{fig:HZ-13TeV_H-ptlarge_NLO_QCD-EW-sud} we show distributions of the
invariant mass of the reconstructed \HZ{} pair and of the transverse momentum
of the Higgs boson.  Similarly as for $\HW(j)$ production, the inclusion of
EW corrections is essential in the tails of these distributions, where the
NLL Sudakov approximation agrees well with the exact NLO EW predictions.

\section{Results for $\boldsymbol{\HV}$ production at NLO+PS QCD+EW}
\label{sec:results}
\label{sec:results_hv}
In this section we present NLO QCD+EW predictions for \HV{} production
completed by the \PythiaEightPone{} QCD+QED parton shower using the ``Monash
2013'' tune~\cite{Skands:2014pea}.
All predictions are subject to the cuts and physics object definitions
specified in Sec.~\ref{sec:cuts}, and NLO EW corrections are treated exactly
throughout, except for  photon-initiated processes, that have been neglected.  The NLL Sudakov approximation is only used in order to speed up
the Monte Carlo integration, as detailed in App.~\ref{sec:fast_ew}.

In Sec.~\ref{sec:FOvsNLOPS} we compare predictions at
fixed-order NLO QCD+EW against corresponding predictions at the level of Les
Houches events, which include only the hardest emission generated in the
\RES{} framework, and at NLO+PS level, where the full QCD+QED parton shower
is applied.
The effect of EW corrections is studied in Sec.~\ref{sec:EWcorrNLOPS} 
in the case of fully showered NLO+PS simulations.

By default, at NLO+PS level, the full QCD+QED parton shower is applied, both
for NLO QCD+EW and for pure NLO QCD simulations. Occasionally, we also
present NLO QCD simulations with a pure QCD shower, where QED radiation is
switched off.  Such predictions are labeled ``QCD (no QED shower)''.

The consistent combination of the NLO radiation to the parton shower requires
the vetoing of shower emissions that are harder than the radiation generated
in the \RES{} framework. 
Since no standard interface is available in a multi-radiation scheme, we have
implemented a dedicated veto procedure on the \PythiaEightPone{} showered
events, as described in App.~\ref{app:py8_interface}.
This veto procedure is applied in case of NLO QCD+EW simulations.  Instead,
in case of NLO QCD simulations combined with the \PythiaEightPone{} QCD+QED
shower, only QCD radiation is restricted by the \RES{} hardest scale, while
arbitrarily hard QED radiation can be generated by the shower.

We have verified that inclusive cross sections at NLO+PS QCD and NLO+PS QCD+EW
agree within statistical uncertainties with the corresponding fixed-order
results reported in Tabs.~\ref{tab:sigtot_NLO_W} and \ref{tab:sigtot_NLO_Z}.
Thus, in the following we will focus on differential distributions.

\newcommand\comparNLP{Results at NLO QCD+EW are compared at fixed order, at
  the level of Les Houches events~(LHE), and including also the full QCD+QED
  parton shower of \PythiaEightPone{}~(NLO+PS).  }

\newcommand\comparNLPminlo{Improved \MINLO{} results for $pp\to \HVJ$ at NLO
  QCD+EW are compared at fixed order, at the level of Les Houches
  events~(LHE), and including also the full QCD+QED parton shower of
  \PythiaEightPone{}~(NLO+PS).  }

\subsection{From fixed NLO QCD+EW to NLO+PS QCD+EW}
\label{sec:FOvsNLOPS}
In this section we compare NLO QCD+EW predictions at fixed order with NLO+PS
ones at LHE level and completed with the \PythiaEightPone{} shower. Since the
various Higgsstrahlung processes behave in a very similar way, we will focus
on \HWm{} production.

\begin{figure}[htb]
  \begin{center}
    \includegraphics[width=\wsmall]{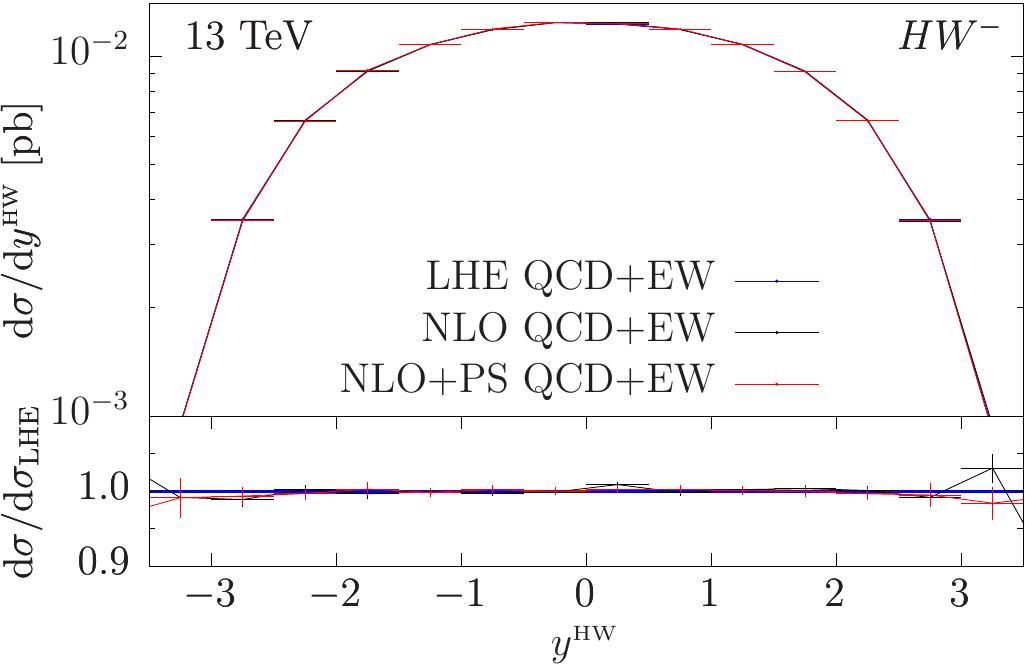}
    \includegraphics[width=\wsmall]{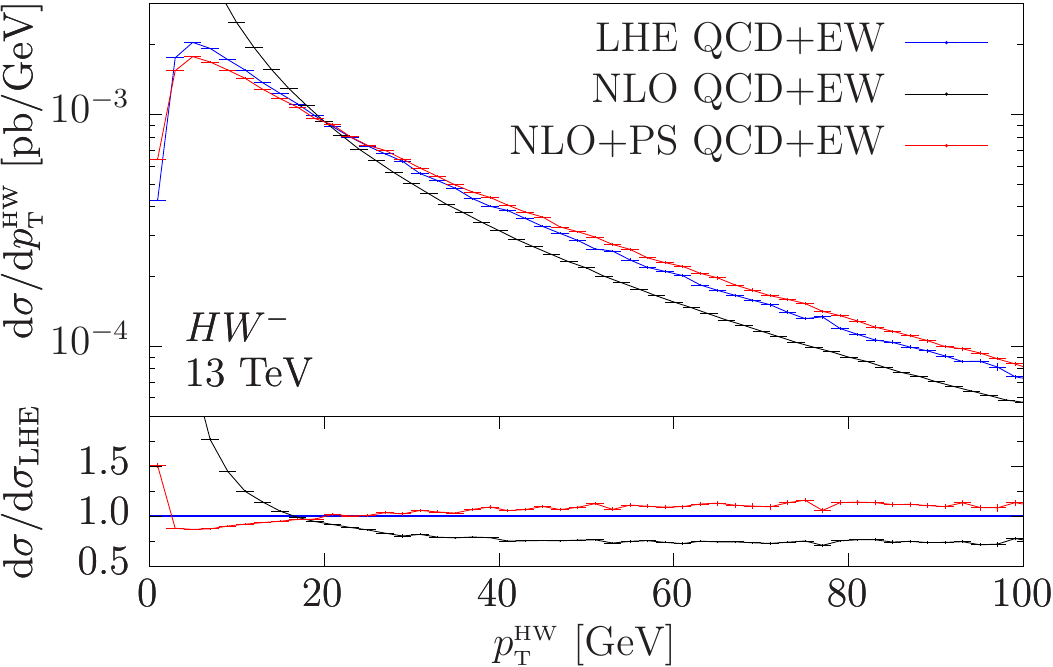}
    \caption{Rapidity~(left) and transverse-momentum distribution~(right) of
      the \HWm{} pair in \HWm{} production. \comparNLP{} In the ratio plot
      results are normalized with respect to the LHE level prediction.}
    \label{fig:W13_HW-y_NLOvsLHvsPY8}
  \end{center}
\end{figure}

In Fig.~\ref{fig:W13_HW-y_NLOvsLHvsPY8} we plot the rapidity of the
reconstructed \HWm{} pair, which is NLO accurate, and its transverse
momentum, which is only LO accurate. Due to the inclusiveness of the rapidity
of the \HWm{} pair, we find, as expected, very good agreement, within the
integration errors, among the three predictions. The fixed-order curve for
the transverse momentum displays the typical divergent behavior at low $\pt$.
At LHE level, instead, the divergence is tamed by the Sudakov form
factor. The effect of the parton shower is modest in the tail of this
distribution, while at low $\pt$ it slightly shifts the position of the
Sudakov peak.

\newcommand\comparLP{Comparison among NLO QCD+EW results at NLO+PS accuracy
  at the level of \RES{} Les Houches events (LHE) and combined with the
  \PythiaEightPone{} shower (NLO+PS).}

\begin{figure}[htb]
 \begin{center}
   \includegraphics[width=\wsmall]{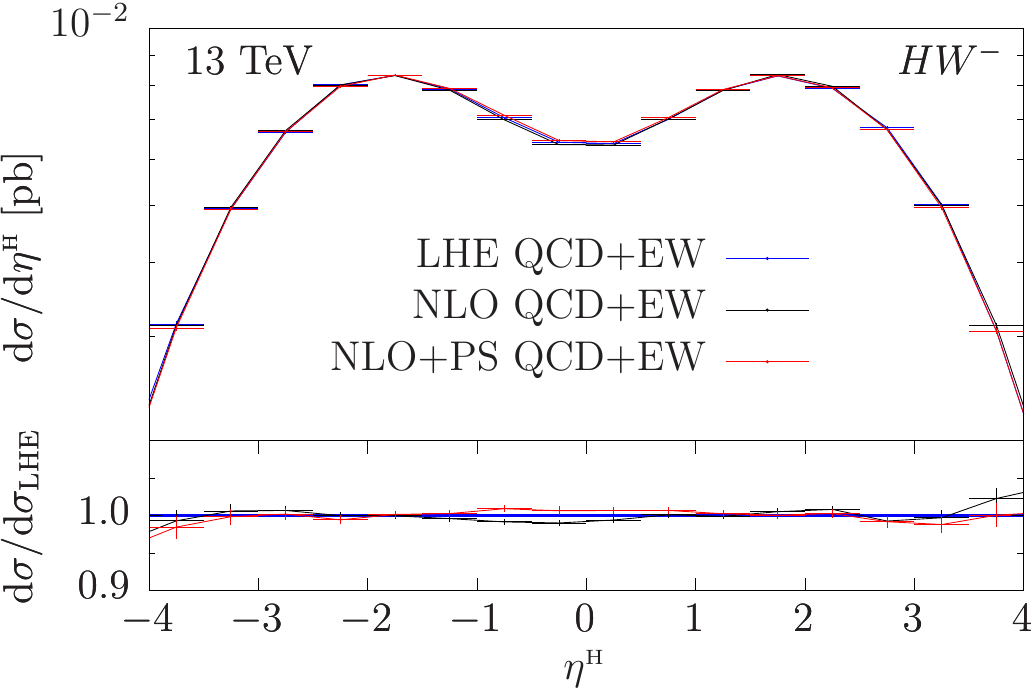}
   \includegraphics[width=\wsmall]{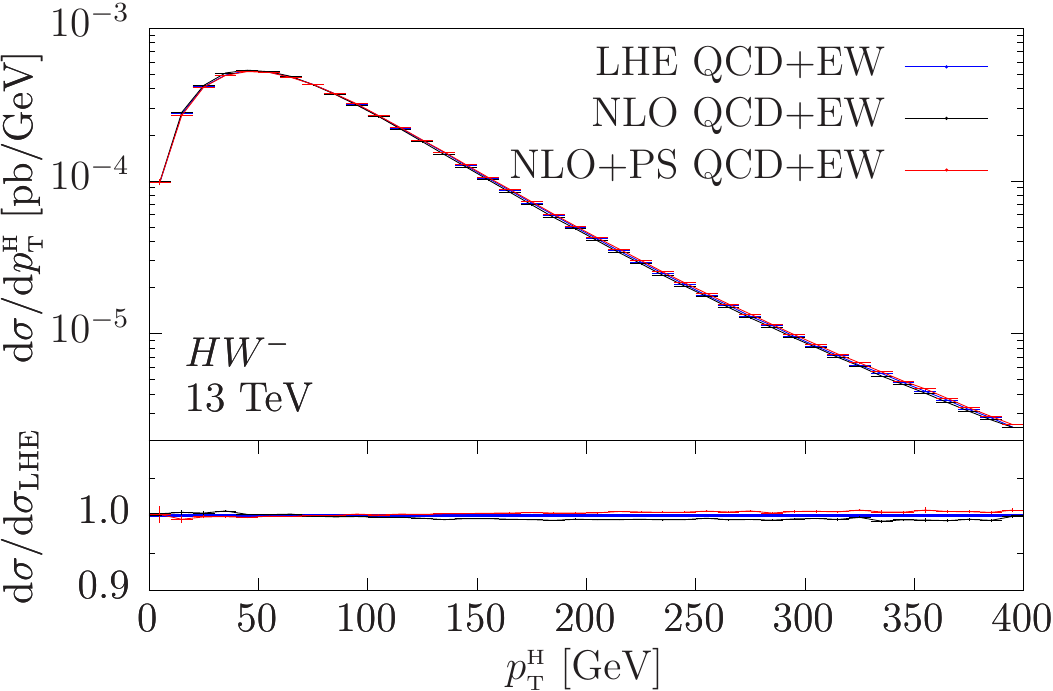}
   \caption{Pseudorapidity~(left) and transverse-momentum
     distribution~(right) of the Higgs boson in \HWm{} production. Same
     curves and labels as in Fig.~\ref{fig:W13_HW-y_NLOvsLHvsPY8}.  }
   \label{fig:W13_H-y_NLOvsLHvsPY8}
 \end{center}
\end{figure}
In Fig.~\ref{fig:W13_H-y_NLOvsLHvsPY8} we plot the pseudorapidity and
the transverse momentum of the Higgs boson: thanks to the inclusiveness of this
variable, we find again very good agreement among the three predictions.

\subsection{Impact of the EW corrections in NLO+PS events}
\label{sec:EWcorrNLOPS}

In this section we investigate EW correction effects at the level of fully
showered NLO+PS predictions.

\begin{figure}[htb]
  \begin{center}
    \includegraphics[width=\wsmall]{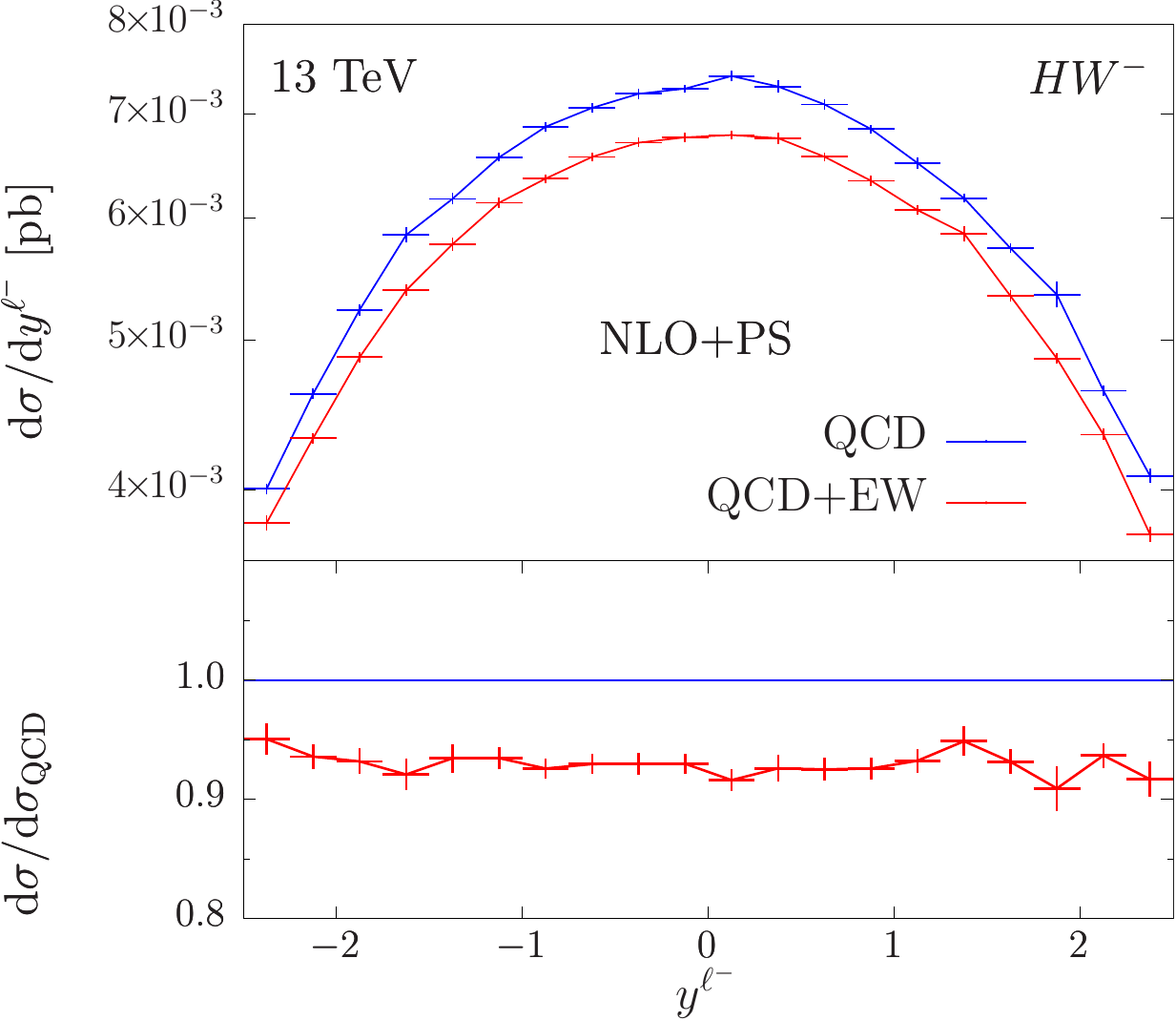}
    \includegraphics[width=\wsmall]{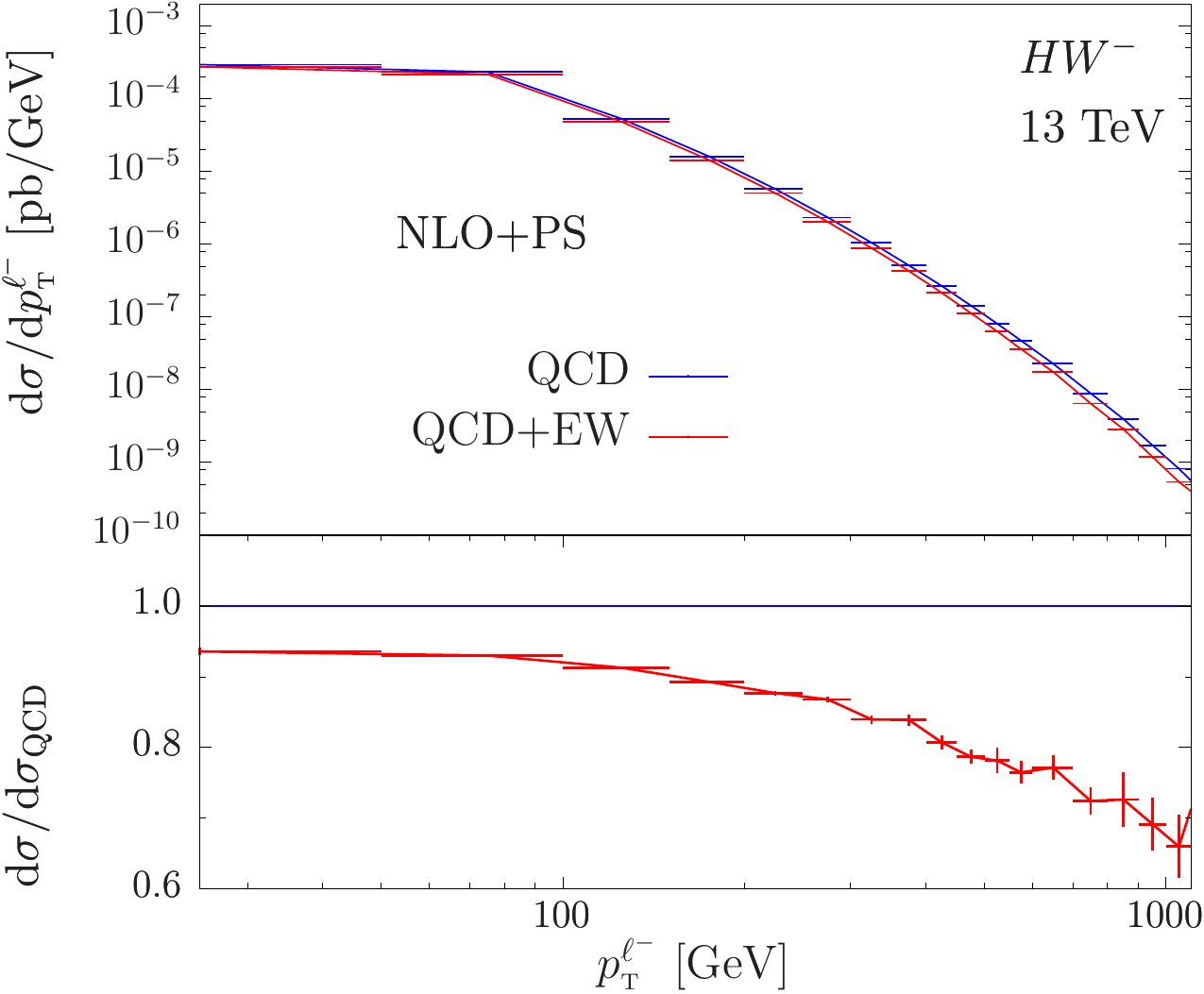}
    \caption{NLO+PS predictions for the distributions in the rapidity~(left)
      and the transverse momentum~(right) of the charged dressed lepton in
      \HWm{} production. Comparison between the full QCD+EW results and the
      QCD ones after the \PythiaEightPone{} QCD+QED shower.}
    \label{fig:W13_lept-pt_PY8_QCD-EW}
  \end{center}
\end{figure}
In Fig.~\ref{fig:W13_lept-pt_PY8_QCD-EW} we show the rapidity~(left) and the
transverse momentum~(right) of the charged dressed lepton in \HWm{}
production.
In the rapidity distribution, the impact of NLO EW effects is constant and
amounts to about $-7\%$.  The shape of the $\pt$ distribution, instead,
changes drastically due to EW Sudakov logarithms in the high-$\pt$ region,
where differences with respect to the pure QCD predictions reach $-30\%$
around 1~TeV.

\begin{figure}[htb]
  \begin{center}
    \includegraphics[width=\wmedium]{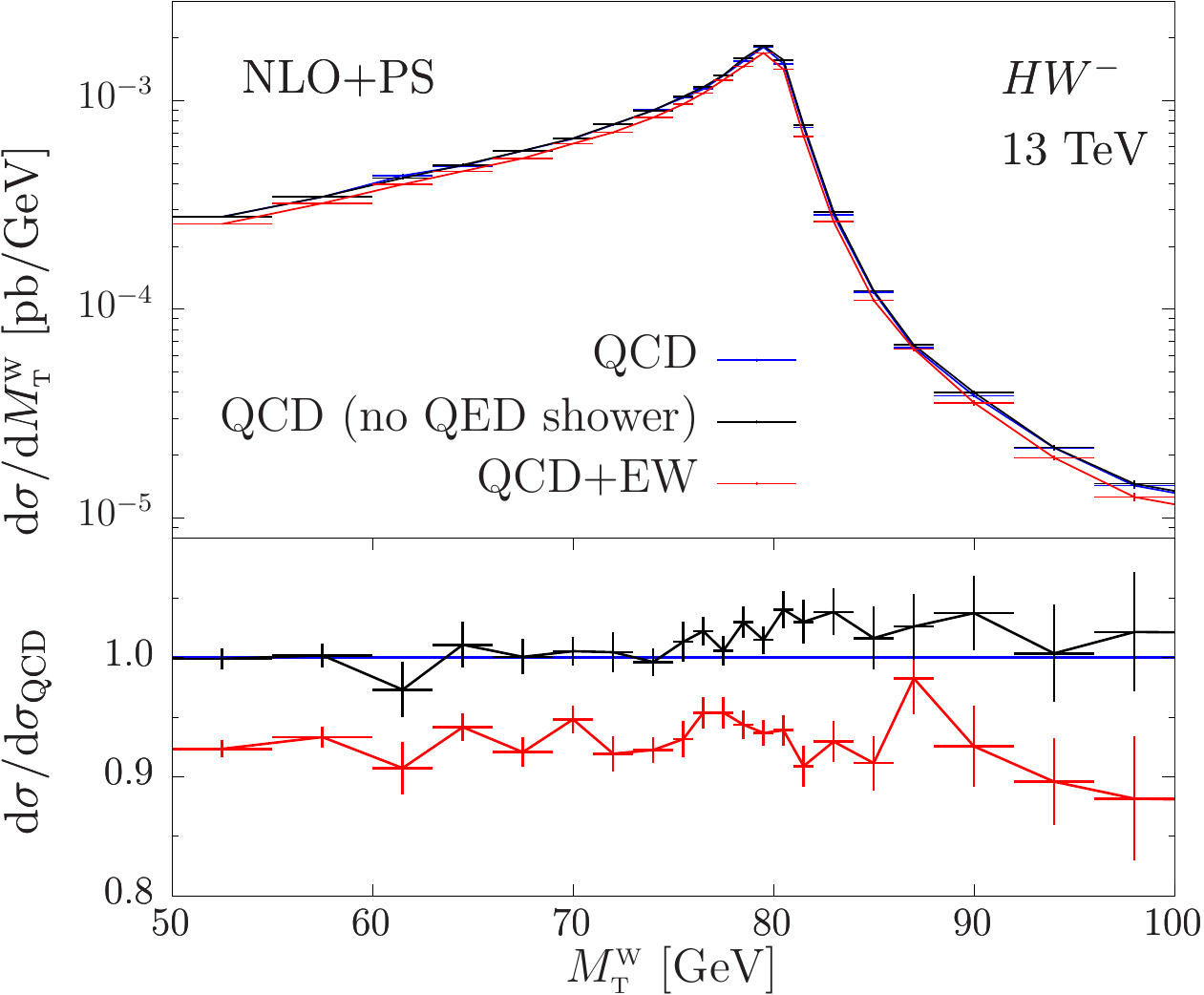}
    \caption{NLO+PS predictions for the transverse mass of the reconstructed $W^-$ boson
      in \HWm{} production. Same curves and labels as in
      Fig.~\ref{fig:W13_lept-pt_PY8_QCD-EW}. To illustrate the 
effect of the QED shower, we also show results obtained by showering QCD-corrected events 
with the QED shower
      switched off  in \PythiaEightPone{} (``no QED shower'').}
    \label{fig:W13_W-mt_PY8_QCD-EW}
  \end{center}
\end{figure}
In Fig.~\ref{fig:W13_W-mt_PY8_QCD-EW} we plot the transverse mass of
the reconstructed $W^-$ boson
\begin{equation}
M_{\sss\rm  T}^{\sss\rm W} = \sqrt{2\, \pt^\ell \, \slashed{E}_{\sss T}
  \l(1-\cos\Delta\phi \r)}\,,
\end{equation}
where $\Delta\phi$ is the azimuthal angle between the charged lepton and the
missing transverse momentum. Similarly, as for the lepton rapidity, the EW
corrections do not change the shape, but lower the differential cross section
by about~7\% with respect to the pure QCD corrections. If no QED shower is
activated when \PythiaEightPone{} showers QCD-corrected events, the curve
that is obtained is very similar to the QCD one, i.e.~the impact of the QED
shower is small for this distribution and no radiative tail can be observed.

\begin{figure}[htb]
  \begin{center}
    \includegraphics[width=\wsmall]{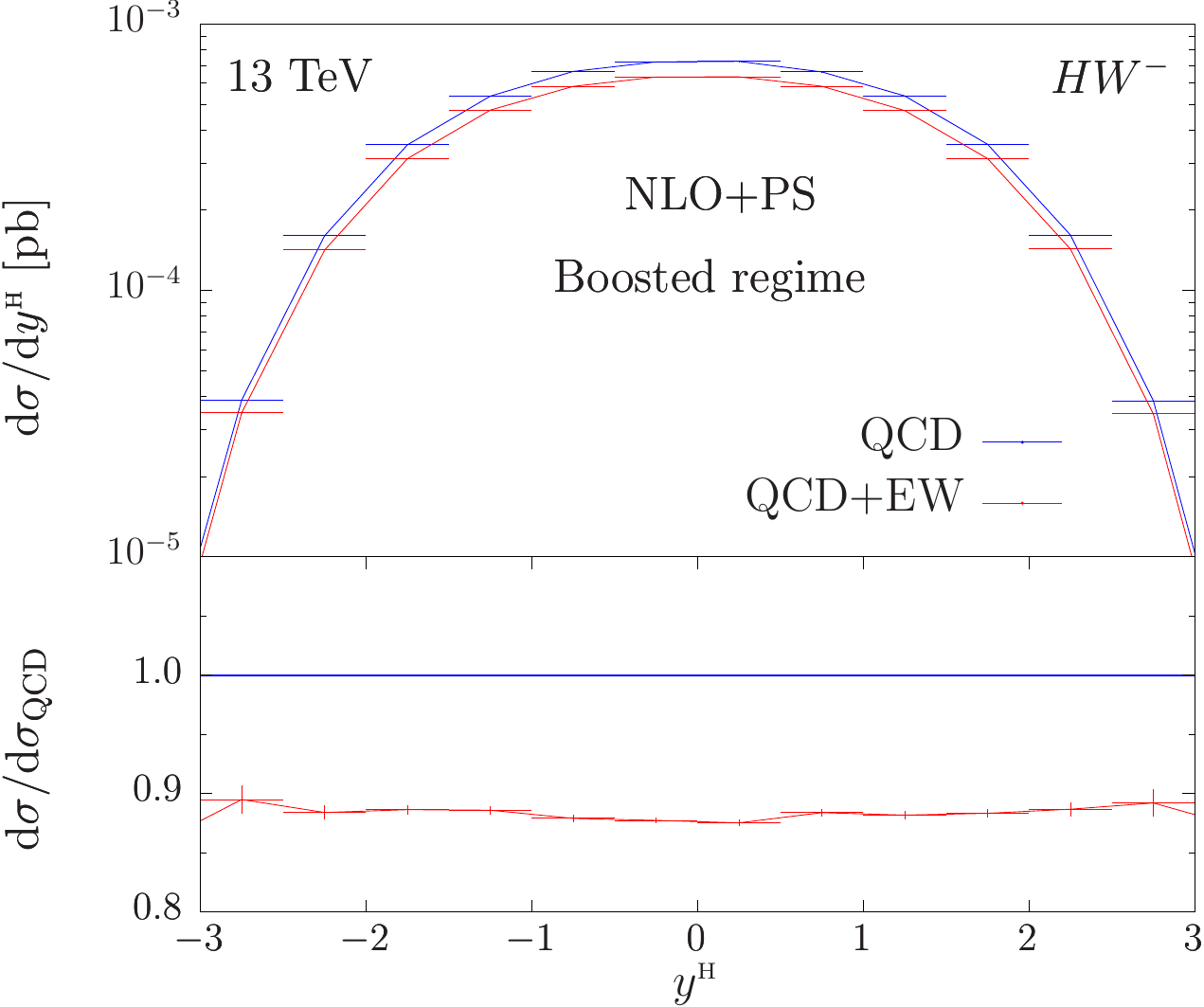}
    \includegraphics[width=\wsmall]{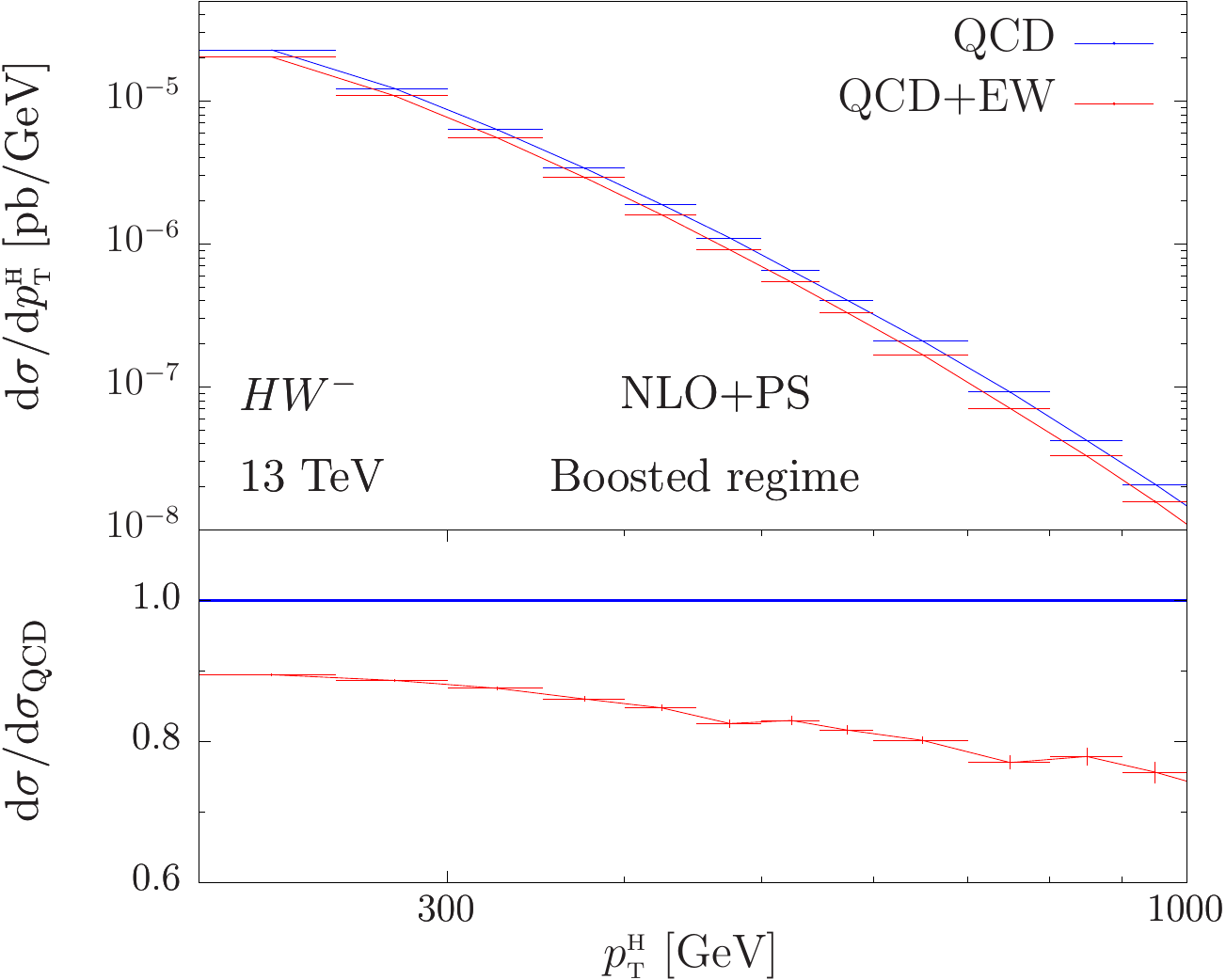}
    \caption{NLO+PS predictions for the rapidity~(left) and
      the transverse-momentum distribution~(right) of the Higgs boson in
      the boosted regime of Eq.~(\ref{eq:boosted_regime}) for \HWm{}
      production. Same curves and labels as in
      Fig.~\ref{fig:W13_lept-pt_PY8_QCD-EW}.}
    \label{fig:W13_H-pt_PY8_QCD-EW}
  \end{center}
\end{figure}
In Fig.~\ref{fig:W13_H-pt_PY8_QCD-EW} we show the rapidity and the transverse
momentum of the Higgs boson in the boosted regime, as defined by the cuts of
Eq.~(\ref{eq:boosted_regime}). The EW corrections have a constant negative
impact around 10\% on the rapidity distribution, and reach $-25\%$ around
1~TeV.  Similar conclusions can be drawn for the rapidity and transverse
momentum of the $W^-$ boson.

\begin{figure}[htb]
  \begin{center}
    \includegraphics[width=\wsmall]{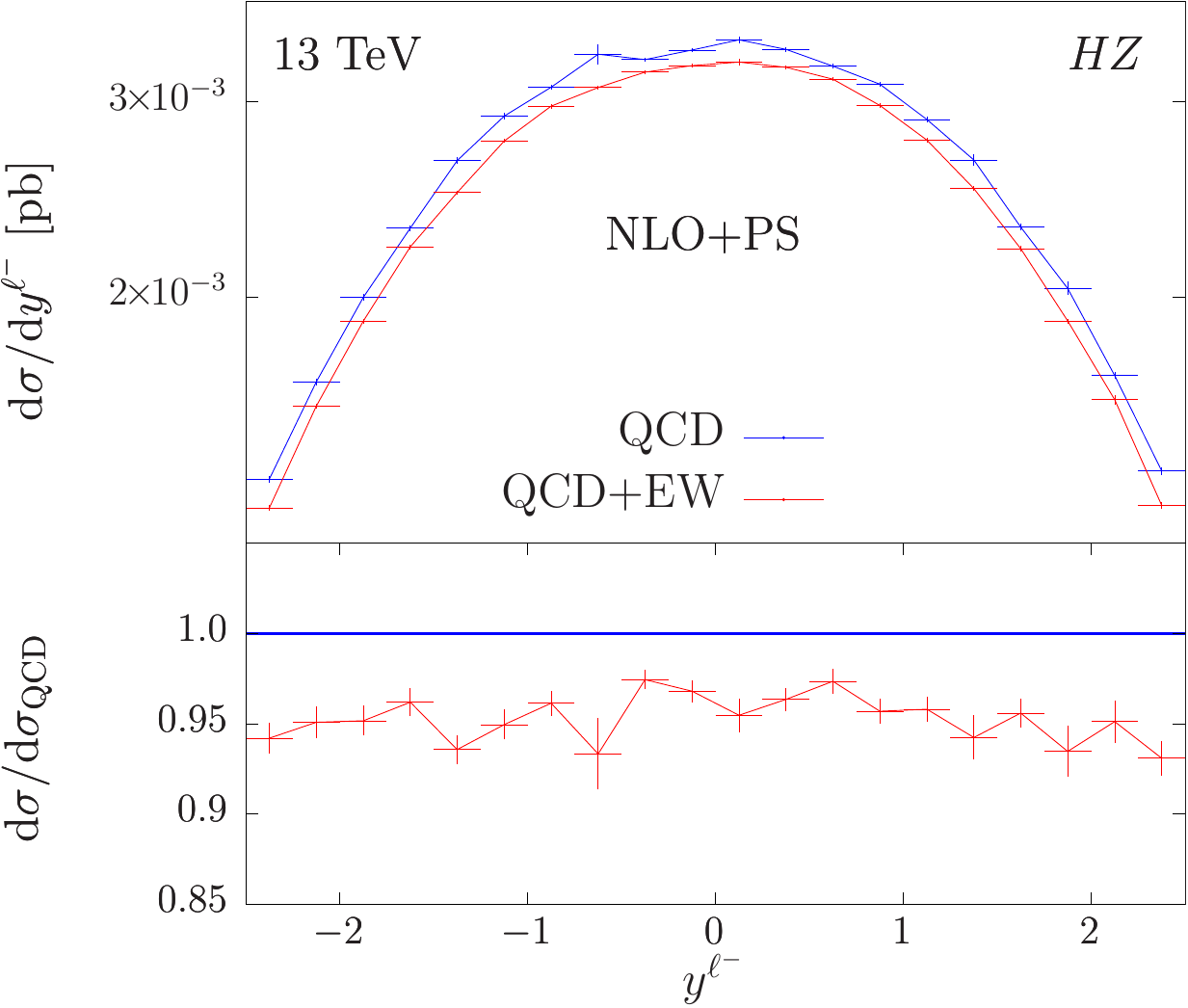}
    \includegraphics[width=\wsmall]{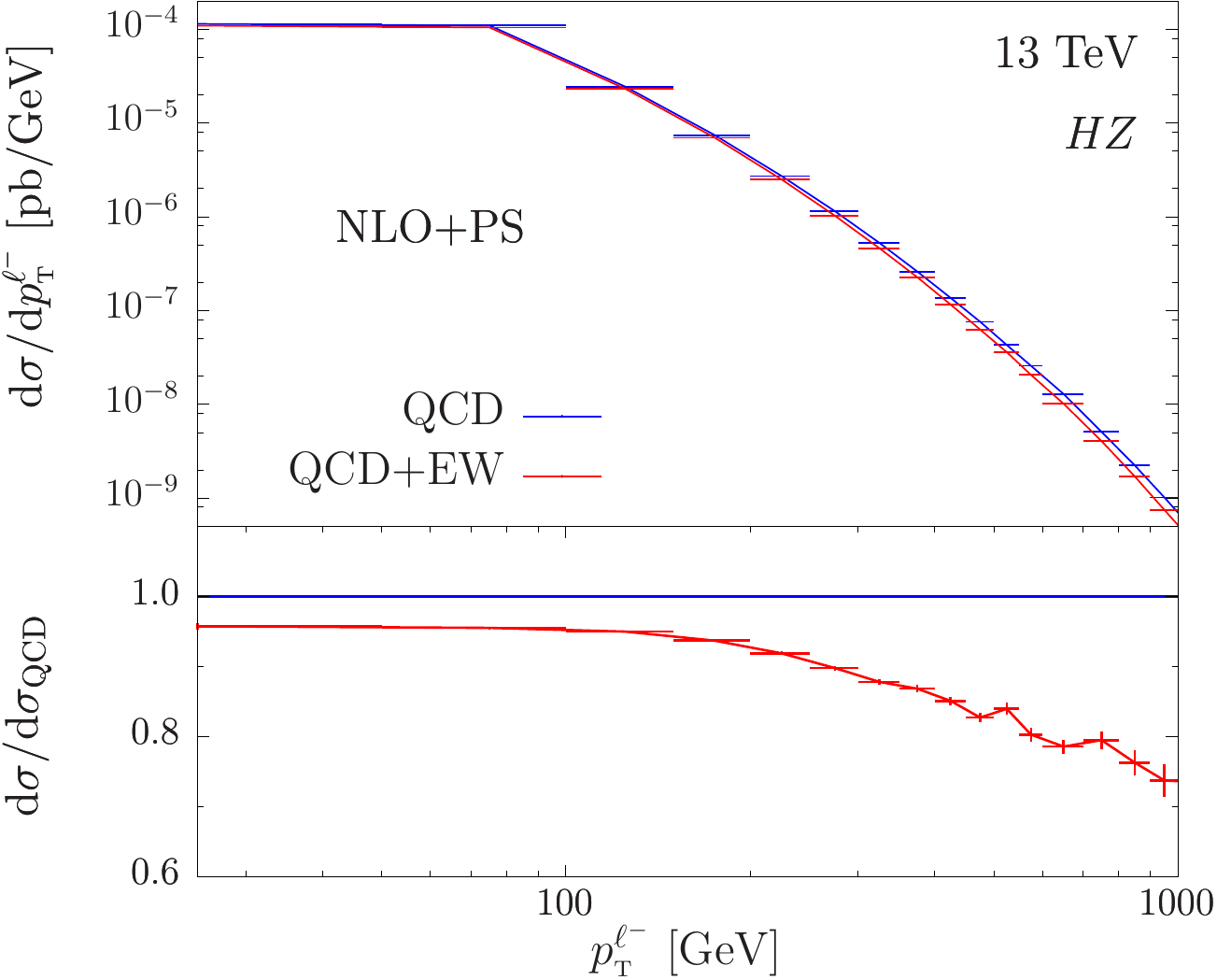}
    \caption{NLO+PS predictions for the rapidity~(left) and for the
      transverse momentum~(right) of the dressed electron in \HZ{}
      production. Comparison between the full QCD+EW results and the
      QCD ones after the \PythiaEightPone{} QCD+QED shower.}
    \label{fig:Z13_lept-pt_PY8_QCD-EW}
  \end{center}
\end{figure}

\begin{figure}[htb]
  \begin{center}
    \includegraphics[width=\wmedium]{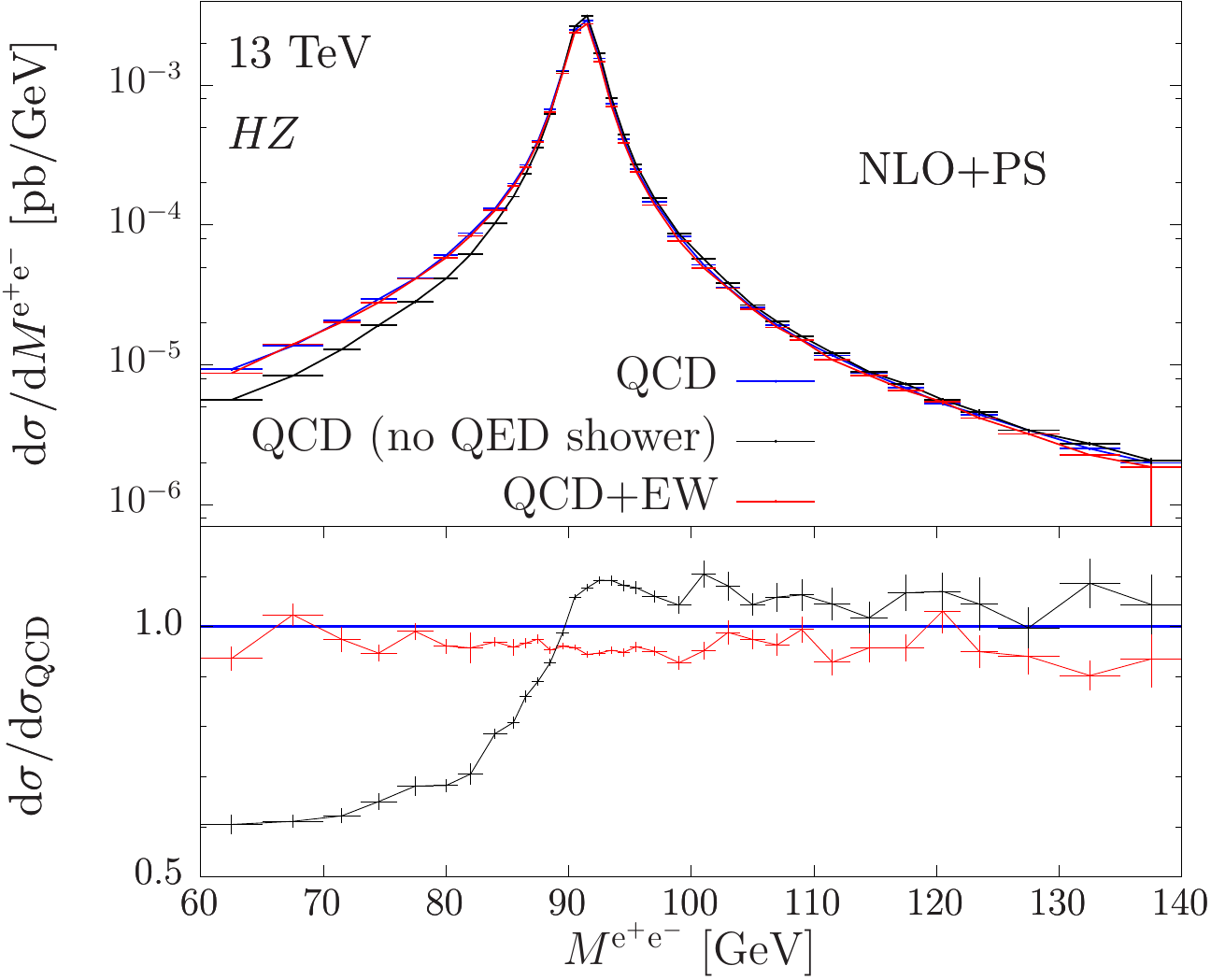}
    \caption{NLO+PS predictions for the invariant mass of the reconstructed leptonic pair
      in \HZ{} production. Same curves and labels as in
      Fig.~\ref{fig:Z13_lept-pt_PY8_QCD-EW}. For comparison, the result
      obtained by showering the QCD-corrected events without QED shower
      in \PythiaEightPone{} is also  plotted.}
    \label{fig:Z13_Z-m_PY8_QCD-EW}
  \end{center}
\end{figure}

\begin{figure}[htb]
  \begin{center}
    \includegraphics[width=\wsmall]{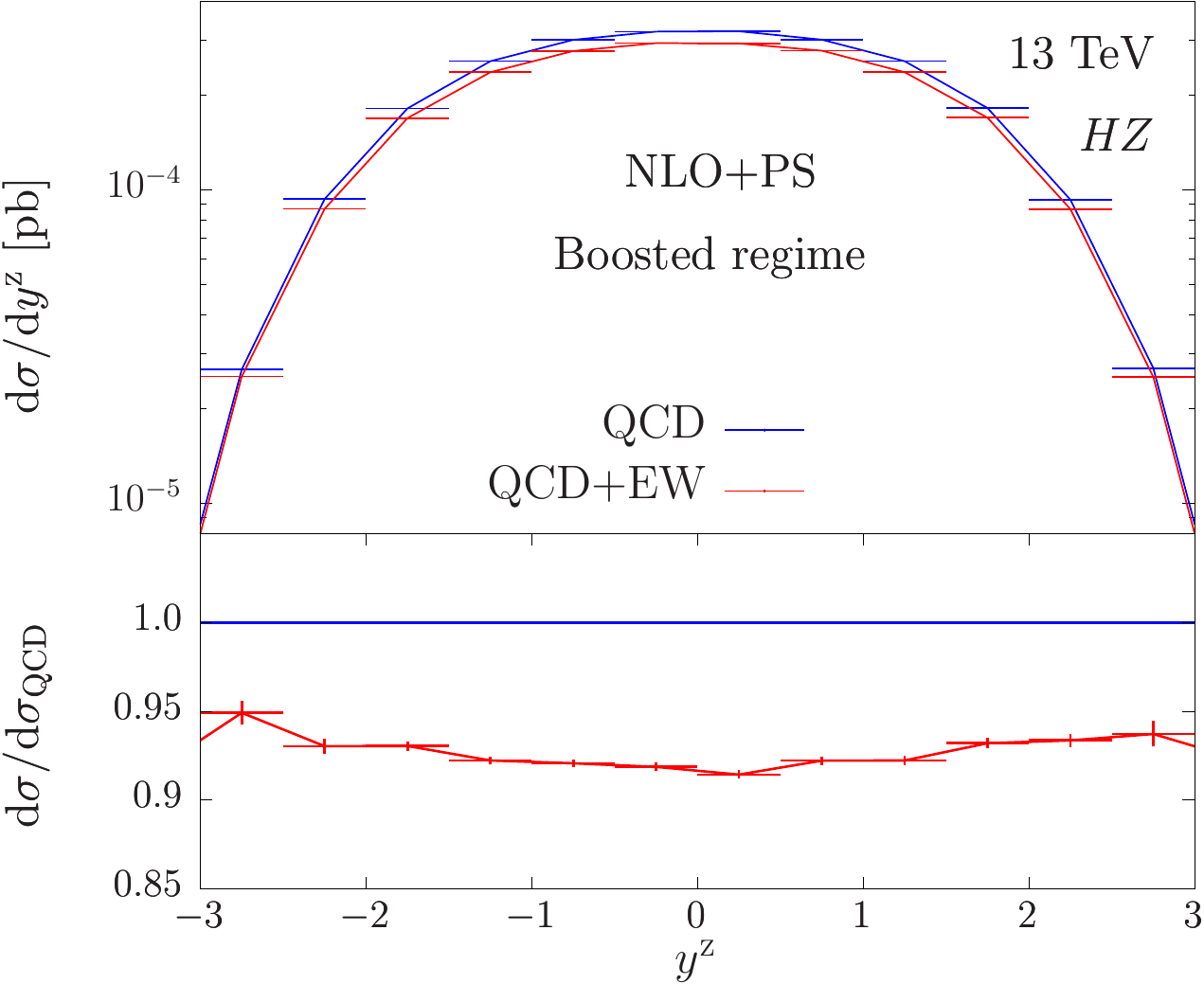}
    \includegraphics[width=\wsmall]{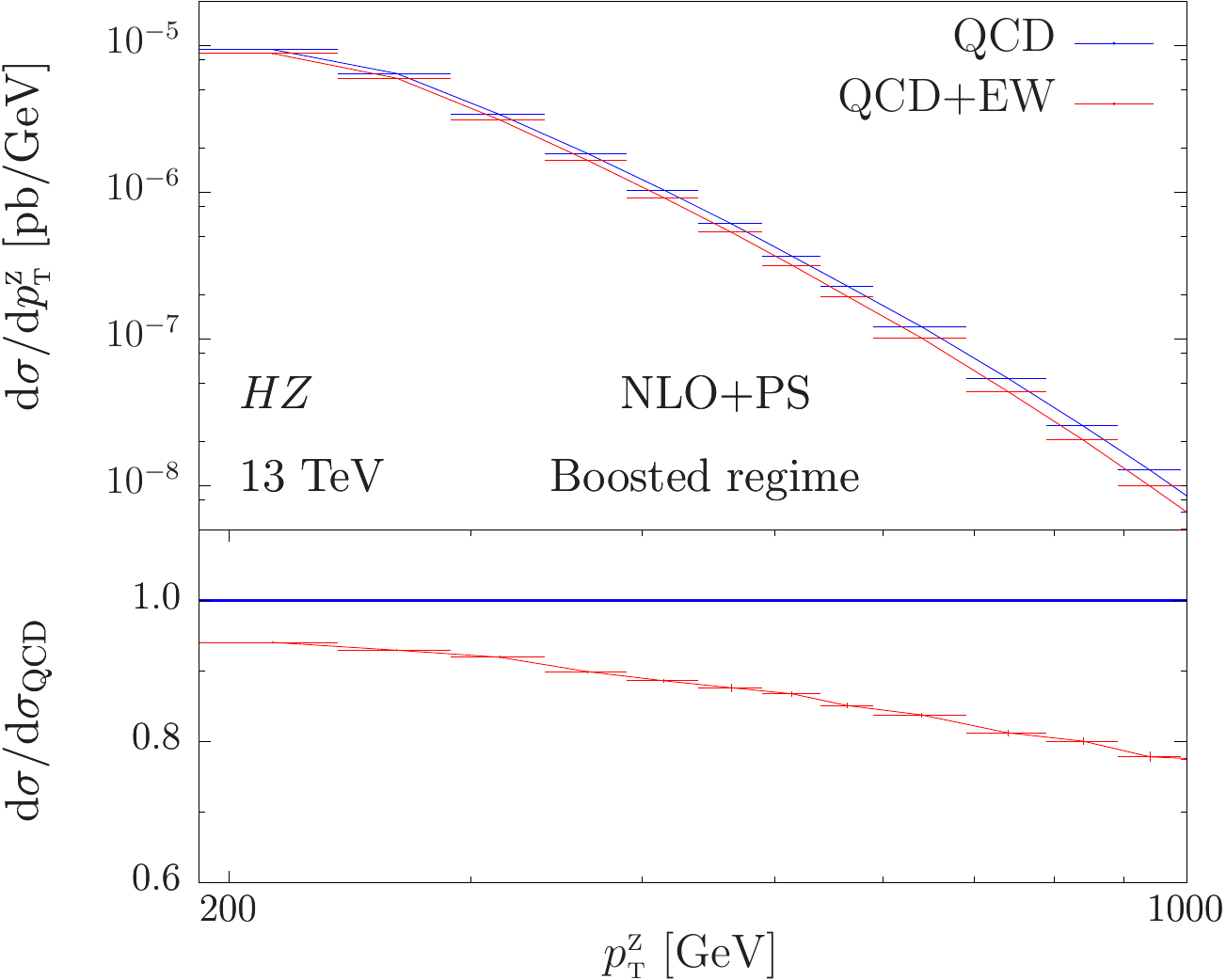}
    \caption{NLO+PS predictions for the rapidity~(left) and the transverse
      momentum~(right) of the $Z$ boson  for \HZ{}
      production in the boosted regime. 
Same curves and labels as in Fig.~\ref{fig:Z13_lept-pt_PY8_QCD-EW}.}
    \label{fig:Z13_Z-pt_PY8_QCD-EW}
  \end{center}
\end{figure}

We conclude this section by presenting kinematic distributions for \HZ{}
production in
Figs.~\ref{fig:Z13_lept-pt_PY8_QCD-EW}--\ref{fig:Z13_Z-pt_PY8_QCD-EW}.
In Fig.~\ref{fig:Z13_lept-pt_PY8_QCD-EW} we show the distribution in the
rapidity and the transverse momentum of the dressed electron.  The EW
corrections give a constant contribution of about~$-5\%$ in the plotted
rapidity range, while in the high-energy tail of the $\pt$ distribution the
EW corrections decrease the differential cross section by roughly~30\% due to
Sudakov logarithms.

In Fig.~\ref{fig:Z13_Z-m_PY8_QCD-EW} we plot the invariant mass of the
reconstructed leptonic pair in the region around the $Z$ resonance. In spite
of the fact that the shape of the $Z$ resonance it known to receive very
large $\mathcal{O}(\aem)$ radiative corrections (see
e.g. Refs.~\cite{CarloniCalame:2005vc, Barze':2013yca}), NLO EW effects turn
out to be almost constant and as small as $-5$\% when we compare showered
NLO+PS predictions at NLO QCD+EW versus NLO QCD.
This is due to the fact that the bulk of the $\mathcal{O}(\aem)$ radiation is
correctly described by the QED shower in \PythiaEightPone{}.
The importance of $\mathcal{O}(\aem)$ radiation becomes evident when we
switch off the QED shower (``no QED shower'') in the NLO QCD simulation.
This results in a radiative tail with distortions of up to $40\%$ in the region below the $Z$ peak.

In \HZ{} production, the momentum of the vector boson can be fully
reconstructed. Thus, in Fig.~\ref{fig:Z13_Z-pt_PY8_QCD-EW} we display the
rapidity and transverse-momentum distributions of the $Z$ boson in the
boosted regime, as defined by the cuts of
Eq.~(\ref{eq:boosted_regime}). These results are very similar to the ones
obtained for the Higgs boson in \HWm{} production
in~Fig.~\ref{fig:W13_H-pt_PY8_QCD-EW}. While EW corrections have a constant
impact of about~$-8\%$ on the rapidity distribution, the tail of the $\pt$
distribution is dominated by large negative EW Sudakov logarithms, and we
observe differences with respect to the pure QCD result of the order
of~$-25\%$ for $\pt \sim 1$~TeV.

\section{Results for $\boldsymbol{\HVJ}$ production at NLO QCD+EW with \MINLO+PS}
\label{sec:results_hvj}

In this section we study $pp\to \HVJ{}$ at NLO+PS accuracy in the \MINLO{}
approach, denoted in the following as \MINLO+PS.  Similarly as in the
previous section, in Sec.~\ref{sec:FOvsMINLOPS} we first compare NLO QCD+EW
predictions obtained with \MINLO{} at fixed order against corresponding
results at the LHE level or including also the full QCD+QED parton shower.
The effect of EW corrections is studied in Sec.~\ref{sec:EWcorrMINLOPS} in
the case of fully showered \MINLO+PS simulations.  The cuts and physics
object definitions of Sec.~\ref{sec:cuts} are applied throughout, and we do
not impose any cut that requires the presence of jets.

\subsection{From fixed-order \MINLO{} to \MINLO+PS at NLO QCD+EW}
\label{sec:FOvsMINLOPS}

\begin{figure}[htb]
  \begin{center}
    \includegraphics[width=\wsmall]{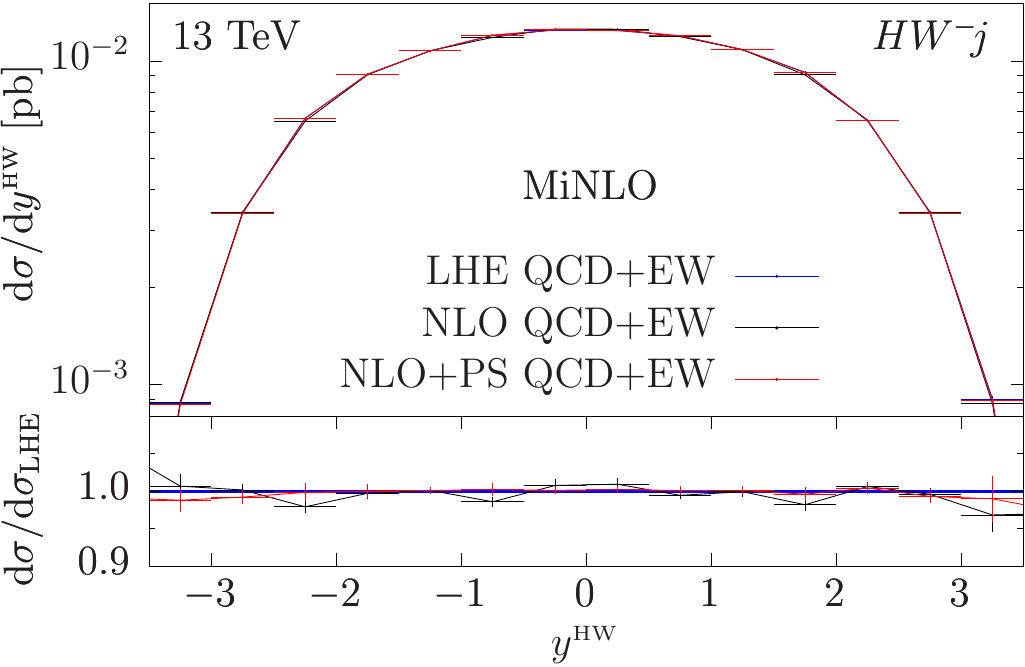}
    \includegraphics[width=\wsmall]{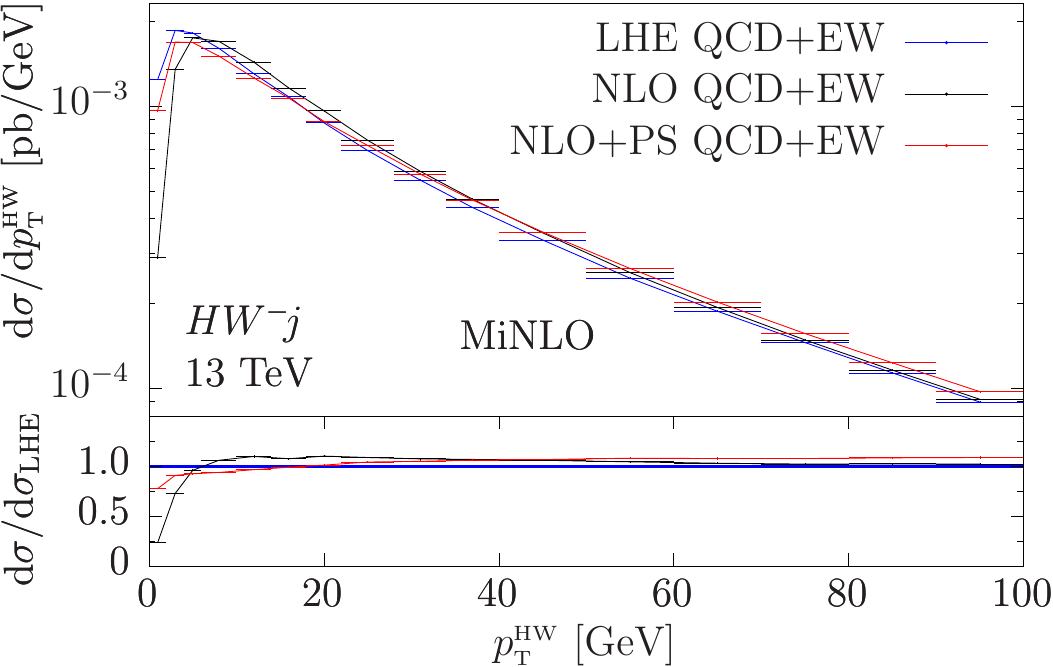}
    \caption{Distributions in the rapidity~(left) and transverse
      momentum~(right) of the reconstructed \HWm{} pair. \comparNLPminlo{} In the
      ratio plot results are normalized with respect to the LHE level
      prediction.}
    \label{fig:WJ13_HW-y_NLOvsLHvsPY8}
  \end{center}
\end{figure}

In Fig.~\ref{fig:WJ13_HW-y_NLOvsLHvsPY8} we analyze the rapidity and the
transverse momentum of the reconstructed \HWm{} system. As a result of the
\MINLO{} prescription, the rapidity distribution, as well as any other
inclusive observable, is finite.  For the rapidity distribution we observe
that the three predictions are very close to each other. At variance with
Fig.~\ref{fig:W13_HW-y_NLOvsLHvsPY8}, here fixed-order predictions for the
$\pt$ distribution are finite at small $\pt$, since soft and collinear
divergences are suppressed by the \MINLO{} Sudakov form factors.  Moreover,
the NLO accuracy in the spectrum of the \HWm{} system leads to an improved
agreement between fixed-order and NLO+PS results.

\begin{figure}[htb]
  \begin{center}
    \includegraphics[width=\wsmall]{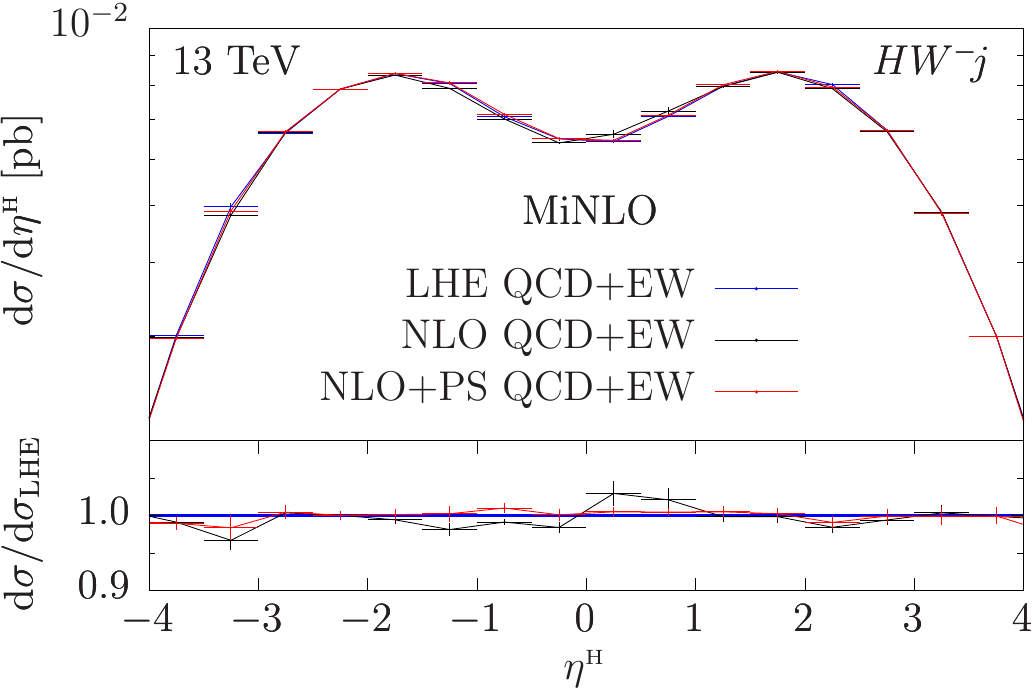}
    \includegraphics[width=\wsmall]{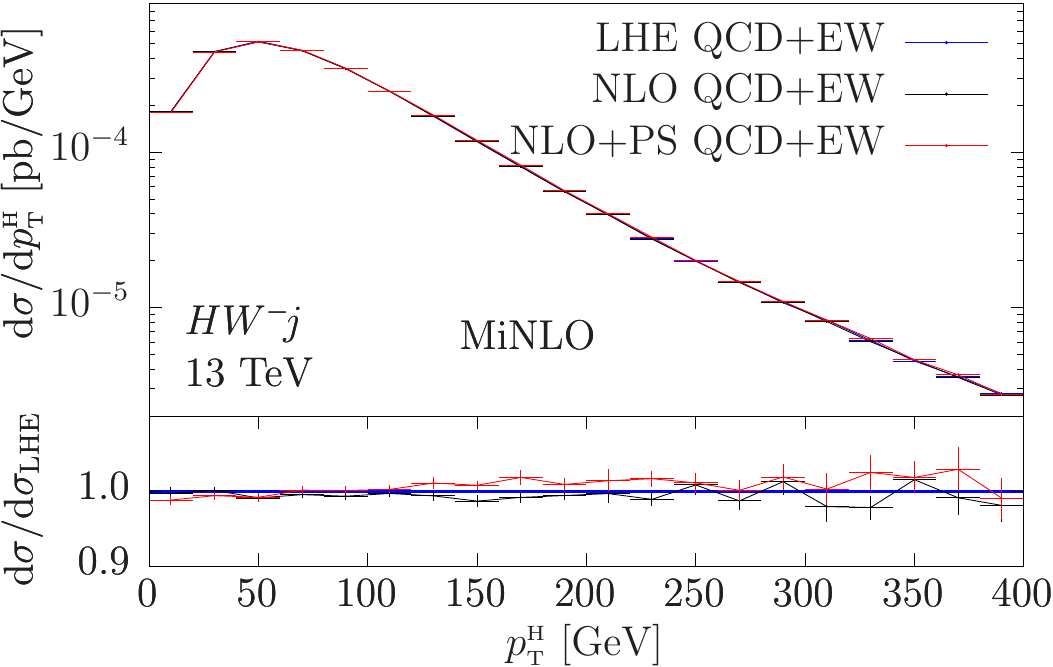}
    \caption{Distributions in the pseudorapidity~(left) and transverse
      momentum~(right) of the Higgs boson in the \MINLO{} improved
      $pp\to\HWmJ$ simulation.  Same curves and labels as in
      \reffi{fig:WJ13_HW-y_NLOvsLHvsPY8}.}
    \label{fig:WJ13_H-eta_LHvsPY8}
  \end{center}
\end{figure}
In Fig.~\ref{fig:WJ13_H-eta_LHvsPY8} we show the pseudorapidity and the
transverse-momentum distributions of the Higgs boson, finding again very good
agreement among the three predictions.

We refrain from presenting results for \HWpJ{} and \HZJ{} production since
they behave qualitatively very similar as the results shown here for \HWmJ{}
production.

\subsection{Impact of EW corrections at \MINLO+PS level}
\label{sec:EWcorrMINLOPS} 

\begin{figure}[htb]
  \begin{center}
    \includegraphics[width=\wsmall]{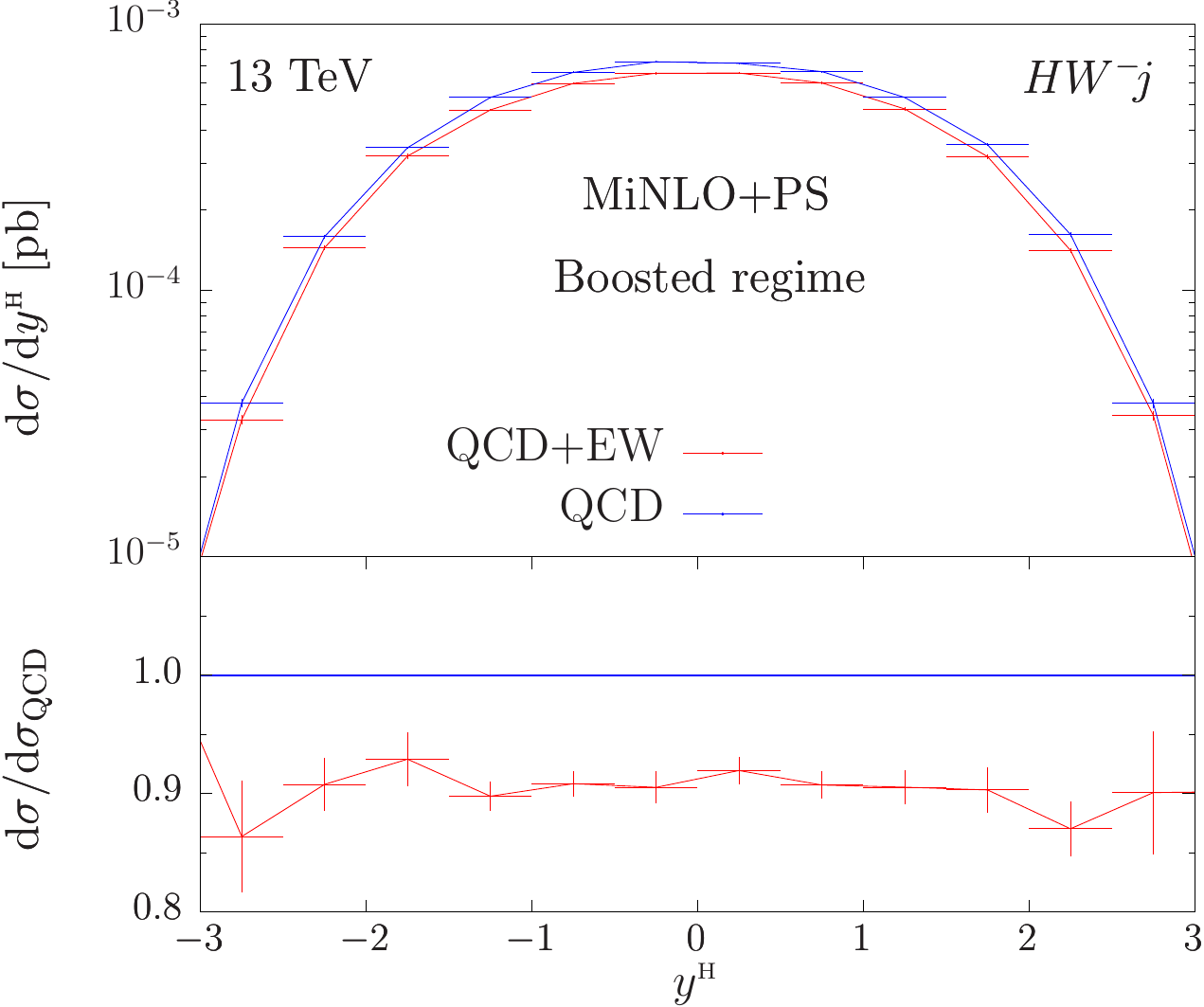}
    \includegraphics[width=\wsmall]{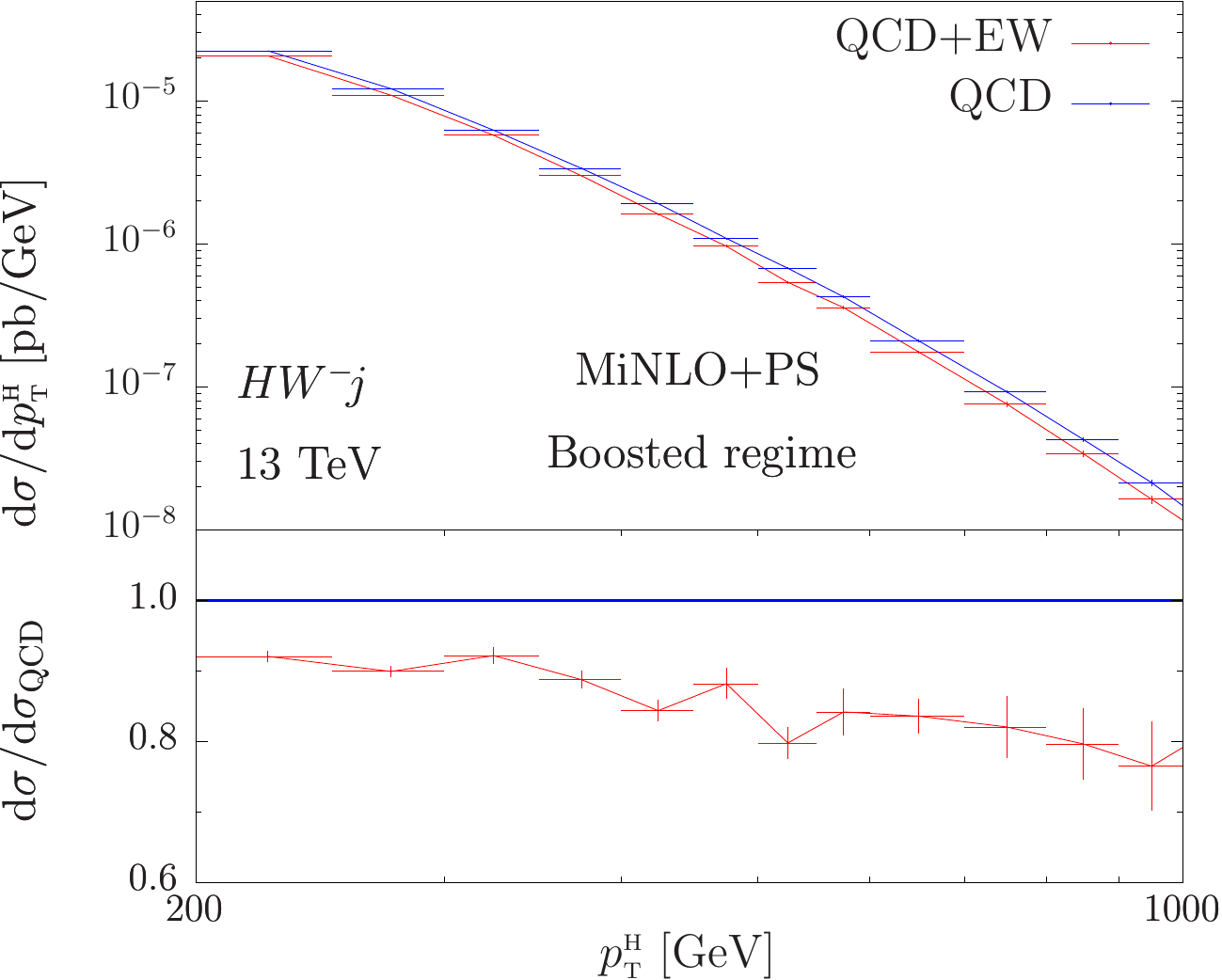}
    \caption{\MINLO+PS predictions for the rapidity~(left) and
      transverse-momentum distribution~(right) of the Higgs boson in the
      boosted regime, in \HWmJ{} production.  Comparison between the full
      QCD+EW results and the QCD ones after the \PythiaEightPone{} QCD+QED
      shower.}
    \label{fig:WJ13_H-pt_PY8_QCD-EW}
  \end{center}
\end{figure}

The impact of EW corrections at the level of fully showered \MINLO+PS
predictions for \HWmJ{} production is illustrated in
Figs.~\ref{fig:WJ13_H-pt_PY8_QCD-EW}--\ref{fig:WJ13_j1-pt_PY8_QCD-EW}.

For the distributions in the rapidity and transverse momentum of the Higgs
boson in the boosted regime (Fig.~\ref{fig:WJ13_H-pt_PY8_QCD-EW}) we find
that the EW corrections induced by the boosted cut, $\pt^{\sss \rm H} \ge
200~\GeV$, are nearly independent of $y^{\sss \rm H}$ and around $-10\%$,
while they grow up to $-20\%$ and beyond when $\pt^{\sss \rm H}$ enters the
TeV regime. These results closely agree with the corresponding ones shown in
Fig.~\ref{fig:W13_H-pt_PY8_QCD-EW} for the NLO+PS simulation of inclusive
\HWm{} production.  Consistently with the fixed-order findings discussed in
Sec.~\ref{sec:foresults}, also this observation supports the theoretical
considerations of Sec.~\ref{se:MINLO}, where we have argued that \MINLO{}
improved predictions for \HVJ{} production should preserve NLO QCD+EW
accuracy when the extra jet is integrated out.
Also other inclusive observables, such as the distribution in the missing
transverse momentum shown in Fig.~\ref{fig:WJ13_miss-pt_PY8_QCD-EW}, confirm
this observation.

The EW corrections to the leading-jet $\pt$ distribution, shown in
Fig.~\ref{fig:WJ13_j1-pt_PY8_QCD-EW}, do not feature the standard Sudakov
behavior.  In this distribution, EW effects remain rather small, at the level
of $-5\%$, in the entire plotted range, i.e.~from very low jet-$\pt$ up to
400~GeV.  This is not surprising, since a similar ``non-Sudakov'' behavior
for inclusive jet spectra was already observed in
Ref.~\cite{Denner:2009gj, Kallweit:2015dum} for the case of $V+$\,jets production. Another
important feature of Fig.~\ref{fig:WJ13_j1-pt_PY8_QCD-EW} is that EW
corrections are nearly constant in the region where the jet $\pt$ approaches
zero.  Again, this confirms the considerations made in Sec.~\ref{se:MINLO}
regarding the factorization of EW corrections in the presence of soft or
collinear QCD radiation, and the NLO QCD+EW accuracy of inclusive \MINLO{}
simulations.
To be more precise, in the left panel of
Fig.~\ref{fig:WJ13_j1-pt_PY8_QCD-EW} we see that EW corrections effects are
nearly constant at small $\pt$ with the exception of the first bin.  This
effect can be attributed to photonic contributions to the jet transverse
momentum, and to the fact that the Sudakov peak associated with the damping
of QCD radiation is located well above the one associated with the damping of
QED radiation.  This mismatch tends to enhance the relative importance of QED
radiation in the region between the QCD and QED Sudakov peaks. In any case,
this effect cancels upon integration over the soft region of the jet
spectrum. Thus, it should not spoil the expected NLO QCD+EW accuracy of
inclusive \MINLO{} predictions.

\newcommand{\samelabminloew}{Same curves and labels as in
  Fig.~\ref{fig:WJ13_H-pt_PY8_QCD-EW}.}
\begin{figure}[htb]
  \begin{center}
    \includegraphics[width=\wmedium]{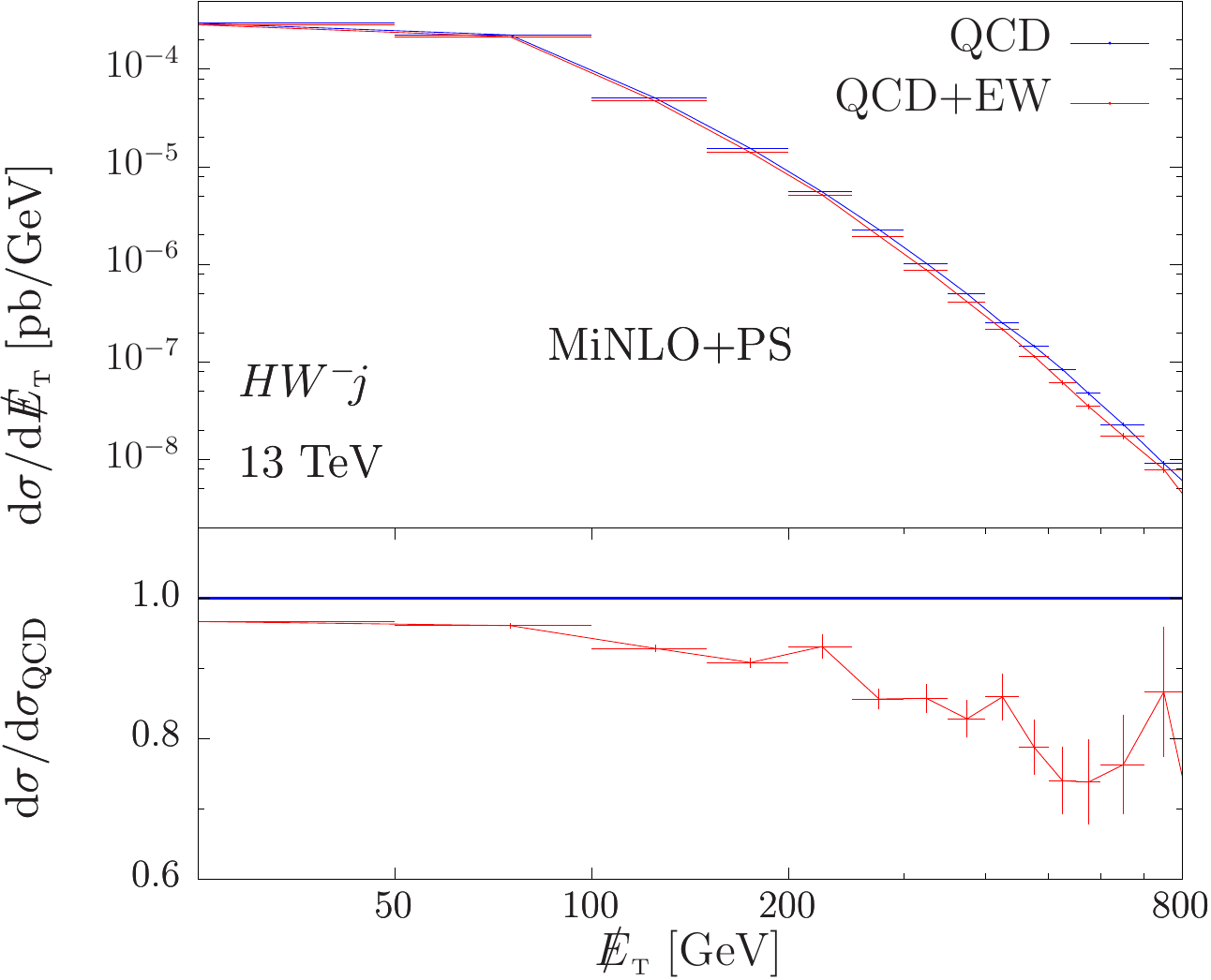}
    \caption{\MINLO+PS predictions for the missing transverse momentum in
      \HWmJ{} production.  \samelabminloew }
\label{fig:WJ13_miss-pt_PY8_QCD-EW}
  \end{center}
\end{figure}
\begin{figure}[htb]
  \begin{center}
    \includegraphics[width=\wsmall]{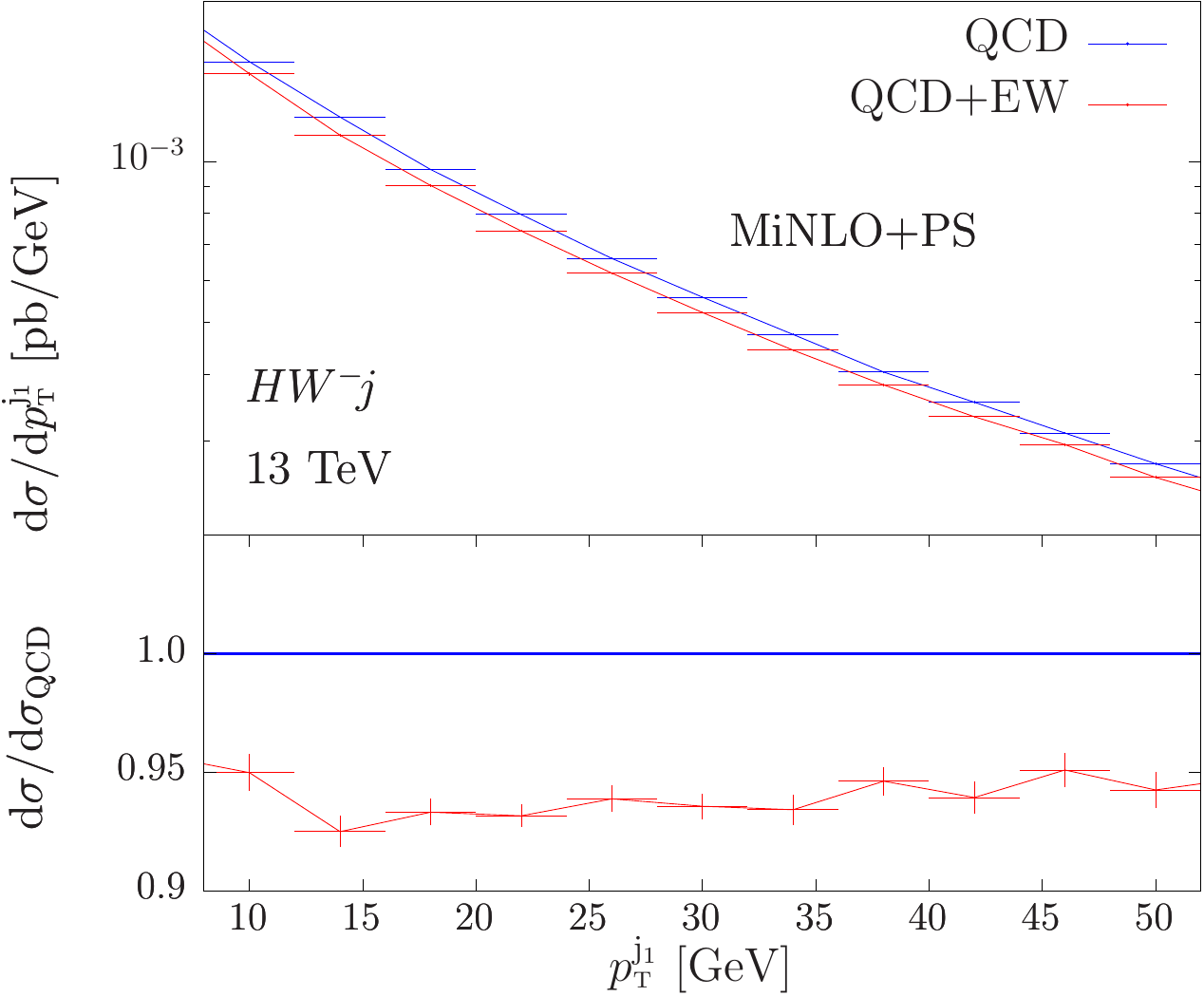}
    \includegraphics[width=\wsmall]{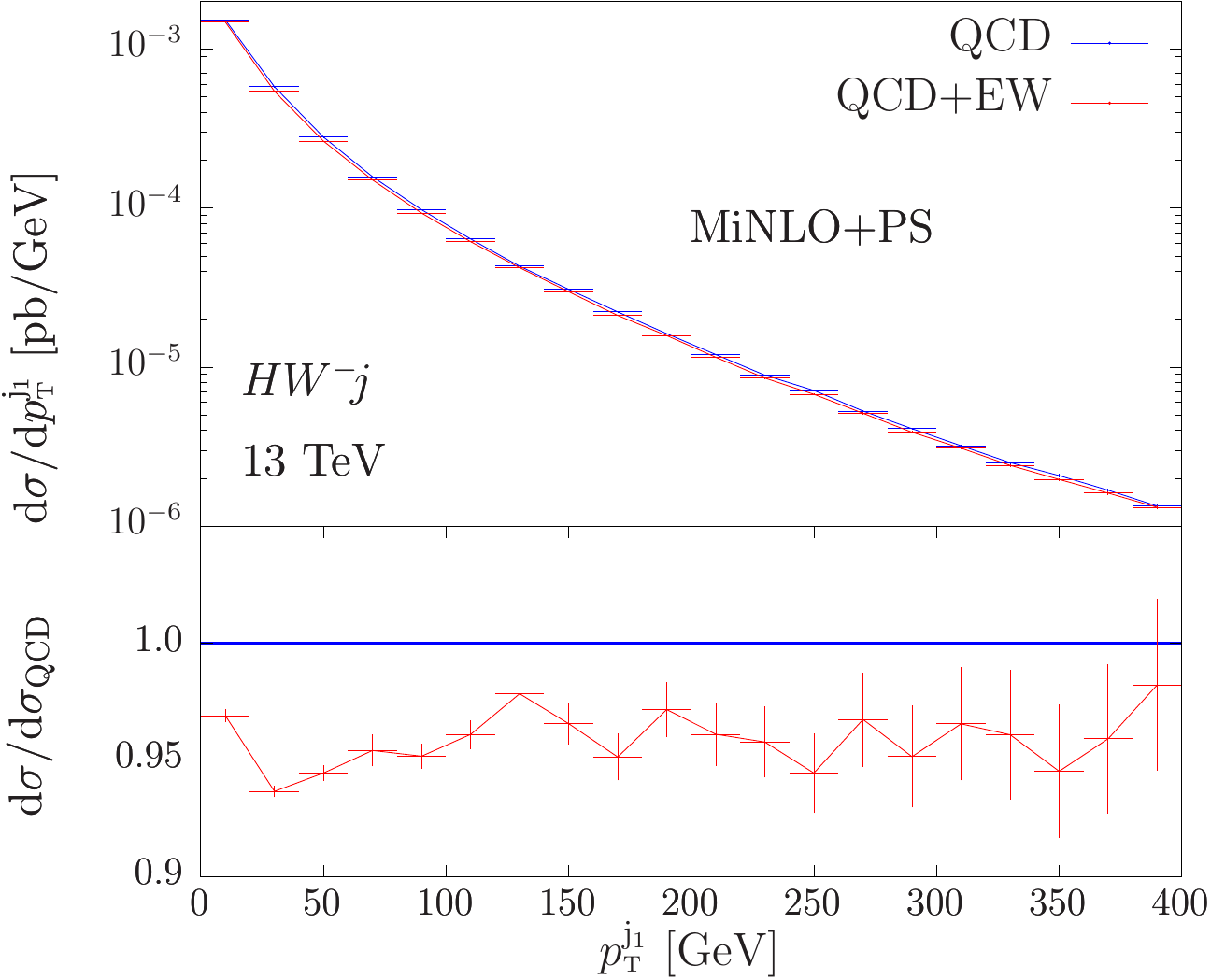}
    \caption{\MINLO+PS predictions for the transverse momentum of the leading
      jet in two transverse-momentum ranges in \HWmJ{} production.
      \samelabminloew }
    \label{fig:WJ13_j1-pt_PY8_QCD-EW}
  \end{center}
\end{figure}

\begin{figure}[htb]
  \begin{center}
    \includegraphics[width=\wmedium]{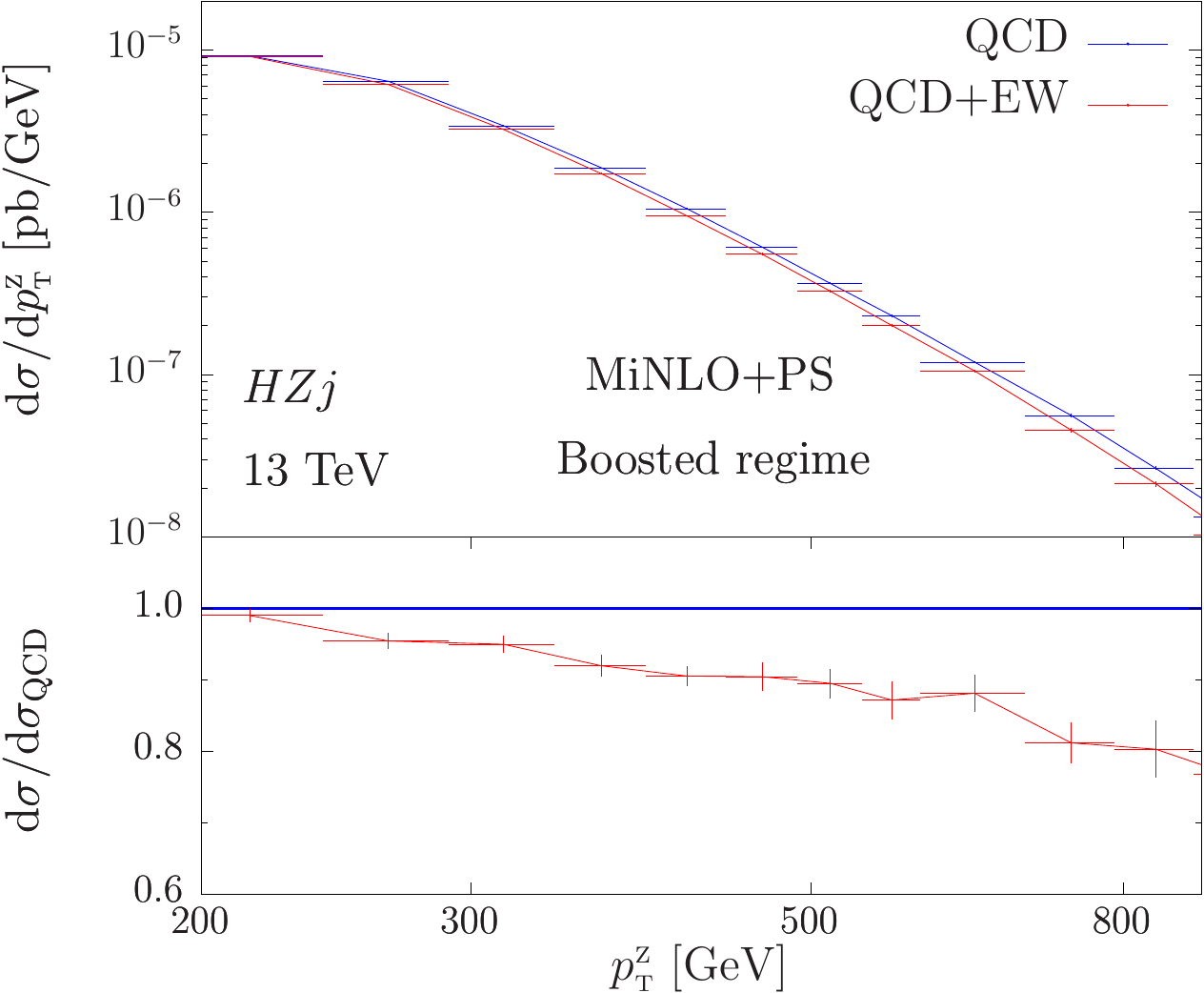}   
    \caption{\MINLO+PS predictions for the transverse momentum of the $Z$ boson
      for \HZJ{} production in the boosted regime.  \samelabminloew  }
    \label{fig:ZJ13_Z-pt_PY8_QCD-EW}
  \end{center}
\end{figure}
\begin{figure}[htb]
  \begin{center}
    \includegraphics[width=\wsmall]{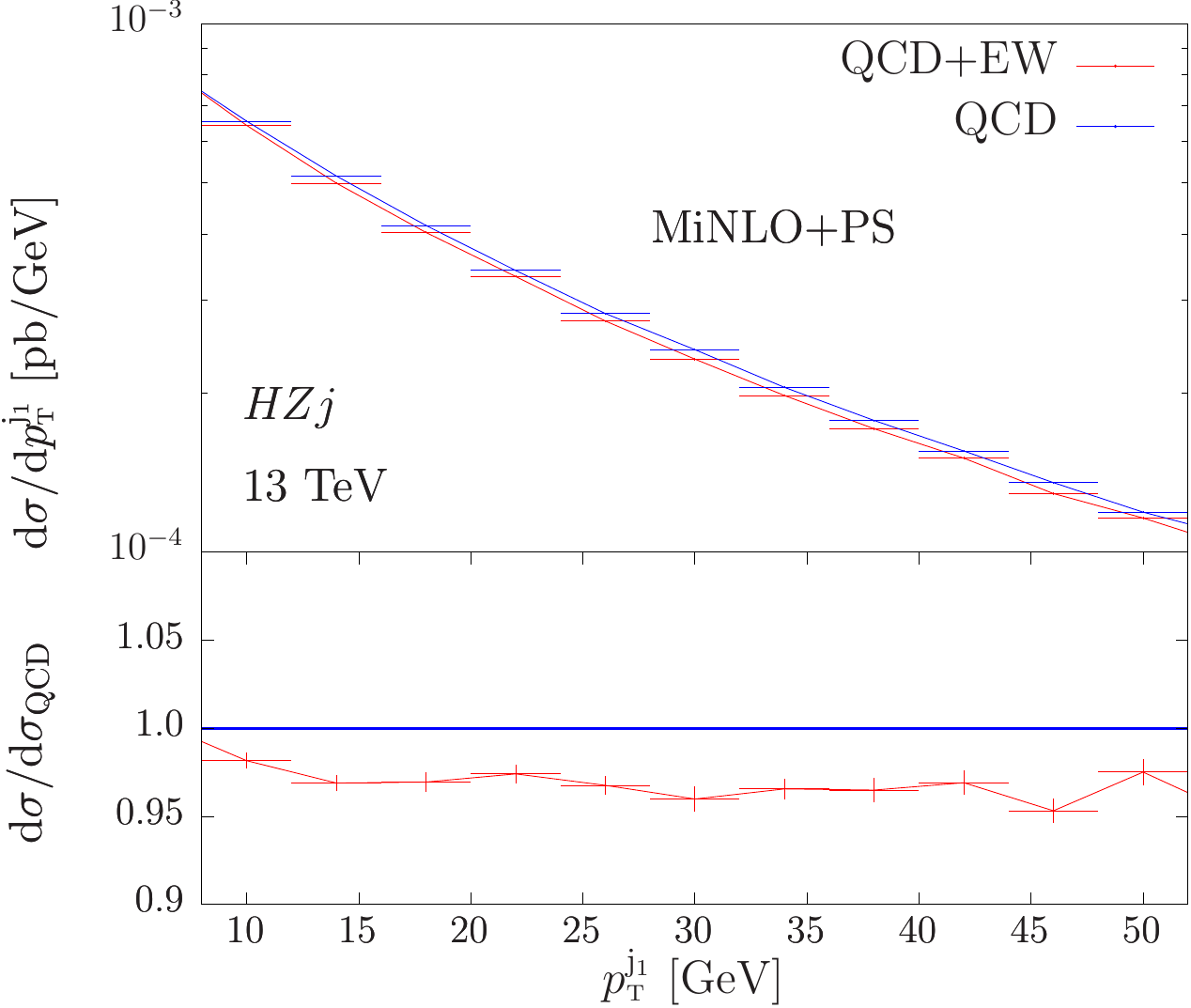}   
    \includegraphics[width=\wsmall]{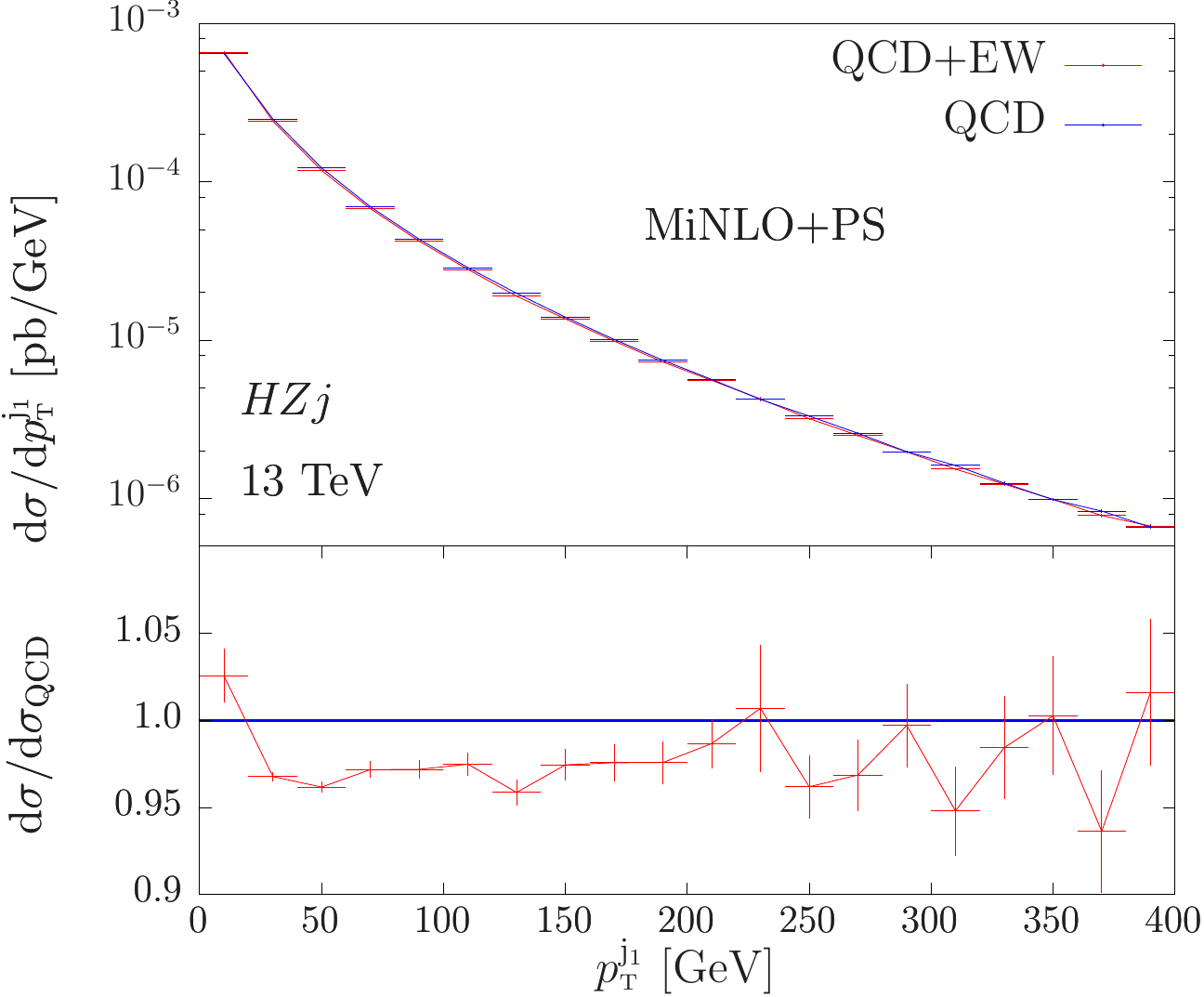}
    \caption{\MINLO+PS predictions for the transverse momentum of the leading
      jet in two transverse-momentum ranges in \HZJ{}
      production. \samelabminloew }
    \label{fig:ZJ13_j1-pt_PY8_QCD-EW}
  \end{center}
\end{figure}

We conclude this section by discussing the impact of NLO EW effects in \HZJ{}
production, illustrated in Figs.~\ref{fig:ZJ13_Z-pt_PY8_QCD-EW}
and~\ref{fig:ZJ13_j1-pt_PY8_QCD-EW}.  The distribution in the $Z$-boson $\pt$
in the boosted regime (Fig.~\ref{fig:ZJ13_Z-pt_PY8_QCD-EW}) features the
typical Sudakov EW behavior, with negative EW corrections that exceed $-20\%$
in the tail.  In the leading-jet $\pt$ distribution
(Fig.~\ref{fig:ZJ13_j1-pt_PY8_QCD-EW}) we observe relatively small and rather
constant EW corrections.  Both distributions behave similarly as the
corresponding distributions for \HWmJ{} production.

We refrain from showing further plots for \HZJ{} or \HWpJ{} production, since
EW correction effects are quite similar to the ones already discussed.

\section{Comparison between the $\boldsymbol{\HV}$
  and $\boldsymbol{\HVJ}$ generators}
\label{sec:generators}

In this section, we discuss and compare NLO+PS predictions for $pp\to\HV$
against \MINLO+PS predictions for $pp\to\HVJ$, both at NLO QCD+EW accuracy. A
similar comparison at NLO QCD accuracy was presented in
Ref.~\cite{Luisoni:2013cuh}.  Since the various Higgsstrahlung processes
behave in a very similar way, we will focus on the associated \HWm{}
production.
The comparison between the \HV{} and \HVJ{} generators is motivated by the
fact that the improved \MINLO{} prescription~\cite{Hamilton:2012rf} applied
to \HVJ{} production provides NLO accuracy also for inclusive \HV{}
quantities, i.e.~for observables where the associated jet is not resolved.
While this is well known at NLO QCD level, in Sec.~\ref{se:MINLO} we have
argued that also NLO EW accuracy should be preserved when the jet is
integrated out.

We also study the dependence of our results on scale variations. To this end
we apply standard seven-point variations obtained by multiplying the central
value of the renormalization and factorization scales $\mu_{\sss R}$ and
$\mu_{\sss F}$, defined in Eq.~(\ref{eq:mur_muf}) for \HV{} production, by
the factors $K_{\sss R}$ and $K_{\sss F}$, respectively, chosen among the
seven pairs
\begin{equation} 
\label{eq:scale_variations}
\l(K_{\sss R}, K_{\sss F}\r) = \l(\tfrac{1}{2},\tfrac{1}{2}\r), \l(\tfrac{1}{2},1\r),
 \l(1,\tfrac{1}{2}\r), \l(1,1\r), \l(2,1\r), \l(1,2\r), \l(2,2\r).
\end{equation}
Scale-variation bands in the following plots are based on the envelope of the
seven-point variations.  In \HVJ{} production, where the scale setting is
based on the improved \MINLO{} prescription, the scaling factors of
Eq.~(\ref{eq:scale_variations}) are applied to the coupling constants at each
interaction vertex and to the scale entering the Sudakov form factor.

For the fully inclusive NLO+PS and \MINLO+PS cross sections at NLO QCD+EW we
find
\begin{eqnarray}
&&  {\displaystyle\s_{\HWm}^{\sss \rm NLO+PS} = 55.29^{+0.80}_{-0.74} {\rm \ fb}\,, \qquad \qquad
\s_{\HWmJ}^{\sss \rm \MINLO+PS} = 55.25^{+1.25}_{-2.57}{\rm \ fb}\,,}
  \nonumber \\[-1mm]
\\  [-1mm]
&& {\displaystyle \s_{\HZ}^{\sss \rm NLO+PS} =  24.41^{+0.27}_{-0.38} {\rm \ fb}\,, \qquad\qquad
\s_{\HZJ}^{\sss \rm \MINLO+PS} = 24.9^{+0.6}_{-1.1} {\rm\  fb}\,, }\nonumber
\end{eqnarray}
where uncertainties correspond to scale variations.  These results are well
consistent, within statistics, to the corresponding ones reported in
Tabs.~\ref{tab:sigtot_NLO_W} and~\ref{tab:sigtot_NLO_Z} at fixed-order NLO.
Moreover, it turns out that, in the presence of EW corrections, cross
sections obtained from \HV{} and \HVJ{} simulations agree at the one-percent
level, confirming again the expectation of inclusive NLO QCD+EW accuracy for
\MINLO{} improved \HVJ{} simulations.

Scale variations are in general larger in \HVJ{} with respect to \HV{}
production.  This is due to the fact that, in standard \POWHEGBOX{}
simulations, the scale associated to the emission of the hardest jet is kept
fixed at the corresponding transverse momentum, while scale variations are
applied only at the level of the so-called $\bar{B}$ term, where QCD and QED
radiation are integrated out.  For this reason, scale variations in
\MINLO{}+PS simulations provide a more realistic estimate of scale
uncertainties associated with QCD radiation.

\begin{figure}[htb]
  \begin{center}
    \includegraphics[width=\wsmall]{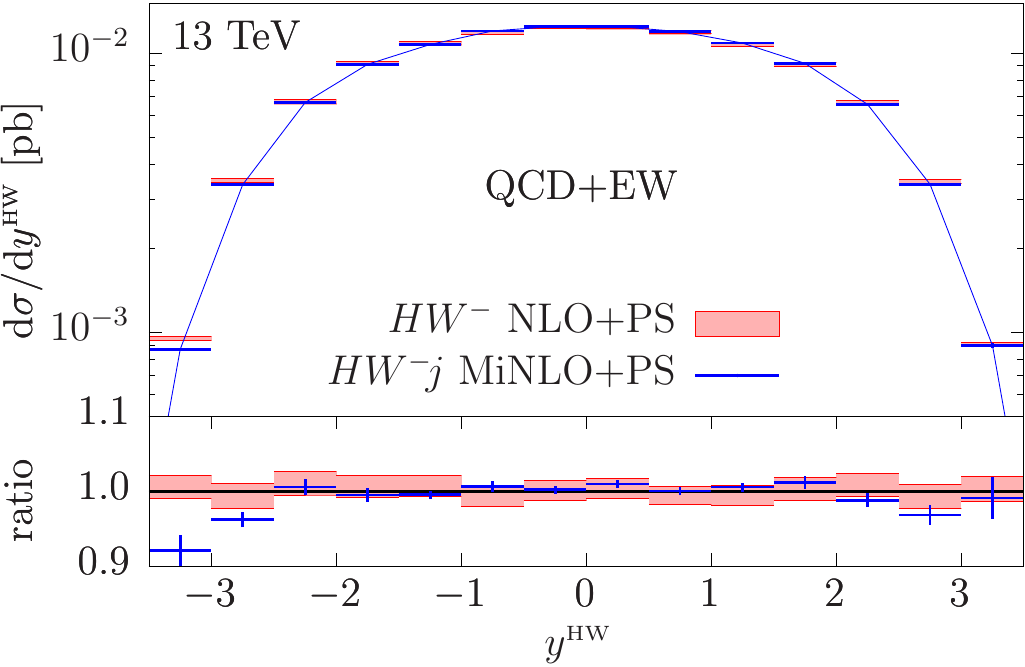}
    \includegraphics[width=\wsmall]{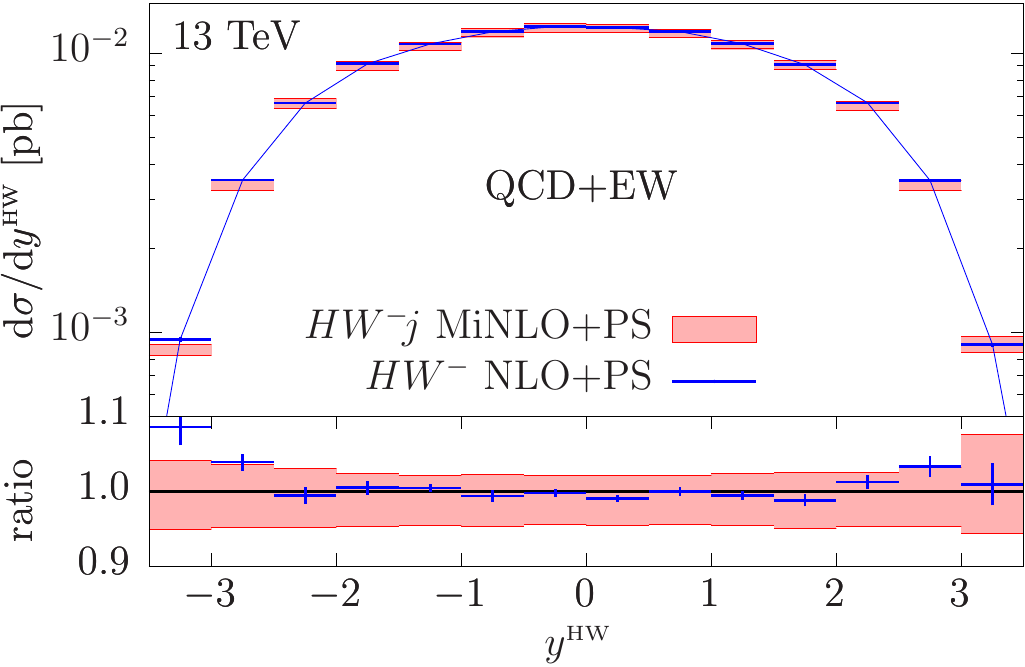}
    \caption{Comparison of NLO+PS and \MINLO+PS predictions for the
      distributions in the rapidity of the \HWm{} system in \HWm{}
      production.  Corrections at NLO QCD+EW are included throughout.  The
      red band is the envelope of the seven-point scale variations for the
      NLO+PS simulation, in the left panel, and for the \MINLO+PS one, in the
      right panel.  The lower panels show the ratio plot with respect to the
      central-scale value of the band.}
    \label{fig:HWvsHWJ-13TeV_HW-y_PY8-bande}
  \end{center}
\end{figure}

\begin{figure}[htb]
  \begin{center}
    \includegraphics[width=\wsmall]{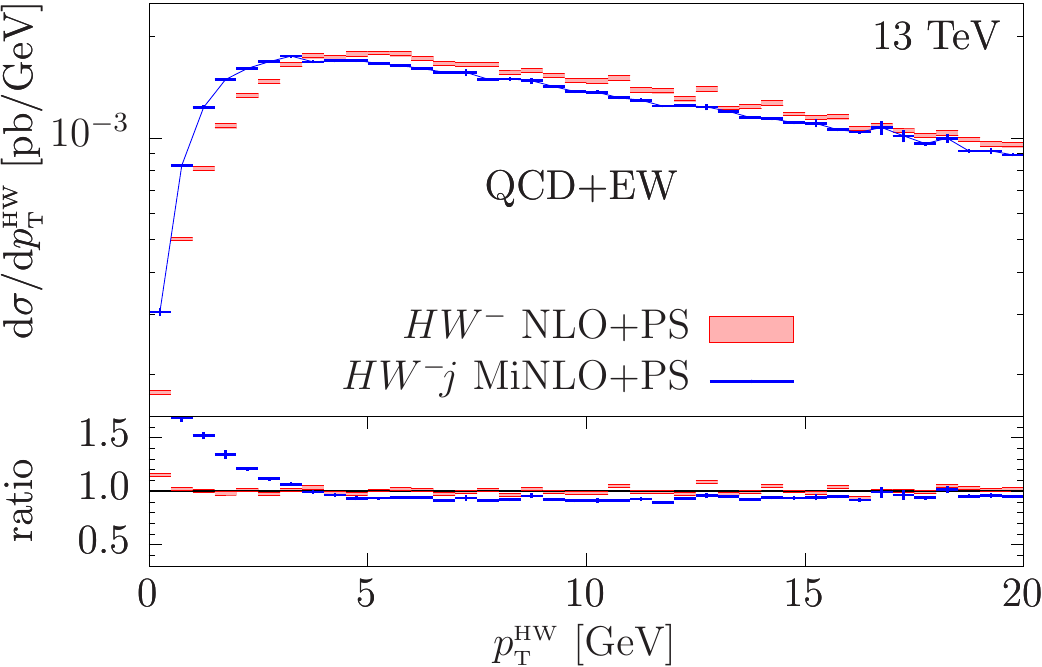}
    \includegraphics[width=\wsmall]{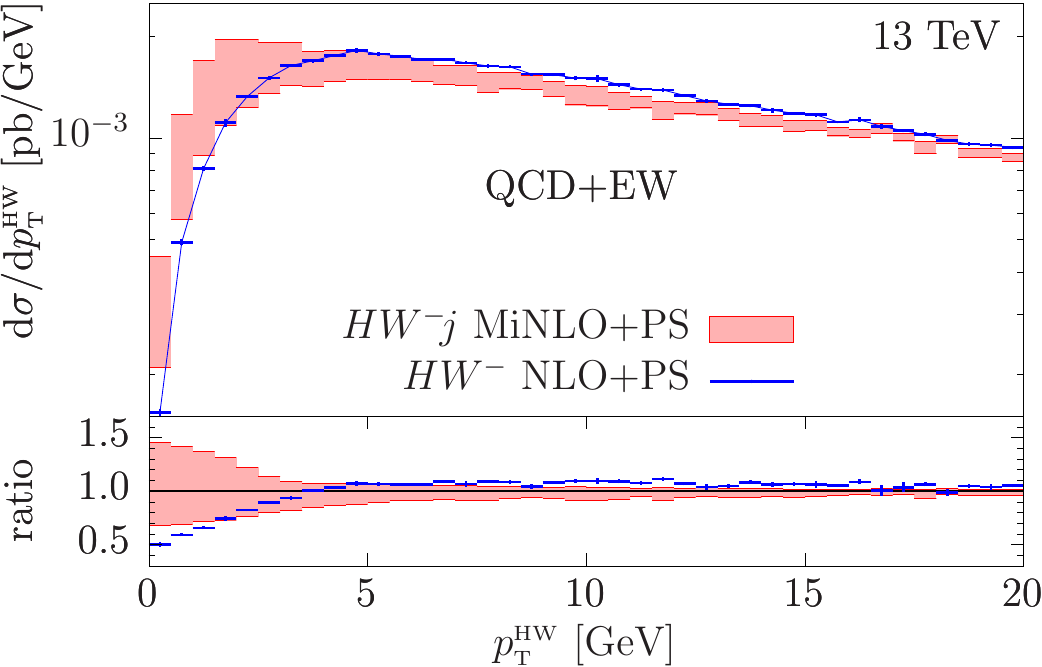}
    \caption{Comparison of NLO+PS and \MINLO+PS predictions for the
      distributions in transverse momentum of the \HWm{} system in \HWm{}
      production. Same curves and labels as in
      Fig.~\ref{fig:HWvsHWJ-13TeV_HW-y_PY8-bande}.}
    \label{fig:HWvsHWJ-13TeV_HW-ptzoom_PY8-bande}
  \end{center}
\end{figure}

\begin{figure}[htb]
  \begin{center}
    \includegraphics[width=\wsmall]{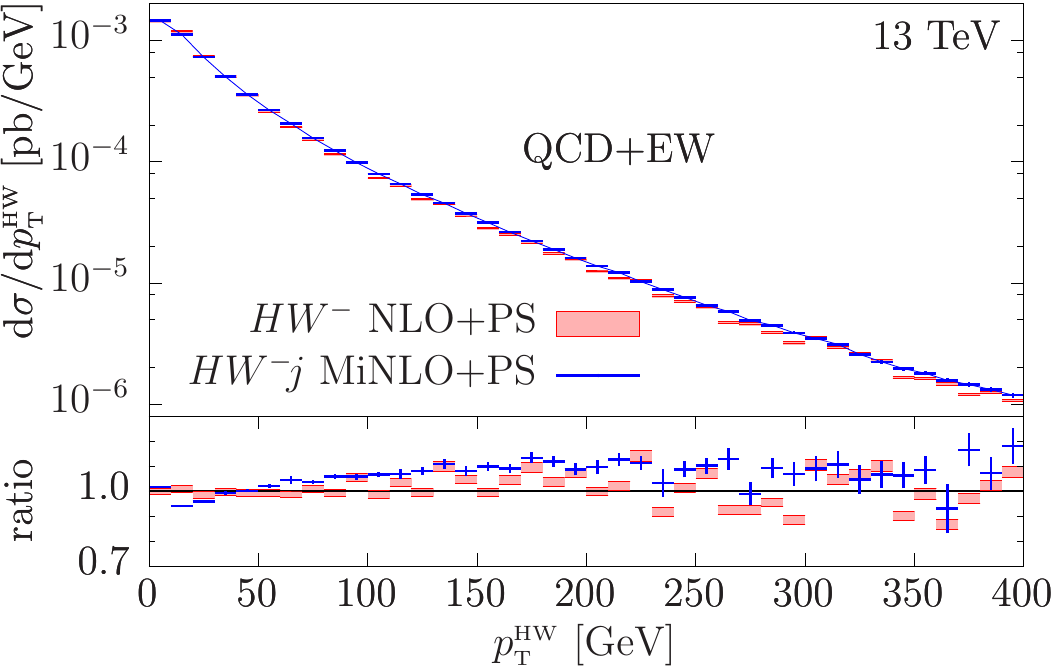}
    \includegraphics[width=\wsmall]{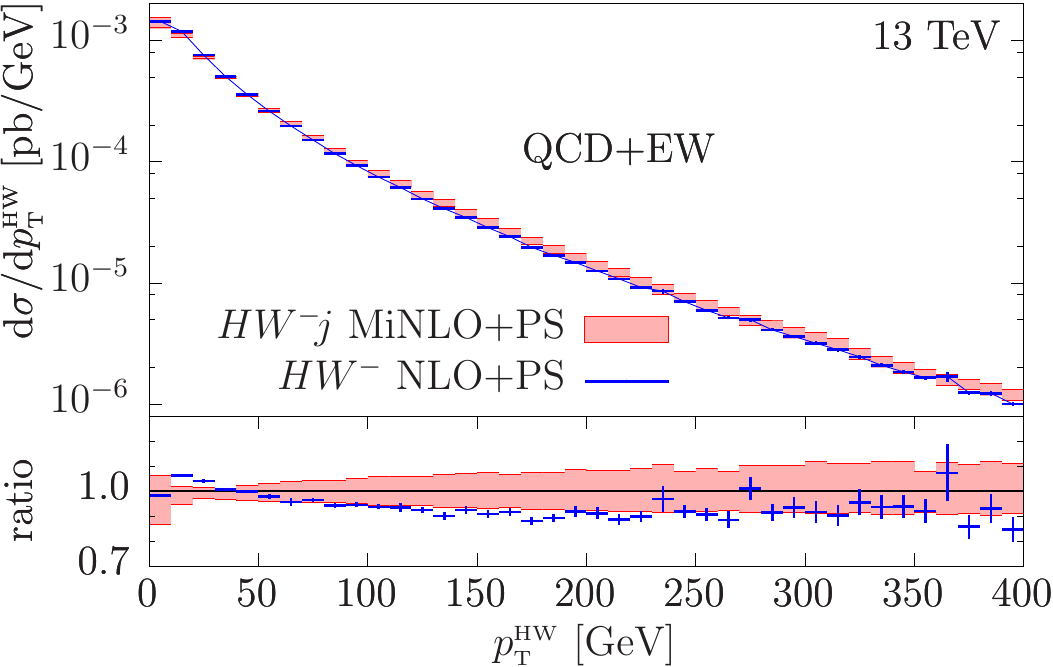}
    \caption{Same comparison as in
      Fig.~\ref{fig:HWvsHWJ-13TeV_HW-ptzoom_PY8-bande}, in a wider $\pt^{\rm
        \sss HW}$ range.}
    \label{fig:HWvsHWJ-13TeV_HW-pt_PY8-bande}
  \end{center}
\end{figure}

Figures~\ref{fig:HWvsHWJ-13TeV_HW-y_PY8-bande}--\ref{fig:HWvsHWJ-13TeV_HW-pt_PY8-bande}
display differential distributions subject to the cuts of
Sec.~\ref{sec:cuts}.  Red bands correspond to scale variations for \HV{} and
\HVJ{} production. We do not show the statistical uncertainties associated to
the integration procedure on these bands, since they are much smaller than
the width of the bands. When plotting instead the blue curves for the
distributions computed at the central scales, we display the statistical
uncertainties of the integration procedure as an error bar.  The plots on the
left-hand side show the uncertainty band for the \HV{} process, while the
ones on the right-hand-side show the uncertainty band for \HVJ{} production.

In Fig.~\ref{fig:HWvsHWJ-13TeV_HW-y_PY8-bande} we display the rapidity
distribution of the \HWm{} system.  Since this inclusive quantity is
predicted at NLO QCD accuracy by both simulations, we find very good
numerical agreement between the two curves at NLO QCD+EW level. The
uncertainty band is larger in the \HWmJ{} case. This is due to the fact that
for \HWm{} production there is no renormalization-scale dependence at LO,
while in \HWmJ{} such dependence is already present at leading order.

In Figs.~\ref{fig:HWvsHWJ-13TeV_HW-ptzoom_PY8-bande}
and~\ref{fig:HWvsHWJ-13TeV_HW-pt_PY8-bande} we compare the transverse
momentum of the \HWm{} pair in two different $\pt$ ranges. Here we observe
significant differences due to the fact that this distribution is only
computed at leading order in the \HWm{} simulation, while it is NLO accurate
in the \HWmJ{} case. Since we included also EW corrections, in our plots
these differences are slightly more pronounced than in the pure QCD
implementation of Ref.~\cite{Luisoni:2013cuh}. The fact that such differences
emerge in the region below the QCD Sudakov peak
(Fig.~\ref{fig:HWvsHWJ-13TeV_HW-ptzoom_PY8-bande}) is consistent with the
observation of enhanced EW effects in that region
(Fig.~\ref{fig:WJ13_j1-pt_PY8_QCD-EW}) as discussed in
Sec.~\ref{sec:EWcorrMINLOPS}.
We also note that the uncertainty band for the \HWm{} generator is smaller
than the \HWmJ{} one. This is due again to the fact that, at Born level,
\HWm{} production does not depend upon $\as$, while \HWmJ{} production does,
and this dependence amplifies the scale-variation band.


\section{Summary and conclusions}
\label{sec:conclusions}

In this paper we have presented the first NLO QCD+EW calculations for \HV{}
and \HVJ{} production, with $V=W^\pm,Z$, at the LHC.  Specifically, we have
considered complete Higgsstrahlung processes corresponding to Higgs boson
production in association with off-shell \leplv{} or \lepll{} leptonic pairs
plus zero or one jet.
In addition to fixed-order predictions we have presented realistic
simulations obtained by combining NLO QCD+EW calculations with a QCD+QED
parton shower.
This was achieved by means of the \RES{} generator, a recent extension of the
\VTWO{} framework, that allows for consistent NLO+PS simulations in the
presence of resonances.
In the case of \HVJ{} production, using the improved \MINLO{} approach, we
have extended the applicability of NLO QCD+EW predictions to the full phase
space, including regions where the hardest jet is unresolved.
This is the first application of the \MINLO{} and \RES{} approaches in
combination with NLO EW corrections.

We have studied several kinematic distributions for \HV{} and \HVJ{}
production in proton-proton collisions at 13~TeV, and we have discussed
predictions at fixed-order NLO, at the level of \RES{} Les Houches events,
and at NLO+PS level using \PythiaEightPone.
Particular care has been taken in combining the QCD+QED shower of
\PythiaEightPone{} with the \POWHEGBOX-generated events, since no standard
interface is available, at present, to deal with multiple NLO emissions that
can arise at production and decay level in resonant processes.

Electroweak corrections typically lower NLO+PS QCD predictions by 5 to 10\%
at the level of integrated cross sections and in angular distributions.  We
have observed quantitatively similar and rather constant EW corrections also
in the jet-$\pt$ spectrum, as well as in the reconstructed $Z$-mass and
transverse $W$-mass in the vicinity of the corresponding resonances.  In
contrast, due to Sudakov logarithms, EW corrections can be much more sizable
in the tails of transverse-momentum and invariant-mass distributions.  For
example, in the Higgs and vector-boson $\pt$ distributions, EW corrections
reach up to $-25\%$ around 1~TeV.  In this respect, the \HV{} and
\HVJ{} Higgsstrahlung processes behave similarly, i.e.~the emission of a jet
does not have a sizable impact on EW corrections.

We have studied theoretical uncertainties associated with standard factor-two
variations of the renormalization and factorization scales.  In the context
of the \POWHEG{} formalism, scale variations are performed only at the level
of the underlying-Born cross section, while the scale of the strong coupling
constant associated with NLO radiation is kept fixed at the corresponding
transverse momentum.  Thus the resulting scale-variation bands are typically
smaller as compared to the ones obtained in fixed-order NLO calculations.  In
the total cross sections for \HV{} and \HVJ{} production we have found scale
uncertainties around 1-2\% and 2-4\%, respectively, while scale variations in
kinematic distributions are typically at the 10\% level.

Thanks to the improved \MINLO{} prescription, simulations based on NLO QCD+EW
matrix elements for \HVJ{} production can be applied to inclusive observables
and compared against more conventional simulations based on NLO QCD+EW matrix
elements for \HV{} production.
At NLO QCD, the observed agreement between \HV{} and \HVJ{} predictions
confirms that, as is well known, the improved \MINLO{} approach guarantees
NLO QCD accuracy also when the extra jet is integrated out.
A similarly good level of agreement was found also at NLO QCD+EW level in a
variety of observables.  In this regard, based on unitarity and factorization
properties of soft and collinear QCD radiation, we have sketched a proof of
the fact that the improved \MINLO{} approach, applied to QCD jet radiation
computed with NLO QCD+EW matrix elements, should provide NLO QCD+EW accuracy
in the full phase space.

All relevant matrix elements at NLO EW have been computed using a recent
interface of the \RES{} framework with the \OpenLoops{} matrix-element
generator.  The other QCD amplitudes have been computed in part analytically
and in part using the standard interface to \MadGraphFour.
We have also presented simple analytic expression that approximate the
virtual EW amplitudes in the Sudakov regime at next-to-leading-logarithmic
accuracy.  This approximation captures the bulk of EW corrections and
reproduces exact NLO EW results with reasonable accuracy.  Moreover it can be
exploited in the combination of the reweighting approach that permits to speed
up NLO QCD+EW simulations while providing full NLO EW accuracy in the final
results.

\vspace{0.5cm}
\noindent
The \RES{} code together with the generators that we have implemented for
\HV{} and \HVJ{} production can be downloaded following the instructions at
the \POWHEGBOX{} web page: {\tt http://powhegbox.mib.infn.it}

\acknowledgments

J.L.~wishes to thank S.~Dittmaier and S.~Kallweit for many detailed answers
on the implementation of EW corrections to Higgsstrahlung processes in \HAWK.
This research was supported in part by the Swiss National Science Foundation
(SNF) under contract PP00P2-128552 and by the Research Executive Agency (REA)
of the European Union under the Grant Agreements PITN--GA--2010--264564 ({\it
  LHCPhenoNet}), and PITN--GA--2012--316704 ({\it HiggsTools}).

\begin{appendix}

\section{Validation of the fixed-order NLO EW corrections in
  $\boldsymbol{\HV{}}$ production } 
\label{sec:validation}

In this section we compare our fixed-order NLO EW predictions for \HW{} and
\HZ{} production with predictions obtained with the Monte Carlo program
\HAWK~\cite{Denner:2014cla}.

\subsection*{Setup for the comparison}

In order to make a comparison between the results generated by \HAWK{} and
our results, we switched off photon-initiated contributions in \HAWK{}, since
these contributions are currently not included in the \RES{} \HV{} generators.
Similarly, $b\bar{b}$-initiated contributions have been discarded in the \RES{},
since this sub-process is not included in \HAWK{}.  The CKM matrix elements
have been set to
\begin{equation}
  \big|\CKM_{ud}\big| = \big|\CKM_{cs}\big| = 0.974\,,  \hspace{2cm}
  \big|\CKM_{us}\big| = \big|\CKM_{cd}\big| = \sqrt{1-\big|\CKM_{ud}\big|^2},
\end{equation}
omitting mixing with third-generation quarks.  The renormalization
and factorization scales are set to the default values used in
\HAWK, i.e.~to the sum of the Higgs and the vector boson masses
\begin{equation}
\mu_{\sss R} = \mu_{\sss F} = M_{\sss V} + \MH, \qquad \qquad V=W,Z\,.
\end{equation}
All other input parameters are chosen in accordance with Sec.~\ref{sec:setup}.

Photons are recombined with collinear charged leptons if $R_{\g \ell} < 0.1$,
where $R_{\g \ell}$ is the angular separation variable in the $(y,\phi)$
plane.  If more than one charged lepton is present in the final state, the
eventual recombination is performed with the lepton having the smallest value
of $R_{\g \ell}$.
After photon recombination, we apply the following cuts on the charged
dressed leptons
\begin{equation}
\pt^\ell > 20\,\GeV{}\,, \qquad  \qquad |y^\ell| < 2.5\,,
\end{equation}
while for \HW{} production we also require a missing transverse momentum of
\begin{equation}
\slashed{E}_{\rm\sss T} > 25\,\GeV{}\,.
\end{equation}

\subsection*{Results}
\begin{figure}[htb]
  \begin{center}
    \includegraphics[width=\wsmall]{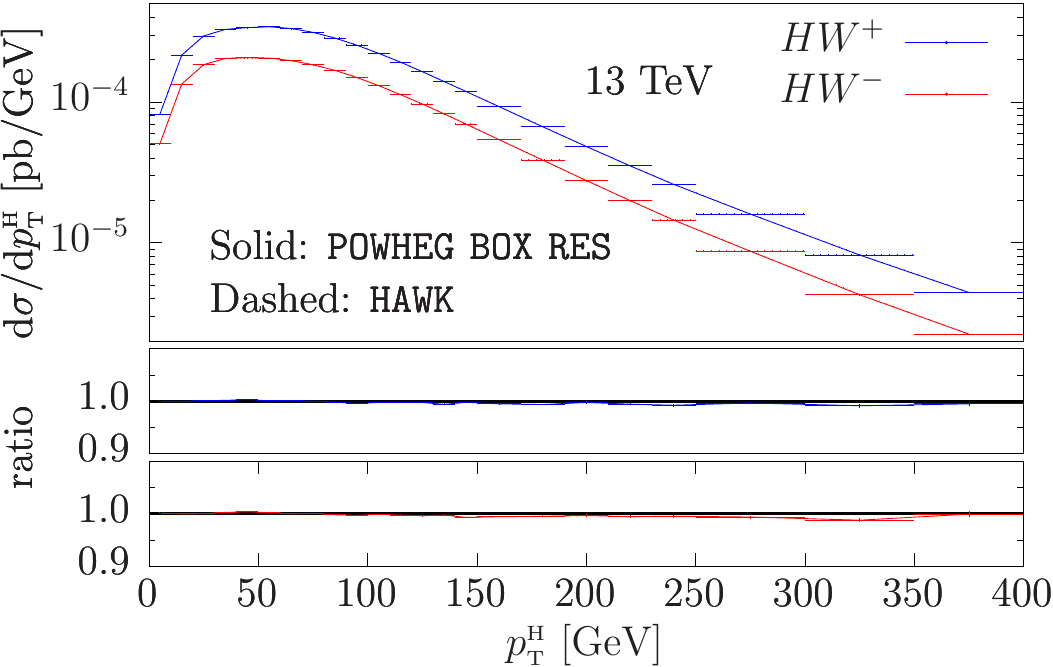}
    \includegraphics[width=\wsmall]{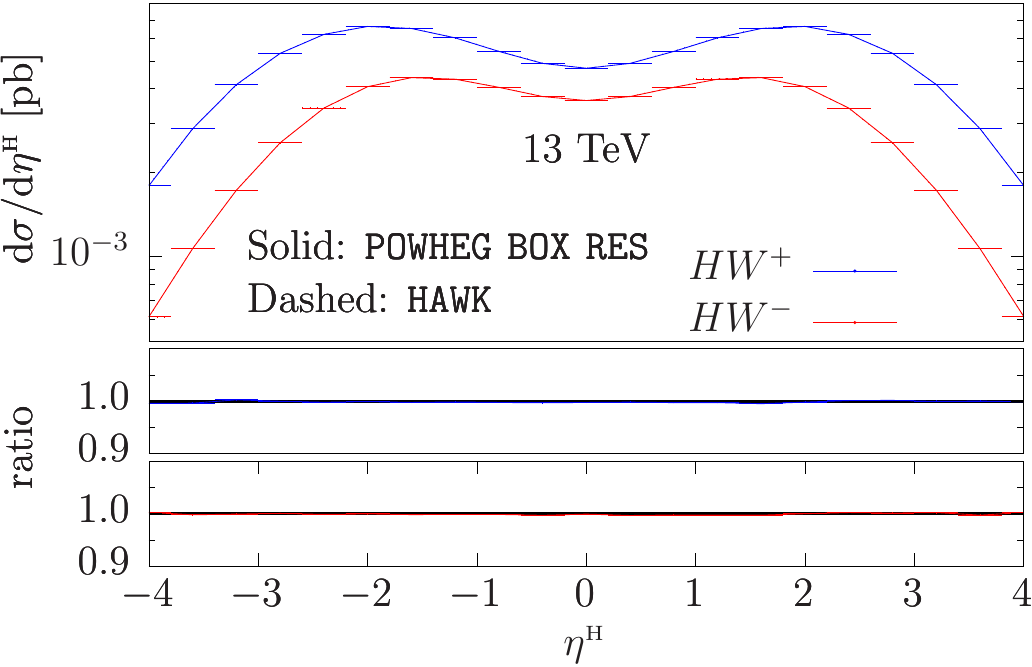}
    \caption{NLO EW predictions for the transverse momentum~(left) and the
      pseudorapidity~(right) of the Higgs boson in \HWpm{} production.  The
      \RES{} and \HAWK{} results are shown with solid and dashed lines
      respectively.  The vertical bars (hardly visible) represent the
      statistical uncertainties associated to the Monte Carlo integration.  }
    \label{fig:hawk_W-H-pt}
  \end{center}
\end{figure}

\begin{figure}[htb]
  \begin{center}
    \includegraphics[width=\wsmall]{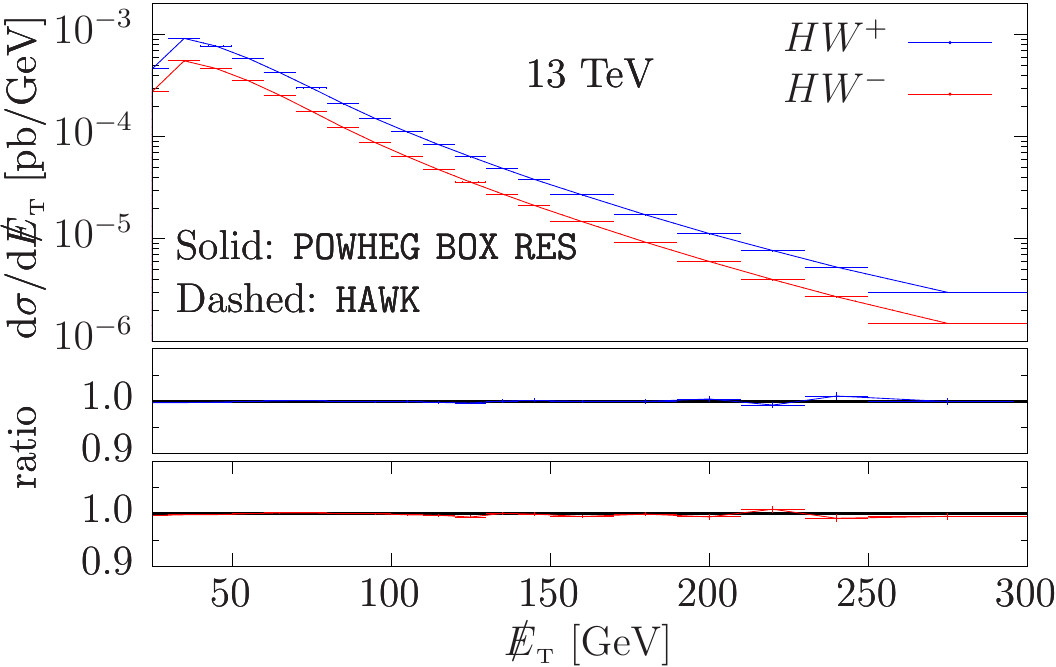}
    \caption{NLO EW predictions for the missing transverse
      momentum. Predictions and labels as in Fig.~\ref{fig:hawk_W-H-pt}.}
    \label{fig:hawk_W-miss-pt}
  \end{center}
\end{figure}

In Figs.~\ref{fig:hawk_W-H-pt} and~\ref{fig:hawk_W-miss-pt} we compare NLO EW
predictions obtained with \RES{}~(solid line) and \HAWK{}~(dashed line) for
selected observables in \HWp{} and \HWm{} production.

Figure \ref{fig:hawk_W-H-pt} displays the Higgs boson transverse-momentum
and pseudorapidity distributions. Within statistical uncertainties the two
predictions fully overlap.
As a further example, in \reffi{fig:hawk_W-miss-pt} we plot the transverse
momentum of the neutrino, i.e.~the missing transverse momentum.  Again, we
observe perfect agreement between the fixed-order NLO \RES{} and \HAWK{}
predictions, and a similar level of agreement was found in all considered
observables.

\begin{figure}[htb]
  \begin{center}
    \includegraphics[width=\wsmall]{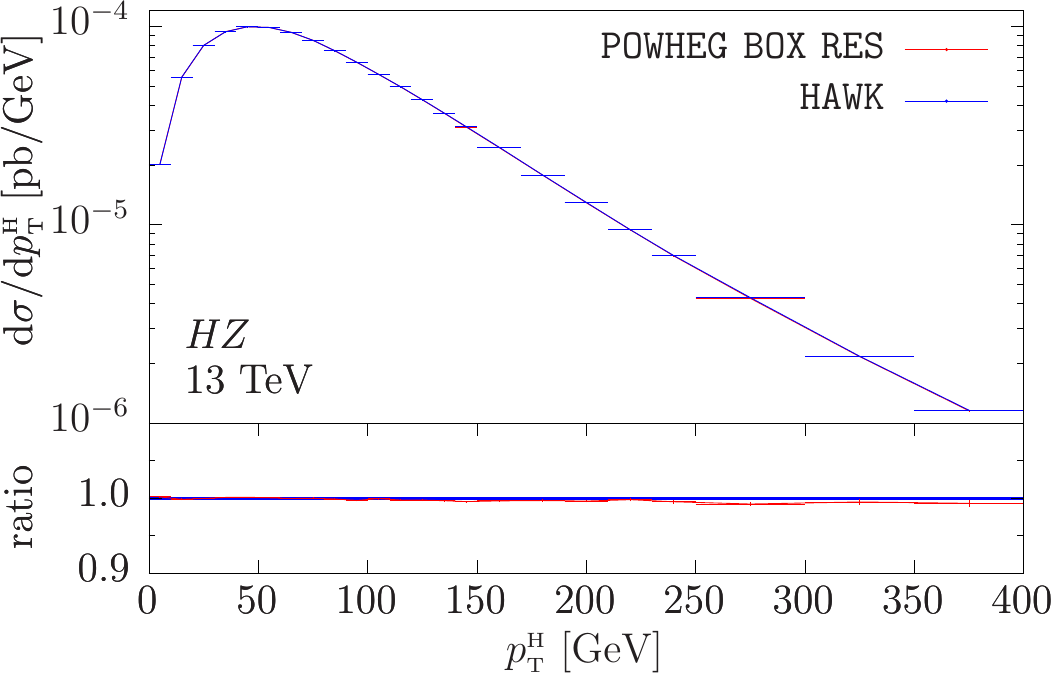}
    \includegraphics[width=\wsmall]{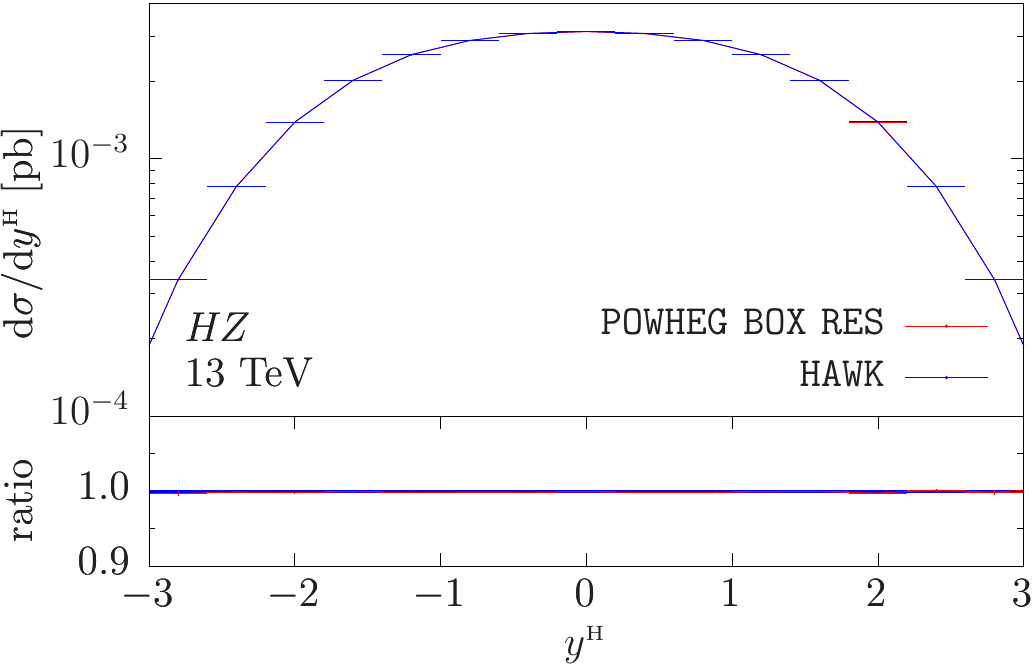}
    \caption{NLO EW predictions for the transverse momentum~(left) and the
      rapidity~(right) of the Higgs boson in \HZ{} production. Comparison
      between the \RES{} and the \HAWK{} predictions.}
    \label{fig:hawk_Z-H-pt}
  \end{center}
\end{figure}

\begin{figure}[htb]
  \begin{center}
    \includegraphics[width=\wsmall]{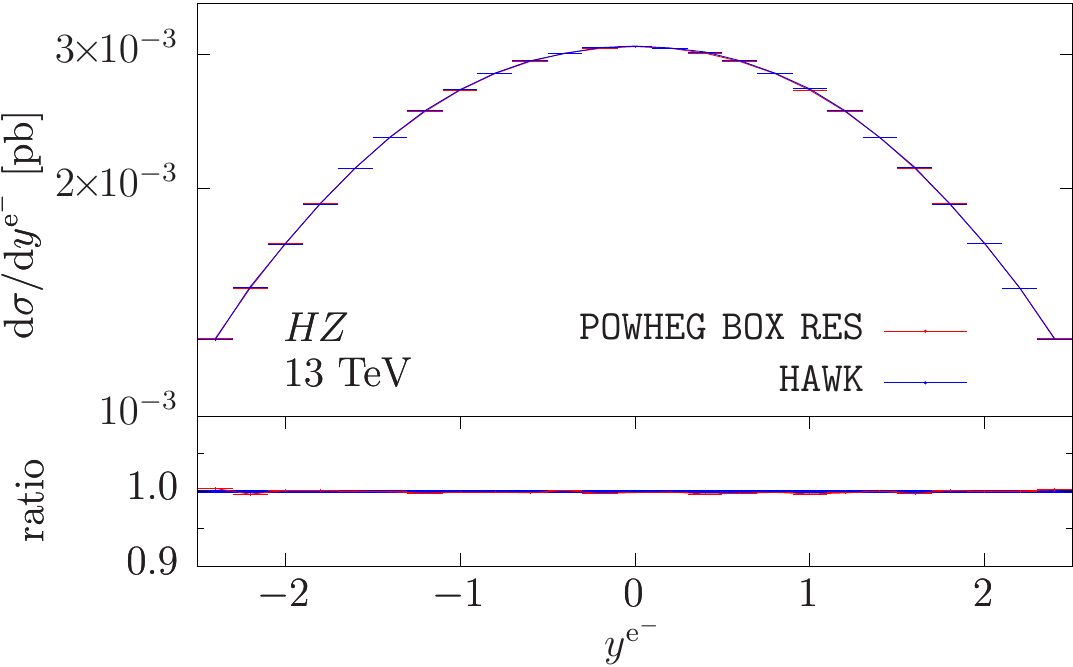}
    \caption{NLO EW predictions for the rapidity of the electron. Same  curves and  labels
      as in Fig.~\ref{fig:hawk_Z-H-pt}.}
    \label{fig:hawk_Z-lept-y}
  \end{center}
\end{figure}

As examples for the validation of \HZ{} production, in
Fig.~\ref{fig:hawk_Z-H-pt} we present the transverse momentum and the
rapidity of the Higgs boson, and in Fig.~\ref{fig:hawk_Z-lept-y} the rapidity
of the produced electron. Again we find a perfect overlap between the \RES{}
and \HAWK{} predictions, and the same level of agreement was found for all
kinematic distributions that we have examined.

\section{The virtual EW Sudakov approximation}
\label{sec:sudakov}
\label{app:sudakov}

The calculation of EW virtual corrections is typically more complex than in
the case of QCD.  This is due to the nontrivial gauge-boson mass spectrum,
the presence of Yukawa and scalar interactions, the fact that EW corrections
enter also in leptonic vector-boson decays, as well as subtleties related to the
treatment of unstable particles.  For these reasons, the numerical evaluation
of NLO EW virtual corrections can be time consuming.  Motivated by this
practical issue, in this appendix we present compact analytic formulas that
provide a decent approximation of the bulk of NLO EW effects, based on the
Sudakov approximation.  Besides speeding up the numerics, this approximation
provides also valuable insights into the origin of the bulk of the EW
corrections.

The largest EW corrections originate in the Sudakov regime, where all
kinematic invariants are of the same order and much larger than the
electroweak scale. In this high-energy regime, the EW corrections are
dominated by Sudakov logarithms~\cite{Sudakov:1954sw, Jackiw:1968zz,
  Ciafaloni:1998xg, Beccaria:1998qe, Fadin:1999bq, Denner:2000jv} of the form
\begin{equation}
\label{eq:Ls_ls}
L(s) = \aofpi\log^2\frac{s}{\MV^2}, \qquad \qquad 
l(s) = \aofpi\log  \frac{s}{\MV^2},
\end{equation}
i.e.~by leading~(LL) and next-to-leading logarithms~(NLL) involving the ratio
of the partonic center-of-mass energy $\sqrt{s}$ to the electroweak-boson
masses, $M_{\sss V}= \MW, \MZ$.  Sudakov EW logarithms originate from virtual
gauge bosons that couple to one or two on-shell external particles in the
soft and/or collinear limits.

General factorization formulas for LLs and NLLs that apply to any Standard
Model process at one loop have been derived in Refs.~\cite{Denner:2000jv,
  Denner:2001gw, Pozzorini:2001rs}.  For a generic $n$-particle scattering
processes with all particles $\f_i$ and momenta $p_i$ incoming\footnote{In
  the following we adopt the notation of Refs.~\cite{Denner:2000jv,
    Denner:2001gw, Pozzorini:2001rs}.}
 \begin{equation}
\f_1 (p_1)\, \f_2(p_2) \ldots \f_n (p_n) \to 0\,,
\end{equation}
high-energy EW logarithms in one-loop matrix elements assume the general factorized form 
\begin{eqnarray}
\label{eq:sudakovstructure}
 \de\M^{\f_1 \ldots \f_n}(\{\lambda_i\},\,p_1 \ldots p_n) &=& \sum_{\lambda_i}
\de\lambda_i \,\frac{\partial\M_0^{\f_1 \ldots
    \f_n}}{\partial\lambda_i}(\{\lambda_i\},\,p_1 \ldots p_n) \nonumber \\
& +& \hspace{-4mm}\sum_{\f_{1'}\ldots \f_{n'}} \hspace{-3mm}\M_0^{\f_{1'} \ldots
  \f_{n'}}(\{\lambda_i\},\,p_1 \ldots p_n)\,\, \de_{\f_{1'}\ldots \f_{n'}}^{\f_1
  \ldots \f_n}(\{\lambda_i\},\,p_1 \ldots p_n).\phantom{aaa}
\end{eqnarray}
Here the first term is related to the running $\de\lambda_i$ of the
dimensionless coupling parameters in the Born amplitude, while the second
term consists of process-independent correction factors $\de_{\f_{1'}\ldots
  \f_{n'}}^{\f_1 \ldots \f_n}$ that contain all LL and NLL terms and multiply
Born matrix elements for the process at hand.  Note that the correction
factors are matrices in ${\rm SU(2)}$ space.  In general they act on one or
two external particles, requiring the evaluation of ${\rm SU(2)}$-transformed
matrix elements $\M_0^{\f_{1'} \ldots \f_{n'}}$.

The logarithmic EW corrections of Eq.~(\ref{eq:sudakovstructure}) can be
schematically split into five contributions
\begin{equation}
\label{eq:corr_fac}
\de\M = \big(\de^{\sss\rm LSC} + \de^{\sss\rm SSC, n} + \de^{\sss\rm SSC, \pm} +
\de^{\sss\rm C} + \de^{\sss\rm PR} \big) \, \M_0 .
\end{equation}
The first three terms are due to double logarithms originating from
soft-collinear gauge bosons exchanged between pairs of external legs. This
gives rise to angular-independent LLs proportional to $L(s)$ ($\de^{\sss\rm
  LSC}$) and subleading angular-dependent logarithms of type
$l(s)\log(|r_{kl}|/s)$, with $r_{kl}=(p_k + p_l)^2$.  The latter are split
into terms associated with neutral~($\de^{\sss\rm SSC, n}$) and
charged~($\de^{\sss\rm SSC, \pm}$) soft-collinear gauge bosons.  The
remaining terms consist of single-logarithmic contributions from
soft/collinear gauge bosons coupling to single external legs~($\de^{\sss\rm
  C}$) and from the usual renormalization-group evolution of dimensionless
coupling parameters~($\de^{\sss\rm PR}$).

In the following, the general results of Refs.~\cite{Denner:2000jv,
  Denner:2001gw, Pozzorini:2001rs} are applied to Higgsstrahlung processes.

\subsection[NLL Sudakov approximation for $\HV$ and
  $\HVJ$ production]{NLL Sudakov approximation for $\boldsymbol{\HV}$ and
  $\boldsymbol{\HVJ}$ production}

In this section we discuss the application of the Sudakov approximation 
to resonant processes.
We focus on vector-boson decays to leptons of the first generation, but our
results are applicable also to $\mu$ and $\tau$ leptons.
With $u$ and $d$ we denote generic up- and down-type quarks, with no
assumptions on their generation, unless specified.

For the leading-order kinematics of \HV{} and \HVJ{} production at particle level we use
the notation
\begin{equation}
\label{eq:qq-Hl1l2}
P_1(p_1) \, P_2(p_2) \to H(p_3) \,V(k)\, \Big( P_6(p_6) \Big) \to H(p_3)
\,\ell_1(p_4)\, \bar{\ell}_2(p_5) \Big( P_6(p_6) \Big),
\end{equation}
where $P_i = q, \bar{q}, g$ are generic partons, and $k=p_4+p_5$ is the
off-shell momentum of the decaying vector boson.  In \HV{} production, $P_6$
is not present, and the two incoming partons are always a quark-antiquark
pair, while for \HVJ{} production an extra gluon can appear, both as an
initial-state parton or in the final state.

In order to apply the Sudakov approximation of Refs.~\cite{Denner:2000jv,
  Denner:2001gw,Pozzorini:2001rs}, the Higgsstrahlung
processes~(\ref{eq:fullprocs}) need to be factorized into separate parts
associated with the production and decay of the vector boson. This is
achieved in a gauge-invariant way by using the leading-pole
approximation~(LPA)~\cite{Beenakker:1998gr, Denner:2000bj}, which corresponds
to the leading term of a systematic expansion in $\Gamma_{\V} /M_{\V}$.  At
leading order, the LPA for Higgsstrahlung processes reads
\begin{equation}
\label{eq:M0_LPA}
\M_{0,\,{\rm {\sss LPA}}}^{P_1 {P}_2 \to H \ell_1 \bar{\ell}_2 (P_6)} = \frac{1}{k^2-M_{\sss V}^2 +
  i\,\Gamma_{\sss V} M_{\sss V}} \sum_{\lambda=0,\pm 1} \M_{ 0} ^{P_1 P_2 \to H
   V_\lambda (P_6)} \M_{ 0}^{V_\lambda \to \ell_1 \bar{\ell}_2},
\end{equation}
where factorized matrix elements for vector-boson production and decay on the
r.h.s.~are summed over the physical polarizations of the vector boson.  The
propagator in Eq.~(\ref{eq:M0_LPA}) depends on the off-shell vector-boson
momentum $k$, while, in the matrix elements on the r.h.s., an on-shell
projected momentum $k'$ must be used in order ensure gauge invariance. This
can be achieved with a mapping that, conserving energy and momentum, projects
on shell the $V$ and $H$ momenta and rescales accordingly the momenta of the
decay products.  In our implementation, we employ such a mapping by keeping
fixed the angles formed by the vector boson and by one of the leptons.

In general, in leading-pole approximation, three types of NLO EW corrections
need to be considered: factorizable corrections to the production and decay
parts, and non-factorizable corrections that connect production and decay.
However, the latter are typically quite small~\cite{Beenakker:1997bp,
  Denner:1997ia, Denner:1998rh, Dittmaier:2014qza}. Moreover, vector-boson
decay do not involve Sudakov EW logarithms. Thus, only the production part
receives Sudakov EW corrections, i.e.
\begin{equation}  
\de\M_{\rm \sss LPA}^{P_1 P_2 \to H \ell_1 \bar{\ell}_2 (P_6)} = \frac{1}{k^2-M_{\sss V}^2 +
 i\,\Gamma_{\sss V} M_{\sss V}} \sum_\lambda\de\M^{P_1 P_2 \to H V_\lambda (P_6)}\,
 \M_0^{V_\lambda \to \ell_1 \bar{\ell}_2},
\end{equation}
and $\de\M^{P_1 P_2 \to H V_\lambda (P_6)}$ as well as its decay counterpart
need to be computed for both transversely- and longitudinally-polarized
vector bosons.  In the framework of Refs.~\cite{Denner:2000jv,
  Denner:2001gw,Pozzorini:2001rs}, tree amplitudes with longitudinal vector
bosons need to be related to corresponding amplitudes with Goldstone bosons
using the Goldstone-boson equivalence theorem~\cite{PhysRevD.10.1145,
  Vayonakis1976, CHANOWITZ1985379}
\begin{equation}
\M_0^{V_{\sss L}^{a_1}\ldots V_{\sss L}^{a_m}\, \varphi_1 \ldots \varphi_n} =
\prod_{k=1}^m i^{(1-Q_{\sss V^{a_k}})} \M_0^{\Phi_{a_1}\ldots \Phi_{a_m}\,
  \varphi_1 \ldots \varphi_n} + \mathcal{O}\l(M E^{d-1}\r),
\end{equation}
where $V_{\sss L}^{a_i}$ are the longitudinal gauge bosons, $\Phi_{a_i}$ the
corresponding Goldstone bosons, $\varphi_i$ are the fermions and scalars in
the process, $M$ and $E$ are typical scale masses and energies involved in
the process, $d$ is the mass dimension of the matrix element and $Q_{\sss
  V^{a_k}}$ is the electric charge of the vector boson $V^{a_k}$.

In the following sections we present analytic results for all relevant NLL EW
correction factors.  These formulae contain group-theoretical quantities such
as the electric charge~$Q$ of the external particles, their weak
isospin~$T^a$, or the electroweak Casimir operator~$\Cew$. Their values can
be found in App.~B of Ref.~\cite{Pozzorini:2001rs}. For the sine and cosine
of the Weinberg angle, we use the shorthand $\sw$ and $\cw$, respectively.

Large logarithms of the light-fermion masses do not need to be included since
we use the~$G_\mu$ scheme, which incorporates the running of the
electromagnetic coupling up to the EW scale, and we regularize QED infrared
singularities of virtual type at the EW scale by using an effective photon
mass $\lambda=\MW$. 
This approach effectively corresponds, in logarithmic approximation,
to the combination of virtual EW corrections with the emission of real photons up to transverse momenta 
of the order of $M_W$.

In the framework of the Sudakov NLL approximation, the Sudakov limit is
applied only to virtual EW effects, while real QED radiation is treated
exactly. More precisely, FKS-subtracted real-emission matrix elements are
treated exactly, while only the finite part of the integrated FKS terms,
defined via $\overline{\mathrm{MS}}$ subtraction of the IR poles at the scale
$\mu=\mu_{\sss R}$, is included.
Concerning IR singularities, this $\overline{\mathrm{MS}}$
subtraction is consistent with the cancellation of virtual QED singularities
through the above mentioned $\lambda=\MW$ regularization approach.  However,
as far as QED logarithms are concerned, we
note that we do not apply a fully consistent matching 
of the (regularized) virtual contributions 
to real QED radiation.  In fact, the former are
effectively cut off at the scale $\MW$, while the latter are subtracted at
the scale $\mu=\mu_{\sss R}$.  This implies missing logarithmic terms of
order $\aem \ln(\mu_{\sss R}/\MW)$.  Nevertheless, as demonstrated by our
numerical results, such uncontrolled logarithms do not jeopardize the
accuracy of the Sudakov approximation at high energies.

\subsection[\HW{} and \HWJ{} production]
           {$\boldsymbol{\HW}$ and $\boldsymbol{\HWJ}$ production}
Here we focus on \HWm{} production and we first consider the partonic process
\begin{eqnarray}
 \label{eq:HW_born_1}  
&& d_{\sss L}(p_1)\, \bar{u}_{\sss L}(p_2) \to H(p_3) \,W^-(k) \to H(p_3)\,
 e^-_{\sss L}(p_4) \,\bar{\nu}_e(p_5),
\end{eqnarray}
which involves only left-chiral quarks and leptons. Matrix elements for the
charge-conjugated process
\begin{eqnarray}
\label{eq:HW_born_2}
&& \bar{d}_{\sss L}(p_1) \,u_{\sss L}(p_2) \to H(p_3)\, W^+(k) \to H(p_3)\,
e^+_{\sss L}(p_4)\, \nu_e(p_5),
\end{eqnarray}
as well as crossing-related matrix elements corresponding to permutations of
the initial-state quarks, can be easily obtained from the ones for the
processes~(\ref{eq:HW_born_1}).  For the crossed production process
\begin{equation}
\label{eq:crossedwmproc}
d_{\sss L}(p_1) \, \bar{u}_{\sss L}(p_2) \, H(-p_3)\, W^+_\lambda(-k) \to 0,
\end{equation}
the Born amplitudes in the high-energy limit read
\begin{eqnarray}
\M_0^{d_L \bar{u}_L H W_T^+}&=& \frac{e^2}{\sqrt{2}\sw^2}\,\MW\,\CKM_{ud} 
\,\frac{A_{{\sss T}-}}{q^2}, \\
\M_0^{d_L \bar{u}_L H W_L^+}&=& \M_0^{d_L \bar{u}_L H \phi^+} =
\frac{e^2}{2\sqrt{2}\sw^2}\,\CKM_{ud}\, \frac{A_{{\sss L}-}}{q^2},
\end{eqnarray}
where $q= p_1 + p_2$. Transverse and longitudinal gauge-boson polarization
states are denoted as $\lambda = \pm 1 \equiv T$ and $\lambda = 0 \equiv L$,
respectively, and
\begin{eqnarray}
A_{{\sss T}-} &=& -i \,\bar{v}_{\sss L}(p_2)\g^\mu u_{\sss L}(p_1)\,
\epsilon_\mu^{\sss T}(-k), \\
A_{{\sss L}-} &=& -i \,\bar{v}_{\sss L}(p_2)\g^\mu u_{\sss L}(p_1)\, (-k + p_3)_\mu.
\end{eqnarray}
For the decay of the polarized $W^-$ boson we have
\begin{equation}
\M_0^{W^-_\lambda e^+ \nu_e} = -i \,\frac{e}{\sqrt{2}\sw}\,\bar{u}_{\sss L}(-p_4) 
\g^\mu v_{\sss L}(-p_5) \, \epsilon_\mu^\lambda(k)\,.
\end{equation}
For the crossed \HWmJ{} production process
\begin{equation}
d_{\sss L}(p_1)\, \bar{u}_{\sss L}(p_2) H(-p_3) \,W^+(-k) \, g(-p_6) \to 0
\,,
\end{equation}
the polarized Born amplitudes at high energy read
\begin{eqnarray}
\label{eq:bornwhT}
\M_0^{d_L \bar{u}_L H W_T^+ g}&=& \frac{e^2}{\sqrt{2}\sw^2}\,\MW\,\CKM_{ud} 
\,g_s \, t^a\, \frac{A'_{{\sss T}-}}{q^2}, \\
\M_0^{d_L \bar{u}_L H W_L^+ g}&=& \M_0^{d_L \bar{u}_L H \phi^+} =
\frac{e^2}{2\sqrt{2}\sw^2}\,\CKM_{ud}\, g_s\, t^a\,\frac{A'_{{\sss L}-}}{q^2},
\end{eqnarray}
where $g_s$ is the strong coupling, $t^a$ is the color matrix, and
\begin{eqnarray}
A'_{{\sss T}-} \!\!&=&\! -i\, \bar{v}_{\sss L}(p_2) \!\l[ \g^\mu \frac{\slashed{p}_1 -
\slashed{p}_6}{(p_1-p_6)^2}\, \g^\nu \!+  \g^\nu \frac{\slashed{p}_6 -
\slashed{p}_2}{(p_6-p_2)^2}\, \g^\mu \r]\!u_{\sss L}(p_1) \epsilon_\mu^{\sss T}(-k)\, 
\epsilon_\nu(-p_6)\,, \\[2mm]
A'_{{\sss L}-} \!\!&=&\! -i\, \bar{v}_{\sss L}(p_2) \!\l[ \g^\mu \frac{\slashed{p}_1 -
\slashed{p}_6}{(p_1-p_6)^2}\, \g^\nu \!+  \g^\nu \frac{\slashed{p}_6 -
\slashed{p}_2}{(p_6-p_2)^2}\, \g^\mu \r]\!u_{\sss L}(p_1) (-k+p_3)_\mu\, 
\epsilon_\nu(-p_6)\,. \phantom{aa}
\end{eqnarray}
The gluon-initiated processes can be obtained via appropriate crossing transformations.

\subsubsection*{Sudakov correction factors for $\boldsymbol{\HW}$ production}

In the following, we list explicit results for the various corrections
factors of Eq.~(\ref{eq:corr_fac}) in the case of the \HW{} production
process~(\ref{eq:crossedwmproc}), using the Mandelstam invariants $s = (p_1 +
p_2)^2$, $t = (p_1 - p_3)^2$ and $u = (p_1 - k)^2$.
For transversely polarized $W$ bosons we obtain
\begin{eqnarray}
\de^{\sss\rm LSC} \!\!&=&\! -\, \frac{1}{2}L(s) \l[2\Cew_q +
  \Cew_{\Phi} + \Cew_{\sss W}\r]\! + l(s)\log\frac{\MZ^2}{\MW^2} 
  \l[(\Iz_{d_L})^2 + (\Iz_{\bar{u}_L})^2 + (\Iz_{\sss H})^2 + (\Iz_{\sss W})^2\r]
 \nonumber \\
 && {} 
+ \de^{\sss\rm LSC,h}_{\sss H},\nonumber\\ 
\de^{\sss\rm SSC,n} \! &=& 2\,l(s)\! \l( R_{d_L \sss{W^+}} \log\frac{|u|}{s} - 
R_{u_L\sss{W^+}} \log\frac{|t|}{s} \r),
\nonumber\\
\de^{\sss\rm SSC,\pm}  \!&=& 2 \,l(s)\,\sw \! \l[\log\frac{|t|}{s} \!
  \l(\!\frac{\Iz_{u_L}}{2\cw} - \frac{Q_u}{2\sw} + \frac{\Iz_{d_L}}{\sw^2\cw}\r)
 \!  - \log\frac{|u|}{s} \!\l(\! \frac{\Iz_{d_L}}{2\cw} - \frac{Q_d}{2\sw} +
  \frac{\Iz_{u_L}}{\sw^2\cw}\r) \! \r], \nonumber \\
  \de^{\sss\rm C}\!\! &=&\! l(s) \!\l[3\Cew_q + 2\Cew_{\Phi} +
    \frac{1}{2}b^{\rm ew}_{\sss WW}\! - \frac{3}{4\sw^2}\,
    \frac{m_t^2}{\MW^2} \r]\! 
\nonumber \\
&& {} +  \aofpi \l[\l(\frac{3}{4\sw^2} \frac{m_t^2}{\MW^2} +
  T_{\sss WW} \r) \log \frac{m_t^2}{\MW^2} + \l(\frac{1}{24\sw^2} -
  2\Cew_{\Phi} \r)  \log\frac{\MH^2}{\MW^2}  \r],\nonumber\\
\de^{\sss\rm PR}&=& \aofpi\l[ \frac{5}{12\sw^2}\log\frac{\MH^2}{\MW^2} - 
\l(\frac{9 + 6\sw^2 - 32\sw^4}{18\sw^4} + T_{\sss WW} -
\,\frac{3}{4\sw^2}\,\frac{m_t^2}{\MW^2} \r)\log\frac{m_t^2}{\MW^2} \r]
\nonumber \\
&& {} \hspace{3cm} + l(s)\l(-\,\frac{3}{2}\,b^{\rm ew}_{\sss WW} + 2\Cew_{\Phi} -
  \frac{3}{4\sw^2}\, \frac{m_t^2}{\MW^2}\r),
\label{eq:sudwhT}
\end{eqnarray}
where $L(s)$ and $l(s)$ are defined in \refeq{eq:Ls_ls}, the factors
$\de^{\sss\rm LSC,h}_{\sss H}$ and $T_{\sss WW}$ are defined, respectively, in
Eqs.~(3.26)--(3.29) and~(5.36)--(5.37) of Ref.~\cite{Pozzorini:2001rs}, and
$\Cew_q = \Cew_{d_L} = \Cew_{\bar{u}_L}$.
The coefficient $R_{\phi_1 \phi_2}$ is related to the charge and to
the weak isospin of the scattering particles via 
\begin{equation}
R_{\phi_1 \phi_2} = Q_{\phi_1}Q_{\phi_2} + \Iz_{\phi_1}\Iz_{\phi_2}\,.
\end{equation}
Note that the parameter-renormalization term, $\delta^{\rm \sss PR}$,
receives contributions from the renormalization of the (dimensionless)
electric charge $e$ and Weinberg angle $\theta_{\sss W}$, as well as from the
renormalization of the $W$-boson mass in \refeq{eq:bornwhT}.  This is due to
the fact that Born matrix elements for transversely-polarized vector bosons
are mass suppressed.\footnote{Note that certain aspects of the derivation of
  the general Sudakov EW formulas of Refs.~\cite{Denner:2000jv,
    Pozzorini:2001rs} are not applicable to mass-suppressed
  processes. Nevertheless, as one can verify by comparison against the exact
  EW corrections, such approximate formulae provide a decent
  approximation. This is also due to the fact that, being mass suppressed,
  transversely-polarized contributions have a minor impact at high energies.}

For longitudinally polarized $W$ bosons, we obtain the Sudakov correction factors
\begin{eqnarray}
\de^{\sss\rm LSC} \!\!&=&\!  -L(s) \l(\Cew_q + \Cew_{\Phi}\r) 
 + l(s)\log\frac{\MZ^2}{\MW^2}\big[(\Iz_{d_L})^2 + (\Iz_{\bar{u}_L})^2
 + (\Iz_{\sss H})^2 +  (\Iz_{\phi^+})^2\big] \nonumber  \\
&& {}+ \de^{\sss\rm LSC,h}_{\sss H} 
+ \de^{\sss\rm LSC,h}_{\phi^\pm},\nonumber\\ 
\de^{\sss\rm SSC,n} \!&=&\! \de^{\sss\rm SSC,\pm} \!=\! 
 2\, l(s) \l[\log\frac{|t|}{s} 
\l( i \Iz_{\sss H \chi}\Iz_{d_L} - R_{u_L \phi^+} \r)- 
\log\frac{|u|}{s}\l( i \Iz_{\sss H \chi}\Iz_{u_L} - 
R_{d_L \phi^+} \r)\r], \nonumber\\
\de^{\sss\rm C} \! &=&  l(s)\! \l[3\Cew_q+4\Cew_{\Phi}-\frac{3}{2\sw^2}\, 
  \frac{m_t^2}{\MW^2} \r]
\nonumber \\ 
&&{}+ \aofpi \l[ \frac{3}{2\sw^2}\, 
\frac{m_t^2}{\MW^2} \log \frac{m_t^2}{\MW^2} + \l(\frac{1}{8\sw^2} - 
2\Cew_{\Phi} \r) \log\frac{\MH^2}{\MW^2} \r],
\nonumber\\
\de^{\sss\rm PR} \! &=& \! -\,b^{\rm ew}_{\sss WW} \, l(s)+ \aofpi
\l(\frac{5}{6\sw^2}\log\frac{\MH^2}{\MW^2} - \frac{9 + 6\sw^2 -
  32\sw^4}{18\sw^4}\log\frac{m_t^2}{\MW^2} \r),
\label{eq:sudwhL}
\end{eqnarray}
where the group theoretical quantities involving the charged Goldstone boson
$\phi^-$ arise from the Goldstone-boson equivalence theorem, and the explicit
expression for the SU(2) $\beta$-function coefficient $b^{\rm ew}_{\sss WW}$
can be found in App.~B of Ref.~\cite{Pozzorini:2001rs}.

The correction factors of Eqs.~(\ref{eq:sudwhT}) and~(\ref{eq:sudwhL}) are
equally valid for the $HW^-$ and $HW^+$ production processes in
Eqs.~(\ref{eq:HW_born_1}) and~(\ref{eq:HW_born_2}). Corresponding results for
processes with the initial-state quarks interchanged are easily obtained by
swapping the Mandelstam variables $t$ and~$u$.

\subsubsection*{Sudakov correction factors for  $\boldsymbol{\HWJ}$ production}
The Sudakov correction factors for \HWJ{} production are quite similar to the
ones for \HW{} production. In fact, the presence of an extra
SU(2)$\times$U(1) singlet gluon has only indirect effects of kinematic type
on the Sudakov EW corrections.  In particular, the $\de^{\rm\sss LSC}$,
$\de^{\rm\sss C}$ and $\de^{\rm\sss PR}$ factors of Eqs.~(\ref{eq:sudwhT})
and~(\ref{eq:sudwhL}) are directly applicable to \HWJ{} production without
any modification.  In contrast, the $\de^{\rm\sss SSC,n}$ and $\de^{\rm\sss
  SSC,\pm}$ factors need to be generalized by including extra
angular-dependent logarithms of type $\log(|r_{12}|/s)$ associated with
vector-boson exchange between the initial-state quarks. For $d\bar u\to HW^-$
this kind of logarithms vanishes due to $r_{12}=s$. However, in the case of
$d\bar u\to HW^-g$, they need to be taken into account, since they give rise
to non-vanishing contributions via crossing transformations of the type
$r_{12}\leftrightarrow r_{16}$, which correspond to the case of quark-gluon initial
states.

In the transverse  case ($\lambda=T$) the SSC correction factors become
\begin{eqnarray}
\de^{\sss\rm SSC,n} &=& 
2\, l(s) \l(\log\frac{|r_{1k}|}{s} \, R_{d_L \sss{W^+}} - 
\log\frac{|r_{2k}|}{s}\,R_{u_L \sss{W^+}}
- \log\frac{|r_{12}|}{s}\, R_{d_L u_L} \r), \nonumber\\
\de^{\sss\rm SSC,\pm} &=&
 2\,l(s)\,  \sw \l[\log\frac{|r_{23}|}{s}\l(
  \frac{Q_d}{2\sw} - \frac{\Iz_{d_L}}{2\cw} \r) + \log\frac{|r_{2k}|}{s}\, 
  \frac{\Iz_{d_L}}{\sw^2\cw} \r. \nonumber \\[2mm]
&& \phantom{2\,l(s)\,  \sw \Big[}\l. - \log\frac{|r_{13}|}{s}\l( \frac{Q_u}{2\sw} -
  \frac{\Iz_{u_L}}{2\cw}  \r) - \log\frac{|r_{1k}|}{s} \, \frac{\Iz_{u_L}}{\sw^2\cw}\r], 
\end{eqnarray}
while for longitudinal $W^-$ bosons ($\lambda=L$) they read
\begin{eqnarray}
\de^{\sss\rm SSC,n}\! &=&\! 2\,l(s)\l[ - \log\frac{|r_{12}|}{s}\,  R_{u_L d_L}   
+ \log\frac{|r_{1k}|}{s} \, R_{d_L \phi^+} - \log\frac{|r_{2k}|}{s} \, 
R_{u_L \phi^+} \r.\nonumber\\
&& \phantom{2\,l(s)\Big[}+ \l. i\Iz_{\sss H\chi}\l(\log\frac{|r_{13}|}{s}\, \Iz_{d_L} -
\log\frac{|r_{23}|}{s}\, \Iz_{u_L} + \log\frac{|r_{3k}|}{s}\, \Iz_{\phi^+} \r)
\r], \\
\de^{\sss\rm SSC,\pm}\! &=& \!2\, l(s)  \l[\log\frac{|r_{23}|}{s}\, R_{d_L \phi^+} \!-
\log\frac{|r_{13}|}{s} \, R_{u_L \phi^+}\!  - 
i \Iz_{\sss H \chi}\l(\log\frac{|r_{1k}|}{s}\, \Iz_{u_L} \!-
  \log\frac{|r_{2k}|}{s}\, \Iz_{d_L} \r)\r]\!, \phantom{aa} \nonumber 
\end{eqnarray}
where
\beq
\label{eq:r_ik}
r_{1k}=(p_1-k)^2,\qquad  r_{2k}=(p_2-k)^2,  \qquad r_{3k}=(p_3+k)^2.
\end{equation}
The above correction factors are directly applicable to \HWpJ{} production as
well, while processes with the initial partons exchanged require the swap
$r_{13} \leftrightarrow r_{23}$ and $r_{1k} \leftrightarrow r_{2k}$.

\subsection[\HZ{} and \HZJ{} production]
           {$\boldsymbol{\HZ}$ and $\boldsymbol{\HZJ}$ production}

One of the main differences between the Sudakov EW corrections for \HZ{} and
\HW{} production is due to the fact that $Z$ bosons couple to both left- and
right-handed currents.  As a consequence, for the \HZ{} production process
\begin{equation}
q(p_1)\,\bar{q}(p_2)\to H(p_3)\,  Z(k) \to H(p_3) \, e^+(p_4)\, e^-(p_5)\,,
\end{equation}
the squared Born amplitude in LPA reads
\begin{equation}
\l|\M_{0,{\sss \rm \,LPA}}^{q \bar{q} \to H e^+ e^-}\r|^2 = \frac{1}{(k^2-\MZ^2)^2 +
  \Gamma^2_{\sss Z} \, \MZ^2} \, \sum_{\ka, \, \ka'} \l|\sum_{\lambda}\M_0 ^{q_{\ka}
  \bar{q}_{\ka} \to H Z_\lambda}\,  \M_0^{Z_\lambda \to e_{\ka'}^+ e_{\ka'}^-}\r|^2, 
\end{equation}
where we have explicitly indicated the incoherent sums over the chiralities
of external quarks and leptons ($\ka, \ka'$), as well as the coherent sum over
intermediate vector-boson helicities.
We thus need the Born elements for the production and decay of both
transverse and longitudinal $Z$ bosons, for different fermion chiralities. 

For the crossed process
\begin{equation}
\label{eq:crossedzproc}
q(p_1)\,  \bar{q}(p_2)\,  H(-p_3) \, Z_\lambda(-k) \to 0\,,
\end{equation}
the Born matrix elements in the Sudakov limit read
\begin{eqnarray}
  \M_0^{q_\ka \bar{q}_\ka H Z_T} &=& \frac{e^2\MZ}{\sw\cw}
\,  \Iz_{q_\ka} \, \frac{A_{\sss TZ}^\ka}{q^2}\,, \\
\M_0^{q_\ka \bar{q}_\ka H Z_L} &=& i\,\M_0^{q_\ka \bar{q}_\ka H \chi} =\frac{e^2}{2\sw\cw}
\,\Iz_{q_\ka}\, \frac{A_{\sss LZ}^{\ka}}{q^2}\,.
\end{eqnarray}
Here and in the following we keep track of the quark chirality $\ka=R,L$ in
the group-theoretical quantities, while $q=p_1+p_2$, and
\begin{eqnarray}
A_{\sss TZ}^\ka &=& -i\,\bar{v}_\ka(p_2)\,\g^\mu \,u_\ka(p_1)\,
\epsilon_\mu^{\sss T}(-k)\,, \\
A_{\sss LZ}^\ka &=& -i\,\bar{v}_\ka(p_2)\,\g^\mu \,u_\ka(p_1)\,(-k + p_3)_\mu\,.
\end{eqnarray}
The interchange of the initial-state quarks modifies only the spinor part,
without changing the structure of the matrix element.  The $Z$-decay matrix
element for generic $\lambda$ reads
\begin{equation}
  \M_0^{Z_\lambda e^-_{\ka'} e^+_{\ka'}} = -i\, e\,
  \Iz_{e_{\ka'}}\,  \bar{u}_{\ka'}(-p_5) \,\g^\mu\, v_{\ka'}(-p_4)\, \epsilon_\mu^\lambda(k)\,. 
\end{equation}
For \HZJ{} production the quark-initiated process is given by
\begin{equation}
q(p_1) \,\bar{q}(p_2) \to H(p_3)\, Z(k)\, g(p_6) \to H(p_3)\,
e^+(p_4)\, e^-(p_5)\, g(p_6)\,.
\end{equation}
The matrix elements for the production of a transverse and a longitudinal $Z$
boson are similar to the ones for \HZ{} production, with the insertion of a
gluon
\begin{eqnarray}
\M_0^{q_\ka \bar{q}_\ka H Z_T g} &=& 
\frac{e^2\MZ}{\sw\cw}\, g_s \, t^a\,
\Iz_{q_\ka}\,  \frac{A_{\sss TZ}^{'\ka}}{q^2}\,,  \\
\M_0^{q_\ka \bar{q}_\ka H Z_L g} &=& i\, \M_0^{q_\ka \bar{q}_\ka H \chi g} = 
\frac{e^2}{2\sw\cw}\, g_s \, t^a\,
\Iz_{q_\ka} \, \frac{A_{\sss LZ}^{'\ka}}{q^2}\,,
\end{eqnarray}
where the spinor parts are given by
\begin{eqnarray}
A_{\sss TZ}^{'\ka}  \!&=&  \!-i\, \bar{v}_{\ka}(p_2) \!\l[ \g^\mu \frac{\slashed{p}_1 -
\slashed{p}_6}{(p_1-p_6)^2}\, \g^\nu +  \g^\nu \frac{\slashed{p}_6 -
\slashed{p}_2}{(p_6-p_2)^2}\, \g^\mu \r] \!  u_{\ka}(p_1) \,\epsilon_\mu^{\sss T}(-k)\, 
\epsilon_\nu(-p_6)\,, \\
A_{\sss LZ}^{'\ka} \! &=&  \!-i\, \bar{v}_{\ka}(p_2) \! \l[ \g^\mu \frac{\slashed{p}_1 -
\slashed{p}_6}{(p_1-p_6)^2}\, \g^\nu +  \g^\nu \frac{\slashed{p}_6 -
\slashed{p}_2}{(p_6-p_2)^2}\, \g^\mu \r] \! u_{\ka}(p_1) (-k+p_3)_\mu\, 
\epsilon_\nu(-p_6)\,.\phantom{aa}
\end{eqnarray}
Related amplitudes with an initial-state gluon can be obtained via crossing
symmetry.

\subsubsection*{Sudakov correction factors for $\boldsymbol{\HZ}$ production}

In the following, we present Sudakov EW correction factors for the \HZ{}
production process~(\ref{eq:crossedzproc}) for generic initial-state quark
chirality~($\ka=R,L$) and flavour~($q=u,d$).

For the transverse case they read
\begin{eqnarray}
\de^{\sss\rm LSC} &=& -\,\frac{1}{2}\,L(s)
\l[2\Cew_{q_\ka} +  \Cew_{\Phi} + \Cew_{\sss ZZ}\r] +
 l(s)\log\frac{\MZ^2}{\MW^2} \l[2\l(\Iz_{q_\ka}\r)^2 + 
\l(\Iz_{\sss H}\r)^2 \r]
+ \de^{\sss\rm LSC,h}_{\sss H},\nonumber\\ 
\de^{\sss\rm SSC,n} &=& 0,\nonumber\\
\de^{\sss\rm SSC,\pm}_u &=& \de_{\ka {\sss L}}\, l(s) \,
\frac{F^{\sss\rm SSC}_{\sss T}}{\Iz_{u_\ka}}\,, \nonumber\\
\de^{\sss\rm SSC,\pm}_d &=& -\,\de_{\ka {\sss L}}\, l(s)
\sum_{u_i = u,c} \big|\CKM_{u_i d}\big|^2 \, \frac{F^{\sss\rm SSC}_{\sss T}}{\Iz_{d_\ka}}\,,\nonumber\\
\de^{\sss\rm C} &=& \aofpi \l[\l(\frac{3}{4\sw^2} \,\frac{m_t^2}{\MW^2} + 
T_{\sss ZZ} \r) \log \frac{m_t^2}{\MW^2} + \l(\frac{\MZ^2}{24\sw^2\MW^2} - 
2\Cew_{\Phi} \r)  \log\frac{\MH^2}{\MW^2} \r] \nonumber \\
&& {}+  l(s) \l[3\Cew_{q_\ka} +  2\Cew_{\Phi} + 
\frac{1}{2}b^{\rm ew}_{\sss ZZ} - \frac{3}{4\sw^2}\,  \frac{m_t^2}{\MW^2} \r] , \nonumber\\
\de^{\sss\rm PR} &=& l(s)\l[-b^{\rm ew}_{\sss WW} + \rho_{q_\ka}\frac{\sw}{\cw}\, 
b^{\rm ew}_{\sss AZ} +2\Cew_{\Phi} - \frac{1}{2}b^{\rm ew}_{\sss ZZ} - \frac{3}{4\sw^2}
  \frac{m_t^2}{\MW^2} \r] \nonumber \\
&& {}+  \aofpi \l\{\l(  \frac{5}{6\sw^2} + \frac{5\rho_{q_\ka}}{6\cw^2} - 
  \frac{5\MZ^2}{12\sw^2\MW^2}\r)\log\frac{\MH^2}{\MW^2} \r. \nonumber \\
&& \l.- \l[\frac{9 + 6\sw^2 - 32\sw^4}{18\sw^2}\l(\frac{1}{\sw^2} +
  \frac{\rho_{q_\ka}}{\cw^2} \r)  +T_{\sss ZZ} - \frac{3}{4\sw^2}\,
\frac{m_t^2}{\MW^2}\r] \log\frac{m_t^2}{\MW^2}\r\}. \phantom{aa}
\label{eq:sudzhT}
\end{eqnarray}
Note that the charged-current SSC contributions take a different
form for up- and down-type quarks, and
\begin{equation}
\label{eq:delta_SSC_ZT}
F^{\sss\rm SSC}_{\sss T} = -\,\frac{\cw(1+\cw^2)}{2\sw^3}
\l[\log\frac{|t|}{s} + \log\frac{|u|}{s} \r].
\end{equation}
The PR corrections depend on $T_{\sss ZZ}$, defined in Eqs.~(5.36)--(5.37) of 
Ref.~\cite{Pozzorini:2001rs}, and
\begin{equation}
\rho_{q_\ka} = \frac{Q_q - T^3_{q_\ka}}{T^3_{q_\ka} - Q_q\sw^2}.
\end{equation}
Similarly as for the $W$ case, they receive contributions from the
renormalization of $e$, $\cw$ and $\MZ$. The renormalization of the 
Weinberg angle also affects the couplings~$\Iz_{q_\ka}$ of the quark to the $Z$
boson.

For longitudinally polarized $Z$ bosons we obtain

\begin{eqnarray}
\de^{\sss\rm LSC} &=& -\,L(s)\l[\Cew_{q_\ka} + \Cew_{\Phi} \r] + 2\,l(s)
\log\frac{\MZ^2}{\MW^2}\l[(\Iz_{q_\ka})^2 + (\Iz_{\sss H})^2 \r] 
+ \, \de^{\sss\rm LSC,h}_{\sss H} + \de^{\sss\rm LSC,h}_{\chi},\nonumber\\ 
\de^{\sss\rm SSC,n} &=& 0,\nonumber\\
\de^{\sss\rm SSC,\pm}_u &=& \de_{\ka {\sss L}} \, l(s)\,
\frac{F^{\sss\rm SSC}_{\sss L}}{\Iz_{u_\ka}}, \nonumber\\
\de^{\sss\rm SSC,\pm}_d &=& -\,\de_{\ka {\sss L}}\, l(s)\,
\sum_{u_i} \big|\CKM_{u_i d}\big|^2\, \frac{F^{\sss\rm SSC}_{\sss L}}{\Iz_{d_\ka}},\nonumber\\
\de^{\sss\rm C} &=&\aofpi \l[\frac{3}{2\sw^2}\, \frac{m_t^2}{\MW^2}
 \log\frac{m_t^2}{\MW^2}+\l(\frac{\MZ^2}{8\sw^2\MW^2} - 2\Cew_{\Phi} \r)
 \log\frac{\MH^2}{\MW^2}  \r]  \nonumber \\
&& {} + l(s) \l[3\Cew_{q_\ka} +
  4\Cew_{\Phi} - \frac{3}{2\sw^2}\, \frac{m_t^2}{\MW^2} \r]\,,\nonumber\\  
\de^{\sss\rm PR}  &=& \aofpi \l(\frac{1}{\sw^2} + \frac{\rho_{q_\ka}}{\cw^2} \r)\!
\l[\frac{5}{6} \log\frac{\MH^2}{\MW^2} - \frac{9 + 6\sw^2 -
    32\sw^4}{18\sw^2}\,\log\frac{m_t^2}{\MW^2} \r]
\nonumber\\
 && {} \hspace{2cm} +  l(s)\l[-b^{\rm ew}_{\sss WW} + 
\rho_{q_\ka}\frac{\sw}{\cw}\,  b^{\rm ew}_{\sss AZ} \r],
\label{eq:sudzhL}
\end{eqnarray}
where
\begin{equation}
F^{\sss\rm SSC}_{\sss L} = -\,\frac{\cw}{\sw^3}\l[\log\frac{|t|}{s} +
  \log\frac{|u|}{s} \r].
\end{equation}
The above correction factors~(\ref{eq:sudzhT}) and~(\ref{eq:sudzhL}) are $t
\leftrightarrow u$ invariant and thus directly applicable also to processes
with exchanged initial-state quarks.

\subsubsection*{Sudakov correction factors for $\boldsymbol{\HZJ}$ production}

Similarly as for \HWJ{} production, also in the case of \HZJ{} production
only the SSC correction factors need to be generalized.  For the transverse
case we get
\begin{eqnarray}
 \de^{\sss\rm SSC,n} &=& -2 \, R_{q_\ka q_\ka}\,l(s) 
\log\frac{|r_{12}|}{s}\,,\nonumber\\
\de^{\sss\rm SSC, \pm}_u &=& \frac{l(s)\, \de_{\ka {\sss L}}} {\Iz_{u_L}} 
\l\{- \,\frac{1}{\sw^2} \l[ \frac{1}{2}\sw\cw \!\l(\log\frac{|r_{13}|}{s} +
  \log\frac{|r_{23}|}{s}\r) + \Iz_{d_L}\log\frac{|r_{12}|}{s} \r.\r.\nonumber\\ 
&& {} + \l.\l.\!\!\frac{\cw^3}{\sw}
\l(\log\frac{|r_{1k}|}{s} +  \log\frac{|r_{2k}|}{s}\r) \r] + 
2\cw^3 \l(\frac{Q_u}{\sw} - \frac{\Iz_{u_L}}{\cw} \r)  \log\frac{|r_{3k}|}{s}
\r\}, \nonumber\\
 \de^{\sss\rm SSC,\pm}_d &=& \frac{l(s)\, \de_{\ka {\sss L}}}{\Iz_{d_L}} 
\l\{ \frac{1}{\sw^2} \sum_{u_i}\big|\CKM_{u_i d}\big|^2 \l[ 
\frac{1}{2}\sw\cw \l(\log\frac{|r_{13}|}{s} +   \log\frac{|r_{23}|}{s}\r)
-\Iz_{u_L}\log\frac{|r_{12}|}{s}\r.\r. \nonumber \\
&& {} +\l.\l.\!\! \frac{\cw^3}{\sw}\l(\log\frac{|r_{1k}|}{s} +
\log\frac{|r_{2k}|}{s}\r) \r]  + 2\cw^3 \l(\frac{Q_d}{\sw} -
\frac{\Iz_{d_L}}{\cw} \r)\log\frac{|r_{3k}|}{s} \r\},
\end{eqnarray}
while for the longitudinal case,
\begin{eqnarray}
 \de^{\sss\rm SSC,n} &=& -2\,l(s) \!\l[R_{q_\ka q_\ka} \log\frac{|r_{12}|}{s} +
\l(i \Iz_{\sss H \chi} \r)^2 \log\frac{|r_{3k}|}{s}  \r],\nonumber\\
\de^{\sss\rm SSC,\pm}_u &=& -\,\frac{l(s)\,\de_{\ka {\sss L}}}
{\sw^2 \Iz_{u_L}} \l[ \Iz_{d_L}\log\frac{|r_{12}|}{s}
 + \frac{R_{u_L \phi^-}}{i \Iz_{\sss H \chi}} \log\frac{|r_{3k}|}{s} 
 \r. \nonumber \\
&& {} + \l.\frac{\cw}{2\sw} \l(\log\frac{|r_{13}|}{s} + \log\frac{|r_{1k}|}{s} +
  \log\frac{|r_{23}|}{s} + \log\frac{|r_{2k}|}{s} \r) \r],  \nonumber  \\
\de^{\sss\rm SSC,\pm}_d &=& \frac{l(s)\, \de_{\ka {\sss L}}}
{\sw^2\Iz_{d_L}} \l\{ - \log \frac{|r_{3k}|}{s} \frac{R_{d_L \phi^-}}
{i \Iz_{\sss H \chi}}  + \sum_i |\CKM_{u_id}|^2 
\l[ -\Iz_{u_L} \log\frac{|r_{12}|}{s} \r.\r. \nonumber \\
&&  {} +\l.\l.\!\frac{\cw}{2\sw} 
\l( \log\frac{|r_{13}|}{s} + \log\frac{|r_{1k}|}{s}
 + \log\frac{|r_{23}|}{s} + \log\frac{|r_{2k}|}{s}\r) \r]\r\},
\end{eqnarray}
where $r_{1k}$, $r_{2k}$ and $r_{3k}$ are defined in Eq.~(\ref{eq:r_ik}).
Due to the fact that these expressions are symmetric under the $r_{13}
\leftrightarrow r_{23}$ and $r_{1k} \leftrightarrow r_{2k}$ permutations,
they also hold for processes with exchanged initial quarks.

\section{Fast evaluation of the virtual electroweak corrections}
\label{sec:fast_ew}
%
The evaluation of EW virtual corrections can be relatively time demanding, in
particular for \HVJ{} production, and the \RES{} framework disposes of a few
options to speed up this part of the calculation.

\subsection*{Fixed-order NLO results}
If one is interested in fixed-order results, the \RES{} code can be run twice
in the following way:
\begin{itemize}
\item[-] In the first high-statistics run, the user sets the flag {\tt
  select\_EW\_virt} to 0 in the input file, thus including only the QCD part
  of the virtual contribution, which are fast to be evaluated. This has the
  advantage that the bulk of the inclusive cross section is computed with
  high statistics at a reduced computational cost.
  
\item[-] Then, the code is run again with lower statistics by using the same
  importance-sampling integration grids generated in the first run, and
  computing only the missing EW part of the virtual contributions. This is
  done by setting the flags {\tt virtonly} and {\tt qed\_qcd} to~1 in the
  input file. In this way, if the flag {\tt select\_EW\_virt} is set to~1,
  the complete virtual corrections are included. If instead the user is
  interested in obtaining the Sudakov approximated results, the code can be
  run by setting {\tt select\_EW\_virt} to~2.

\end{itemize}
Finally, the kinematic distributions obtained in the two previous steps
should be combined, by summing them together.
  
\subsection*{Les-Houches-level Monte Carlo events}
If one is interested in generating Monte Carlo events at the Les Houches
level, then the code can be run with some approximation of the virtual
contributions (or even with the virtual corrections set to zero). At the end,
the generated events need only to be reweighted, using the \RES{} reweighting
feature, with the full virtual corrections activated.
In this way, when the event is generated, the use of an approximated virtual
contribution (whose evaluation could be requested several times per event)
considerably speeds up the code. Once the event is generated, the reweighting
procedure calls the full virtual contribution only once per event. This
reduces the running time in a drastic way.

The different options available are the following:
\begin{itemize}
\item[-] The user can generate the events omitting virtual contributions of
  QCD and EW kind.  This is obtained by setting the \RES{} flag {\tt
    novirtual} to~1 in the input file.  Since the inclusive cross section
  used to generate the weight associated to the event is computed without the
  finite part of the virtual corrections, the weight associated with a single
  event can be very different with respect to the weight obtained after the
  reweighting procedure applied on that event. This can give rise to
  statistical fluctuations in the kinematic distributions, that would need a
  higher number of events to be smoothed.

\item[-] The user can generate the events including only the QCD part of the
  virtual contributions, that are quite fast to be evaluated. This is done by
  setting the \RES{} flag {\tt select\_EW\_virt} to 0 in the input file. The
  difference with the previous case is that an important part of the virtual
  corrections is included in the calculation of the inclusive cross section,
  and the results after reweighting tend to be smoother.
  
\item[-] The best option is to include the QCD part of the virtual
  corrections together with their EW NLL approximation. This is achieved by
  setting {\tt select\_EW\_virt} to 2. This option is the one we have used to
  generate the events analyzed in Sec.~\ref{sec:results}. Since the EW NLL
  approximation of the virtual corrections captures most of the dominant
  Sudakov logarithms, running the code with this setting generates events
  whose weight is very similar to the final weight associated to each event
  after reweighting.
\end{itemize}
The default value for the {\tt select\_EW\_virt} flag is~1, which
corresponds to the inclusion of exact virtual EW contributions at all stages.

\section{Interface to \PythiaEightPone{} and the veto procedure}
\label{app:py8_interface}
\label{sec:py8_interface}

In order to generate realistic event samples at NLO+PS accuracy, including
both QCD and QED corrections, the radiation of QCD partons and photons
generated at the LHE level by the \RES{} framework has to be completed by a
Monte Carlo showering program.  This is achieved through a dedicated
interface that feeds the LH events to \PythiaEightPone{}. The initialization
requires the following instructions:
\begin{verbatim}
    pythia.readString("SpaceShower:pTmaxMatch = 1");
    pythia.readString("TimeShower:pTmaxMatch = 1");
\end{verbatim}
Photon radiation off quarks and
leptons is activated with:
\begin{verbatim}
    pythia.readString("TimeShower:QEDshowerByL = on");
    pythia.readString("TimeShower:QEDshowerByQ = on");
    pythia.readString("SpaceShower:QEDshowerByQ = on");
    pythia.readString("TimeShower:QEDshowerByGamma = off");
\end{verbatim}
The last instruction prevents photons from further splitting into
fermion--antifermion pairs.

In our analysis we do not include hadronization or
underlying-event effects, and we consider the Higgs boson as stable.

\subsection*{The veto procedure}
In the following we discuss the veto procedure that is applied in order to
guarantee a consistent combination of QCD and QED radiation generated at LHE
level with subsequent parton-shower emission.  Since we apply the
multi-radiation mode described in Sec.~\ref{sec:RES}, each LH event generated
by the \RES{} framework can be accompanied by both QCD and QED radiation.
Radiation of QCD type arises only at the ``production'' level, while photon
radiation can come both from ``production'' and from the charged leptons that
arise from the decays of $Z$ and $W$ resonances.

For QCD radiation, the standard veto shower implemented in any Monte Carlo
program is used. In practice, the highest transverse momentum of the
radiation (of QCD or QED type) generated at the ``production'' level by the
\RES{} framework is passed to the shower Monte Carlo program through the
variable {\tt scalup}, in the Les~Houches interface.

For what concerns QED radiation, since the Les~Houches interface does not
provide a standard mechanism to veto radiation from resonance decays, we have
implemented a dedicated veto procedure.  The \RES{} events can have up to two
photons at LHE level, one associated with the production part, and one with
the decay part of the process, and the shower Monte Carlo has to be
instructed to veto, separately at the level of production and decay, any
photon with transverse momentum higher than the hardness of the emissions
produced by the \RES{} framework.

To this end, we first scan the Les~Houches event to identify the photons that
have been generated by the \RES{} framework, determining if their mother
belongs to the production or to the decay products.\footnote{In \HVJ{}
  production, photons radiated by the hardest final-state quark, already
  present at the LH event level, are considered as coming from the production
  stage.}
We then shower the event with \PythiaEightPone{}, restricting QCD radiation
by means of the {\tt scalup} variable as discussed above, and identifying the
extra photons produced by the shower algorithm.

For photons that are generated by the shower at the production level we apply
a similar veto procedure as for QCD radiation:

\begin{enumerate}

\item we compute the transverse momentum of each photon produced by the
  shower, at production level, and store its maximum value~$\pt^{\rm max}$
  for the event at hand;

\item if $\pt^{\rm max}$ is greater than {\tt scalup} the event is vetoed.
  This procedure effectively amounts to requiring that, at production level,
  the shower does not generate any QED radiation with transverse momentum
  greater than the radiation of QCD or QED type produced by the \RES{}.

\item Since, in order to ensure momentum conservation, the Monte Carlo
  reshuffling procedure during shower generation slightly modifies the
  momenta of the particles, we also check that $\pt^{\rm max}$ does not
  exceed the hardness of the LH photon after reshuffling. If this happens,
  the event is vetoed.

\end{enumerate}

For photons associated to the resonance decay, we proceed as follows: 

\begin{enumerate}

\item if no photon is present at the LHE level, this means that the \RES{}
  has not been able to generate radiation harder than the minimum value of
  $10^{-3}$~GeV, set as a minimum for the transverse-momentum of photon
  radiation.\footnote{The parameter that fixes the square of this value is
    {\tt rad\_ptsqmin\_em = 1e-6}.}  In this case, any shower QED radiation
  harder than 1~MeV is vetoed in the decay.

\item If instead a photon is already present, we compute its transverse
  momentum with respect to the lepton emitter in the center-of-mass frame of
  the mother resonance, and store this value in $p_{{\sss \rm T},\,{\rm
      rel}}^{\rm max}$.
In \HZ{} and \HZJ{} production, at Les~Houches level, it is not possible to
know if the photon has been emitted by the lepton or by the antilepton, and
$p_{{\sss \rm T},\,{\rm rel}}^{\rm max}$ is set to the minimum value between
the two relative transverse momenta.
We then veto the event if, among the photons produced at decay level, the
maximum relative transverse momentum is greater than $p_{{\sss \rm T},\,{\rm
    rel}}^{\rm max}$.

\end{enumerate}

\end{appendix}

\bibliographystyle{JHEP}

\providecommand{\href}[2]{#2}\begingroup\raggedright\endgroup

\end{document}